\documentclass[11pt,a4paper]{article}
\usepackage[utf8]{inputenc}
\usepackage{verbatim}
\usepackage{cprotect}
\usepackage{float}
\usepackage{amsmath}
\usepackage{graphicx}

\makeatletter

\pdfpageheight\paperheight
\pdfpagewidth\paperwidth

\usepackage{jheppub}

\usepackage[toc,page]{appendix}
\usepackage{todonotes}
\setuptodonotes{color=green!20}

\allowdisplaybreaks

\title{Perturbative Dissipation in Minimal Warm Inflation}

\author[1]{Daniel J. H. Chung}
\author[1]{Nidhi Sudhir}
\author[2,3]{Lian-Tao Wang}
\emailAdd{danchung@wisc.edu}
\emailAdd{kandathpatin@wisc.edu}
\emailAdd{liantaow@uchicago.edu}

\affiliation[1]{Department of Physics, University of Wisconsin-Madison, Madison, WI
53706, USA}
\affiliation[2]{Department of Physics, University of Chicago, Chicago, IL 60637, USA}
\affiliation[3]{Enrico Fermi Institute, Kavli Institute for Cosmological Physics, and Leinweber Institute for Theoretical Physics, University of Chicago, Chicago, IL 60637, USA}

\abstract{
We present an analysis of the perturbative modes of dissipation in
the minimal warm inflation model where an axion-inflaton is coupled
to a gluon thermal bath through the axion-gauge coupling. 
The corresponding friction coefficient, which generically is 
governed by IR physics, for the processes we consider is dominated by Landau damping
of soft space-like gluons and the plasmon decay of hard on shell gluons. 
Since both of these processes are IR-dominated, the 
friction coefficient becomes sensitive to the non-perturbative magnetic 
mass scale $m_{g}^{-1} \sim (\alpha T)^{-1}$ which in turn leads to 
the friction coefficient's overall linear proportionality to $\alpha$, 
which is much less suppressed in $\alpha$ powers compared to the 
well understood term from Chern-Simons (CS) diffusion
($\propto\alpha^{5}$). 
On the other hand, this contribution to the friction coefficient is 
suppressed due to slow roll.  
We show that 
this can be viewed as a suppression of 
local spontaneous CPT violation for derivative couplings in general 
and that time correlations of nonlocal variables which do not fall off at large time separations are what allow an unsuppressed friction coefficient.
}


\makeatother

\begin{document}
\maketitle \flushbottom 

\section{Introduction}

\label{sec:Intro}

A period of exponential expansion in the early universe provides a
simple explanation of many interesting aspects of our universe such
as flatness, homogeneity, and also the origin of structure formation.
The simplest models of such a period of inflation belong to the class
of cold inflation models~\cite{Guth:1980zm,Linde:1981mu}, where
an inflaton field with negligible interactions to other particle fields
rolls on a nearly flat potential maintaining a constant energy density
in the universe. In these models the initial conditions for structure
formation are provided by quantum fluctuations during inflation. Matching
these models to observations, however, generically runs into problems
such as super-Planckian field excursions, large tensor-to-scalar ratio,
and extremely flat potentials. In addition, at the end of inflation
reheating mechanisms should take over, transferring the inflaton potential
energy into Standard Model particles.

Some of these shortcomings are targets of mitigation in the alternative
inflationary scenario of warm inflation~\cite{Berera:1995ie,Berera:1995wh,Berera:1996nv,Berera:2000xe,Kamali:2023lzq}.
In these models, the inflaton is taken to be sufficiently strongly
coupled to other particle fields such that as it rolls down a potential
it dissipates its kinetic energy into radiation. Warm inflation is
differentiated from the cold inflationary scenarios when the energy
density in the radiation formed satisfies $\rho^{1/4}_{r}>H$. In
scenarios where the associated interaction rates are much larger than
the rate of evolution of the background (including the Hubble rate),
the radiation bath thermalizes and the effect of dissipation on the
background inflaton evolution can be approximated by a friction term
$\Gamma\dot{\phi}$. For $\Gamma\gg3H$ in the strong regime of warm
inflation, the background dynamics are dictated by dissipation, whereas
in the weak regime when $\Gamma<3H$, Hubble friction dominates the
dynamics. The advantage of the former regime is that the additional
friction contribution relaxes the flatness constraints and super-Planckian
excursions. Warm inflation also presents a scenario where the inflationary
era changes continuously into the radiation era without requiring
additional mechanisms for energy transfer. Finally, in contrast to
the cold inflationary scenarios, it predicts that the initial conditions
for structure formation were set by classical/thermal fluctuations
rather than quantum ones.

While the idea of warm inflation provides many advantages, model building
has been difficult due to large thermal and quantum corrections to
the inflaton effective potential\cite{Yokoyama:1998ju}. In fact,
in the initial attempts where the inflaton was taken to be coupled
to other scalar fields, friction is a subleading effect compared to
these potential corrections. Consequently, models were developed
aimed at suppressing the the potential corrections \cite{Bastero-Gil:2016qru,BasteroGil:2009ec,Bastero-Gil:2019gao,Moss:2006gt,Berera:2002sp},
two such being the two-stage catalyst models and axion-inflaton models
with pseudoscalar couplings~\cite{Visinelli:2011jy,Mishra:2011ph,Kamali:2019ppi,Ferreira:2017lnd}.

A special case of the latter kind is the minimal warm inflation model
(MWI)~\cite{Berghaus:2019whh,Laine:2021ego} where the axion inflaton
is coupled to a thermal bath of gluons (not of QCD but beyond standard
model physics) through the topological coupling 
\begin{equation}
\mathcal{L}_{\mathrm{int}}\ni\frac{\alpha\phi}{16\pi f}\tilde{G}G=\frac{\phi}{f}\partial_{\mu}K^{\mu}\label{eq:IntVert}
\end{equation}
where $\alpha=g^{2}/4\pi$ and $g$ is the gauge coupling. Since the
above preserves the continuous $\phi$ translation symmetry, the inflaton
mass does not receive any perturbative potential corrections from
this interaction. Additionally, for temperatures much larger than
the confinement scale of the gauge group, the mass generated by non-perturbative
effects decays as a power law and may be neglected~\cite{Gross:1980br,Frison:2016vuc}.
The friction, on the other hand, is commonly understood to have contributions
from non-perturbative hot sphaleron transitions (Chern-Simons diffusion)~\cite{McLerran:1990de,Moore:2010jd,Arnold:1987mh,Laine:2016hma}.
From the interaction vertex in eq.~(\ref{eq:IntVert}) we might also
expect energy loss from the inflaton to be mediated by perturbative
processes since we do expect there to be significant gluon contributions
that do not mediate gauge field topology change that can interact
with $\phi$ (see e.g. \cite{Broadberry:2025hep}).

More specifically, although we know the longitudinal gluon components
are Debye screened, the transverse gluon components can build up in
the IR. In this work we compute this contribution $(\Gamma_{\mathrm{eff,pert}})$
to leading order in $\alpha$ on a slow-rolling background in the
framework of linear response theory. We point out $\Gamma_{\mathrm{eff,pert}}$
can be sensitive to the non-perturbative magnetic mass scale $m^{-1}_{g}\sim(\alpha T)^{-1}$,
and consequently $\Gamma_{\mathrm{eff,pert}}$ can obtain a linear
dependence on $\alpha$ instead of $\alpha^{2}$. Unfortunately, although
$\alpha$ is significantly larger than $\alpha^{5}$ factor that resides
in the Chern-Simons (CS) number diffusion generated friction coefficient,
$\Gamma_{\mathrm{eff,pert}}$ remains subdominant because of the additional
powers of slow-roll parameters involved. Indeed, we will explain how
these additional power of slow-roll is a generic feature of derivatively
coupled $\phi$, locality of perturbative propagators, and CPT representation.
This can be used to show that the expected leading local contribution to the effective equation of motion
is proportional to $\dddot{\phi}$,  not
$\dot{\phi}$. 

The bulk of this paper's technical focus is about using the hard thermal loop (HTL)-resummed
EFT with an IR cutoff valid on energy scales $(\alpha T,T)$ to compute
$\gamma_{3}$ and the resulting effective friction term. To further
contrast the perturbative processes considered in this paper and the
often quoted hot sphaleron contributions to $\gamma_{1}$, we will
also explain how diffusion effects allow $\gamma_{1}$ to be generated
whether or not relevant leading approximation correlators come from
gauge field topology change. We note it is the secular buildup of
multiple scattering that can yield a non-local time propagator effect
acting on nonlocal variables such as $\int d^{3}xK^{0}(t)$. In the
process, we speculate there can be other scattering processes that
diffuse and contribute to $\gamma_{1}$ at a larger value than what
is firmly computed in the technical portion of this paper.

As far as warm inflation relevant magnitudes are concerned, we find
that $\gamma_{3}\dddot{\phi}$ induced friction term is slow-roll
suppressed and smaller by a factor of at least $\sim10^{-6}$ compared
to the sphaleron contributions to $\gamma_{1}\dot{\phi}$ when 
\begin{equation}
H\ll\alpha^{2}T,
\end{equation}
a condition necessary for thermalization. This paper also should serve
as a contribution to the literature that corrects aspects of physics
in the previous literature such as \cite{Mishra:2011ph}. We finally
note that scenarios where the perturbative contributions are more
significant may be realized on backgrounds which are not slowly rolling
and/or by introducing fermions coupled to the gluons through the vector
current~\cite{Broadberry:2025hep,Berghaus:2024zfg,Berghaus:2025wdx,Drewes:2023khq}.

The paper is organized as follows. In section \ref{sec:Effective-friction-is},
we explain the CPT classification of general non-conservative terms
in the effective EOM and use it to show why the friction term $\gamma_{1}$
is generically absent when coming from leading perturbative order
local propagator computation for derivative couplings to a single
power of $\phi$ (in contrast with topological transitions). The sections
following this focus on the details of perturbative non-conservative
physics related to eq.~(\ref{eq:IntVert}). In section~\ref{sec:EffE.O.M}
we derive the effective inflaton equation of motion (EOM), identify
the leading order dissipative terms in it, and express them in terms
of the spectral density of the operator $\tilde{G}^{a}_{\mu\nu}G^{a,\mu\nu}$,
denoted as $\rho_{\tilde{G}G}$. In section~\ref{sec:-Euc-corr}
we detail the computation reducing the perturbative contributions
to $\rho_{\tilde{G}G}$ in terms of the transverse gluon spectral
density $\rho^{T}$, and review the latter's analytic structure. Finally,
in section~\ref{sec:TransGluonSpecFunc} we use the results of these
past two sections to estimate the perturbative contributions to friction
and understand their parametrics. Many technically useful thermal
field theory details are summarized in the appendices. The speculation
regarding scattering induced diffusion is given in Appendix \ref{sec:Diffusion-effects}.
An interesting connection between nonlocal spacetime symmetries and
Abelian Chern-Simons charge is also presented in Appendix \ref{sec:Comparison-to-Dissipation}.

In the following, we use the metric sign convention $(+,-,-,-)$.
Also, we use several different types of self-energies in this paper: 
\begin{enumerate}
\item $\Pi^{R}_{\mathcal{O}}$: Fourier transform of the retarded $\left\langle \mathcal{O}\mathcal{O}\right\rangle $
correlator in Lorentzian spacetime (eq.~(\ref{eq:retarded-corr})) 
\item $\Pi^{E}_{\mathcal{O}}$: Fourier transform of the retarded $\left\langle \mathcal{O}\mathcal{O}\right\rangle $
correlator in Euclidean space (eq.~(\ref{eq:euclidean})). 
\item $\Pi_{\nu\beta}:$ exact Euclidean gluon propagator; $\Pi^{T}$ and
$\Pi^{L}$ are the transverse and longitudinal projection pieces (eq.~(\ref{eq:SlavanovTaylor})). 
\item $\bar{\Pi}_{\mu\nu}$: combined 1-loop gluon Euclidean self-energy
computed using the bare propagators; $\bar{\Pi}^{T}$ and $\bar{\Pi}^{L}$
transverse and longitudinal projection pieces (eq.~(\ref{eq:transverse-long})). 
\item $\bar{\Pi}^{R,T}$: retarded transverse Lorentzian gluon self energy
(eq.~(\ref{eq:RetardedSEYM})). 
\end{enumerate}
We also use the integration shorthand $\int_{q}\equiv\int\frac{d^{3}q}{(2\pi)^{3}}$.

\section{Effective friction is a type of spontaneous CPT violation}

\label{sec:Effective-friction-is}

Before we delve into the perturbative computations,  we would
like to explain a general way to categorize and estimate non-conservative
terms in the effective EOM from the viewpoint of spontaneous CPT violation.
This sketch will shed light on the robustness of the HTL-resummed
thermal field theory computations presented in this paper as well
as understanding friction from a symmetry perspective.

Since an effective EOM can capture friction while actions cannot,
we start with an effective EOM of the form in FLRW spacetime

\begin{equation}
\ddot{\bar{\phi}}+3\frac{\dot{a}}{a}\dot{\bar{\phi}}+\sum_{b=\text{odd}}\gamma_{b}\partial^{b}_{t}\bar{\phi}+V_{\text{eff}}'(\bar{\phi})=0\label{eq:EOM}
\end{equation}
where $\gamma_{b}$ are real numbers with $\gamma_{1}$ being a friction
term and we have restricted ourselves to linear non-conservative terms
to illustrate our point in this section without too much overhead.
This represents the EOM for the $\text{Tr}\left(\rho\phi\right)$
for the renormalized quantum field 
\begin{equation}
\phi(x)=\bar{\phi}(t)+\delta\phi(x)
\end{equation}
\begin{equation}
\mathcal{L}=\frac{1}{2}\left(\partial\phi\right)^{2}-V(\phi)+\mathcal{L}_{1}(\phi,A)+\mathcal{L}_{2}(A)+\mathcal{L}_{\text{ct}}\label{eq:Lagrangian}
\end{equation}
\begin{equation}
\rho=\text{``free" theory thermal equilibrium density matrix}\label{eq:free}
\end{equation}
with $\mathcal{L}_{1}(\phi,A)$ being interactions of $\phi$ with
other fields $A$, $\mathcal{L}_{\text{ct}}$ being counter terms,
and $V'(0)=0$ (assumed in this section only without much loss of
generality for notational and conceptual convenience). Eq.~(\ref{eq:EOM})
is supposed to be derived from 
\begin{equation}
\ddot{\bar{\phi}}+3\frac{\dot{a}}{a}\dot{\bar{\phi}}+\partial^{2}_{\phi}V(0)\bar{\phi}=\text{Tr}\left(\rho_{I}U^{\dagger}_{I}\Omega_{I}U_{I}\right)\label{eq:interactionRHS}
\end{equation}
with 
\begin{equation}
\bar \phi(t) = \text{Tr} \left(\rho_{I}U^{\dagger}_{I}\phi_{I}(x)U_{I}\right), \quad \text{Tr}\left(\rho_{I}U^{\dagger}_{I}\delta\phi_{I}(x)U_{I}\right)=0
\end{equation}
where $\Omega_{I}$ are the set of interaction operators coming from
$V(\phi)$, $\mathcal{L}_{1}$, $\mathcal{L}_{2}$, $\mathcal{L}_{\text{ct}}$,
and the $I$ subscript represents the interaction picture perturbative
fields, and $U_{I}$ are the usual propagators 
\begin{equation}
U_{I}(t)=T\left(\exp\left[-i\int^{t}_{t_{0}}dt'H_{I}(t')\right]\right)
\end{equation}
where we have split the Hamiltonian as 
\begin{equation}
H_{\text{tot}}=H_{0}+H_{I}
\end{equation}
in the interaction picture with $H_{0}$ preserving CPT and time-translation
invariance. The ``free'' density matrix $\rho$ in eq.~(\ref{eq:free})
allows chemical potentials for charges defined with free particles.
In addition, we will only be concerned with leading order effects
here such that we drop dissipation corrections, for example, terms of order $\alpha/M_{P}$,  to the Hubble expansion rate from these interactions. 

As is usual, invariance of actions turn into tensor maps in EOM. For
example, we know 
\begin{equation}
\int dtd^{3}x\phi^{2}(t,\vec{x})\overset{T}{\rightarrow}\int dtd^{3}x\phi^{2}(-t,\vec{x})=\int dtd^{3}x\phi^{2}(t,\vec{x})
\end{equation}
after a change of variables which makes $T$ an invariance in actions.
In an EOM, $\phi(t,\vec{x})$ is a component of a tensor representation
of $T$: 
\begin{equation}
\square\phi(t,\vec{x})\rightarrow\square\phi(-t,\vec{x})\label{eq:tensor}
\end{equation}
which shows the transformation in the space $\{\phi(t,\vec{x}),\phi(-t,\vec{x})\}$.
Although we can belabor this point with more notation, we will mod
out the transformation in the argument and simply call $\phi$ a $T$-even
field. Similarly, we can say $\partial_{t}\phi$ is $T$-odd.

Note that what people usually call the friction term $\gamma_{1}\dot{\bar{\phi}}$
is CPT-odd while $3H\dot{\bar{\phi}}$ is CPT-even, 
if we treat $a(t)$ and $\bar{\phi}(t)$ not as spontaneous symmetry
breaking objects but as fields in a field equation such as 
\begin{equation}
\left(\frac{\dot{a}}{a}\right)^{2}=\frac{\frac{1}{2}\dot{\bar{\phi}}^{2}+V(\bar{\phi})+...}{3M^{2}_{P}}\label{eq:Hubble-expansion}
\end{equation}
and eq.~(\ref{eq:EOM}), which transform under CPT. Spontaneous symmetry breaking
of the quantum theory occurs once one restricts the classical background
$\bar{\phi}(t)$ to either a single classical solution or a family
of solutions that do not span the full space of solutions allowed
by the original field equations. We consider CPT instead of $T$ because
$G\tilde{G}$ (which can be in $\Omega_{I}$ and $H_{I}$) is $T$-odd
but CPT-even. This means that $G\tilde{G}$ and $3H\dot{\bar{\phi}}$
terms which can come from an action with conservative dynamics are
CPT-even.\footnote{For example a simple $G\tilde{G}\in\mathcal{L}$ without the $\phi$
coupling is $T$ odd but certainly will not generate friction to leading
order in interactions.} By contrast, the dissipative term $\gamma_1 \dot{\bar \phi}$ is CPT-odd. It cannot be derived from a local action, as expected from the CPT invariance of the underlying conservative theory.
Indeed, that is one of the main lessons: 
\begin{equation}
\boxed{\text{CPT, not \ensuremath{T}, selects non-conservative terms induced by a time-dependent \ensuremath{\phi(t)}.}}
\end{equation}
Since the $\gamma_{b}\partial^{b}_{t}\bar{\phi}$ for $b\in\mathbb{Z}_{\text{odd}}$
is coming from $\text{Tr}\left(\rho_{I}U^{\dagger}_{I}\Omega_{I}U_{I}\right)$
of eq.~(\ref{eq:interactionRHS}), we require $\text{Tr}\left(\rho_{I}U^{\dagger}_{I}\Omega_{I}U_{I}\right)$
is spontaneously CPT violating due to classical backgrounds. Note
in making these statements, it is important to recognize that $\gamma_{b}$
does not transform under CPT. However, since 
\[
\gamma_{1}\dot{\bar{\phi}}\in\text{Tr}\left(\rho_{I}U^{\dagger}_{I}\Omega_{I}U_{I}\right)
\]
$\gamma_{1}\dot{\bar{\phi}}$. The term $\gamma_1 \dot{\bar \phi}$ 
 depends on \emph{nonlocal} functionals of the background history $\bar{\phi}(t')$
 contained in $U_I$, because the interaction-picture evolution operator involves a time-ordered integral over the interaction Hamiltonian at all earlier times $t'<t$.

After writing $\phi=\bar{\phi}(t)+\delta\bar{\phi}$, suppose all
explicit time-dependences of $H_{I}$ come from $\bar{\phi}(t)$ (e.g.~neglect
$\dot{a}$): 
\begin{align}
H_{I}(t') & =\sum_{jb}(t'-t)^{b}\frac{\partial^{b}_{t}\bar{\phi}^{j}(t)}{b!j!}\left(\partial^{j}_{\bar{\phi}}H_{I}(\delta\bar{\phi}(t'))\right)_{\bar{\phi}=0}
\end{align}
will generate the time dependence in $U_{I}$ leading to 
\begin{align}
\text{Tr}\left(\rho_{I}U^{\dagger}_{I}\Omega_{I}U_{I}\right) & =\sum^{\infty}_{n=0}i^{n}\sum_{j_{n}b_{n}...j_{1}b_{1}}\partial^{b_{n}}_{t}\bar{\phi}^{j_{n}}...\partial^{b_{1}}_{t}\bar{\phi}^{j_{1}}\int^{t}_{t_{0}}dt_{1}\cdots\int^{t_{n-1}}_{t_{0}}dt_{n}\frac{(t_{n}-t)^{b_{n}}}{b_{n}!j_{n}!}...\frac{(t_{1}-t)^{b_{1}}}{b_{1}!j_{1}!}\nonumber \\
 & \left\langle [\left(\partial^{j_{n}}_{\bar{\phi}}H_{I}(\delta\phi(t_{n}))\right)_{\bar{\phi}=0},\cdots,[\left(\partial^{j_{1}}_{\bar{\phi}}H_{I}(\delta\phi(t_{1}))\right)_{\bar{\phi}=0},\Omega_{I}(t)]\cdots]]\right\rangle 
\end{align}
in addition to any additional time-dependences in $\Omega_{I}$.\footnote{Here the expectation value is obviously the trace over $\rho$.}
One way CPT violation arises can be from integrating out the time
through ignoring tails of correlations, leaving $\partial^{b_{1}}_{t}\bar{\phi}^{j_{1}}$
to transform as $(-1)^{b_{1}}$.\footnote{As a side technical note, it is important to remember that for $\mathcal{L}_{I}$
with derivative couplings, $H_{I}\neq-\int d^{3}xa^{3}\mathcal{L}_{I}$,
but one can typically use such identification to leading order in
$H_{I}$ insertion.} Note eq.~(\ref{eq:EOM}) tells us 
\begin{equation}
\frac{d}{dt}\left(\frac{1}{2}\dot{\bar{\phi}}^{2}+V_{\text{eff}}(\bar{\phi})\right)+3H\dot{\bar{\phi}}^{2}=-\dot{\bar{\phi}}\sum_{b}\gamma_{b}\frac{d^{b}}{dt^{b}}\bar{\phi}
\end{equation}
which means as far as the energy $\dot{\bar{\phi}}^{2}/2+V_{\text{eff}}(\bar{\phi})$
is concerned, only $\gamma_{1}$ is manifestly dissipative. The other
$\gamma_{b}$ coefficients can be dissipative depending on the particular
boundary conditions chosen for $\bar{\phi}(t)$ similar to the Abraham-Lorentz
force in electrodynamics. All $\gamma_{b}$ terms represent back reaction
in interacting with the thermal bath $\rho$. Some $\gamma_{b}$ can
vanish quite generically depending on the symmetry of the coupling.
We will illustrate this in the next couple of subsections.

\subsection{Linear $\phi$ coupling}

Here, we will show linear coupling of the form 
\begin{align}
H_{I}(t') & =\int d^{3}x\phi h_{I}+...=\bar{\phi}(t')\int d^{3}xh_{I}+\int d^{3}x\delta\phi h_{I}+...
\end{align}
where $h_{I}$ is independent of $\phi$ is of interest in this paper
(see eq.~(\ref{eq:IntVert})) will yield a particularly simple expression
for $\gamma_{b}$. With a single insertion of $H_{I}$, eq.\,(\ref{eq:interactionRHS})
has 
\begin{align}
\text{Tr}\left(\rho_{I}U^{\dagger}_{I}\Omega_{I}U_{I}\right) & =\text{Tr}\left(\rho_{I}\Omega_{I}\right)+i\int^{t}_{t_{0}}dt_{1}\langle[\int d^{3}x\delta\phi h_{I},\Omega_{I}(t)]\rangle\nonumber \\
 & +i\sum_{b}\partial^{b}_{t}\phi(t)\int^{t}_{t_{0}}dt_{1}\frac{(t_{1}-t)^{b}}{b!}\langle[\int d^{3}xh_{I},\Omega_{I}(t)]\rangle+...
\end{align}
Note if $h_{I}$ and $\Omega_{I}(t)$ have no field-independent time
dependences, time translations are \emph{generated by} $H_{0}$, and
as long as 
\begin{equation}
[H_{0},\rho_I]=0\,,\label{eq:requirement}
\end{equation}
then we can use the time-translation representation to write 
\begin{align}
-i\int^{t}_{t_{0}}dt_{1}\frac{(t_{1}-t)^{b}}{b!}\langle[\int d^{3}xh_{I}(t_{1},\vec{x}),\Omega_{I}(t)]\rangle & =-i\int^{t}_{t_{0}}dt_{1}\frac{(t_{1}-t)^{b}}{b!}\langle[\int d^{3}xh_{I}(t_{1}-t,,\vec{x}),\Omega_{I}(0)]\rangle.
\end{align}
This means if $\langle[\int d^{3}xh_{I}(t_{1}-t,\vec{x}),\Omega_{I}(0)]\rangle$
only has support in a time interval 
\begin{equation}
|t_{1}-t|\lesssim\xi
\end{equation}
then as long as $t_{0}\ll t$, this integral will be insensitive to
$t_{0}$ and $t$:  
\begin{equation}
-i\int^{t}_{t_{0}}dt_{1}\frac{(t_{1}-t)^{b}}{b!}\langle[\int d^{3}xh_{I}(t_{1},\vec{x}),\Omega_{I}(t)]\rangle\approx-i\int^{0}_{-\infty}dz\frac{z^{b}}{b!}\langle[\int d^{3}xh_{I}(z,\vec{x}),\Omega_{I}(0)]\rangle\label{eq:simplified}
\end{equation}where $z\equiv t_1-t$.
Hence, we have converted a function that depended on $t_{0}$ and
$t$ to one that does not by restricting the functional space of $\bar{\phi}(t)$
that is slowly varying such that 
\begin{equation}
\left|\frac{\xi^{b}}{b!}\partial^{b}_{t}\bar{\phi}(t)\right|\gg\left|\frac{\xi^{b+1}}{b+1!}\partial^{b+1}_{t}\bar{\phi}(t)\right|
\end{equation}
if the Taylor expansion is truncated at order $b$. Combining this
with the selection rule of CPT violation, we have arrived at selecting
the non-conservative terms in the EOM being 
\begin{equation}
\text{Tr}\left(\rho_{I}U^{\dagger}_{I}\Omega_{I}U_{I}\right)\ni-\sum_{b=\text{odd}}\partial^{b}_{t}\bar{\phi}(t)\gamma_{b}\label{eq:result}
\end{equation}
where 
\begin{equation}
\boxed{\gamma_{b}\equiv-i\int^{0}_{-\infty}dz\frac{z^{b}}{b!}\langle[\int d^{3}xh_{I}(z,\vec{x}),\Omega_{I}(0)]\rangle}\label{eq:gammab}
\end{equation}
with $b$ sum being truncated to leading order in practice. This is
interesting because by using time-translation invariance and correlation
function support in the \emph{approximate} computation of the projection,
we have eliminated CPT transformation properties that is representationally
equivalent to a spontaneous CPT violation.\footnote{The CPT even objects would amount to a renormalization of the effective
theory of $\phi(t)$.}

It is straightforward to show that 
\begin{equation}
\gamma_{b}=\frac{-i}{b!}\left[\left(-i\partial_{\omega}\right)^{b}I(\omega)\right]_{\omega=0}\label{eq:derivative}
\end{equation}
where 
\begin{equation}
R(\omega)+iI(\omega)\equiv-i\int^{\infty}_{-\infty}dze^{-i\omega z}\Theta(-z)\langle[\int d^{3}xh_{I}(z,\vec{x}),\Omega_{I}(0)]\rangle\label{eq:fourier}
\end{equation}
and the isolation of $\gamma_{b}$ into the imaginary part $I(\omega)$
is by design of including a step function $\Theta(-z)$ (manifestly enforcing causality) in the Fourier
transform, analogous to how the absorptive part arises in S-matrix
computations. Hence, from a particle picture perspective, these terms
arise from properties of long wavelength states as expected from the
fact that the thermal bath is exchanging energy with effectively homogeneous slow varying
$\phi(t)$. This is where we expect $h_{I}$ and $\Omega_{I}$ being
in the gauge sector to shine in enhancing non-conservative terms since
there can be pileup of IR modes.

We will use this standard Fourier space method to compute $\gamma_{1,3}$
in the sections below, comprising the main work of this paper. In
particular, we will be using HTL-resummed EFT to accurately account
for the IR physics in the range $\left(\alpha T,T\right)$. Before
turning to this task, we will show in the next subsection that for
systems that possess continuous field translation symmetry (i.e. derivative
coupling), $\gamma_{1}$'s leading perturbative contribution from
eq.~(\ref{eq:gammab}) is zero unless there are obstructions to dropping
boundary terms. Such obstructions arise for example when $\langle[\int d^{3}xh_{I}(z,\vec{x}),\Omega_{I}(0)]\rangle$
is diffusively supported, as we will explain.

\subsection{Derivative coupling}

Consider $b=1$ of eq.~(\ref{eq:gammab}): 
\begin{equation}
\gamma_{1}=-i\int^{0}_{-\infty}dzz\langle[\int d^{3}xh_{I}(z),\Omega_{I}(0)]\rangle.\label{eq:gamma1beg}
\end{equation}
with a divergence term for $h_{I}(w)$ 
\begin{equation}
h_{I}(x)=-\partial_{\mu}J^{\,\mu}(x)\label{eq:hispec}
\end{equation}
corresponding to a derivative coupling $\mathcal{L}\ni\phi\partial_{\mu}J^{\mu}$
and 
\begin{equation}
\Omega_{I}(0)=\partial_{\mu}J^{\mu}(0,\vec{y}).\label{eq:omispec}
\end{equation}
Note that this coupling is perturbatively symmetric under 
\begin{equation}
\phi\rightarrow\phi+c
\end{equation}
for a continuous parameter $c$ (which can be embedded into PQ symmetry
for the case of axion). It is straight forward to show if one freely
integrates by parts dropping all boundary terms 
\begin{align}
\gamma_{1} & =-i\int^{0}_{-\infty}dx^{0}\langle\int d^{3}x[J^{0}(x^{0},\vec{x}),\partial_{\mu}J^{\mu}(0,\vec{y})]\rangle\label{eq:firststep}\\
 & =i\int^{0}_{-\infty}dt\frac{\partial}{\partial x^{0}}\mathcal{Y}(x^{0})\\
 & =i\left(\mathcal{Y}(0)-\mathcal{Y}(-\infty)\right)\label{eq:answer}
\end{align}
where 
\begin{equation}
\mathcal{Y}(x^{0}-y^{0})\equiv\langle[\int d^{3}xJ^{0}(x),J^{0}(y^{0},0)]\rangle.
\end{equation}
We see 
\begin{equation}
\mathcal{Y}(0)=\langle[\int d^{3}xJ^{0}(0,\vec{x}),J^{0}(0,0)]\rangle=0
\end{equation}
vanishes by causality for spacelike separations since $J^{0}$ is
Hermitian and the time variable here is taken to be strictly zero
(thereby avoiding Schwinger terms) while we know $\mathcal{Y}(-\infty)=0$
with local physics with a mass gap (all correlations fall off at infinite
time separations), and therefore $\gamma_{1}=0$. Hence, we have found
that $\gamma_{1}$ vanishes as long as long-time-separation correlation
$\mathcal{Y}(t)$ of nonlocal operators vanishes. In other words,
the vanishing property of $\gamma_{1}$ is not a special to $J^{\mu}$
having anything to do with gauge fields. Note that this type of argument
should be useful in other contexts, not just in the linearly derivatively
coupled case.

What about $\gamma_{3}$? Start with eq.~(\ref{eq:gammab}) 
\begin{equation}
\gamma_{3}=i\int^{0}_{-\infty}dz\frac{z^{3}}{3!}\langle[\int d^{3}x\partial_{\mu}J^{\,\mu}(z,\vec{x}),\partial_{\mu}J^{\mu}(0,\vec{y})]\rangle
\end{equation}
and go through similar manipulation to obtain 
\begin{align}
\gamma_{3} & =-i\int^{0}_{-\infty}dzz\mathcal{Y}(z)\label{eq:linearmoment}
\end{align}
Hence, locality still leaves $\gamma_{3}$ in tact since it is sampling
$\mathcal{Y}(z)$ near $z=0$ instead of sampling only beyond the
correlation length. Obviously, higher $b$ objects $\gamma_{b}\partial^{b}_{t}\phi$
will be sensitive to different moments weighted by $\mathcal{Y}(z)$.

One may wonder why for the hot sphaleron case, such a generic argument
as locality can seemingly be violated. For this hot sphaleron case,
we have $J^{\mu}\propto K^{\mu}$, and it is well known that 
\begin{equation}
\lim_{t\rightarrow\infty}\left\langle \left(\int d^{3}xK^{0}(t,\vec{x})-\int d^{3}yK^{0}(0,\vec{y})\right)^{2}\right\rangle \sim\Gamma_{\text{sphal}}Vt\label{eq:diffusion-build}
\end{equation}
which diverges characteristic of diffusion. This remarkable behavior
implies 
\begin{equation}
\mathcal{Y}(-\infty)=\lim_{x^{0}\rightarrow-\infty}\frac{i}{2TV}\frac{d}{dx^{0}}\left(\left\langle \left\{ \int d^{3}xJ^{0}(x),\int d^{3}yJ^{0}(y^{0},0)\right\} \right\rangle \right)=\frac{i\Gamma_{\text{sphal}}}{2T}
\end{equation}
giving support to eq.~(\ref{eq:answer}), and this can be understood as a result of a combination of secular effects associated with multiple scattering and nonlocal nature of the operator in the correlator.\footnote{Given that this is a multiple scattering effect, our discussion here in the context of the leading operator insertion is only suggestive with appropriate resummation of subdiagram interpretation.} Hence, when $J^{\mu}\propto K^{\mu}$,
this non-decoupling nature of the correlator allows $\gamma_{1}\dot{\phi}$
to be significant. Note in arriving at eq.~(\ref{eq:answer}), we
have neglected all nonlocal effects (e.g.~integrated by parts in
eq.~(\ref{eq:firststep}) and threw out boundary terms on both timelike
and spacelike boundaries, took finite correlation length argument
to throw out tails of integrals, and used the perturbative behavior
$\mathcal{Y}(-\infty)=0$). In Appendix \ref{sec:Diffusion-effects},
we will redo the computation without making time integration by parts
assumptions and show that the same result can be obtained. Additionally,
there it will be argued that ordinary CS number violating scattering
without any topological changes ($\Delta N_{CS}=O(\alpha)$) might contribute competitively with
the $\Gamma_{\text{sph}}$ to $\gamma_{1}$: 
\begin{equation}
\gamma^{\text{scatter}}_{1}\sim\frac{\alpha^{4}T^{3}}{2f^{2}}\left(\frac{n_{g}}{T^{3}}\right)^{2}\left(\frac{\left\langle \sigma_{2\rightarrow3}v\right\rangle T^{2}}{\alpha^{3}}\right)\label{eq:scattergam1}
\end{equation} 
where $n_g$ is the gluon number density and $\langle \sigma_{2\rightarrow3}v\rangle$ is the thermally averaged helicity changing two to three scattering cross section.
This estimate assumes sustained diffusion of the relevant nonlocal charge and negligible correlations between successive scatterings. A first-principles justification of these assumptions is beyond the scope of this work.

This work can thus be seen as filling in the perturbative physics
gap of \cite{Berghaus:2019whh}. What our explicit computations will
lead to is a better understanding of what is required for scattering
to generate $\gamma_{1}$ and the linear moment of $\mathcal{Y}(z)$
given by eq.~(\ref{eq:linearmoment}) will pick up time dependence
contribution from Landau damping and plasmon modes (both of which
are IR sensitive due to the lack of perturbative mass gap for transverse
modes in Yang-Mills), giving 
\begin{equation}
\gamma_{3}\sim-\frac{\alpha}{4\pi^{2}}\frac{T}{f^{2}}\frac{N^{2}_{c}-1}{24}\times\text{function of }N_{c}\text{ and IR cutoff}
\end{equation}
(for $SU(N_{c})$) which is much less suppressed as far as $\alpha$
power is concerned compared to the hot sphaleron contribution's $\alpha^{5}$.
Nonetheless $\gamma_{3}$ has a subdominant effect compared to $\gamma_{1},$
because of the slow-roll nature of $\dddot{\phi}$ multiplying $\gamma_{3}$.
Note that naive dimensional analysis of eq.~(\ref{eq:linearmoment})
would have said $\gamma_{3}$ should behave as 
\begin{equation}
\gamma^{\text{\ensuremath{\text{naive}}}}_{3}\sim\frac{-4\alpha^{2}T\left(N^{2}_{c}-1\right)}{\left(16\pi f\right)^{2}}
\end{equation}
which shows that the IR effects computed in this paper give an enhancement
of order O($1/\alpha$).\footnote{The factor of 4 would be from naive multiplication of polarization.
The minus sign would naively be from 2 additional powers of $z$ compared
to $\gamma_{1}>0$ and noting that would correspond to a frequency
space derivative of $(-i\partial_{\omega})^{2}$ shown in eq.~(\ref{eq:derivative})
with the Fourier convention of $\exp(i\omega t)$ of eq.~(\ref{eq:fourier}).}

Compared to the work of \cite{Broadberry:2025hep}, our scenario starts
the density matrix in equilibrium (e.g.~eq.~(\ref{eq:requirement})).
For example, their equation 3.10 shows the damping rate controlled
by $\alpha\dot{\phi}/(fT)-\mu_{\text{dist}}/T$ where the $\mu_{\text{dist}}$
is a time-evolving chemical potential of the thermal bath that is
initially out of equilibrium. Hence, although they also consider the
perturbative effects, they seem to be on a different thermodynamic
trajectory as they are emphasizing the effects of the chemical potential
while we are treating that effect as a correction to our computation.
Furthermore, the kinetic theory employed there is explicitly valid only for modes with spatial momentum $p \equiv |\bf p| > \sqrt{\alpha} T$, where $\sqrt{\alpha} T$ is identified parametrically with the inverse mean free path. By contrast, our HTL-resummed EFT extends to softer momentum scales that are particularly important for the response to a slowly varying, approximately homogeneous $\phi(t)$:
\begin{enumerate}
\item $p\in(\alpha T, \sqrt{\alpha}T \ll \Lambda_{HTL} \ll T)$ for transverse mode propagators 
\item $p\in(0, \sqrt{\alpha}T \ll \Lambda_{HTL} \ll T)$ for longitudinal mode propagators
\end{enumerate}
and this leads to important IR effects for the $\dddot{\phi}$ coefficient
which \cite{Broadberry:2025hep} does not compute.

\section{Dissipative Terms in the Inflaton EOM}

\label{sec:EffE.O.M}

The minimal warm inflation model described by

\begin{equation}
S\!\left(\phi,A^{a}_{\mu}\right)=\int d^{4}x\sqrt{-g}\,\left(\frac{1}{2}\partial_{\mu}\phi\partial^{\mu}\phi-V\!\left(\phi\right)+\frac{\alpha}{16\pi}\frac{\phi}{f}\tilde{G}^{\mu\nu}_{a}G^{a}_{\mu\nu}-\frac{1}{4}G^{\mu\nu}_{a}G^{a}_{\mu\nu}\right)\label{eq:MWIaction}
\end{equation}
where here 
\begin{equation}
\tilde{G}^{a,\mu\nu}\equiv\frac{\epsilon^{\mu\nu\alpha\beta}G^{a}_{\alpha\beta}}{\sqrt{-g}}
\end{equation}
involves an inflaton coupled to a thermal bath of gluons and perturbative
inflaton modes through the axion-gauge interaction term. We work in
a regime where the background is slow-rolling and the interaction
rates $\sim\alpha^{2}T\gg H$, allowing the system to be close to
equilibrium at all times. The effect of the interactions with the
bath is captured by the effective EOM of the inflaton background.
This is obtained by varying the action in eq.~(\ref{eq:MWIaction}),
re-expressed in terms of $\phi=\bar{\phi}+\delta\phi$ (background
and perturbations), and taking a thermal expectation value of the
resulting operator EOM. Restricting to inflaton potentials of the
form 
\begin{equation}
V(\phi)=\rho_{\Lambda}+a_{1}\phi+a_{2}\phi^{2}
\end{equation}
such that contributions from inflaton self-interactions are absent,
the equation of motion is 
\begin{align}
\ddot{\bar{\phi}}+3H\dot{\bar{\phi}}+V'\!\left(\bar{\phi}\right)-\frac{\alpha}{16\pi f}\langle\tilde{G}^{\mu\nu}G_{\mu\nu}\rangle=0,\label{eq:phibackEOM}
\end{align}
where the last term captures the effect of interactions with the thermal
bath.

The time dependence of $\langle\tilde{G}G\left(t\right)\rangle$ can
be captured using linear response theory (LRT)~\cite{Berera:2002sp,Laine:2016hma,Bellac:2011kqa,Bastero-Gil:2010pb}
which is approximately the same as what was presented in section \ref{sec:Effective-friction-is}.
Here, the bath is taken to be at equilibrium at some time $t_{0}$
such that 
\begin{equation}
\left(\alpha^{2}T\right)^{-1}\ll t-t_{0}\ll H^{-1},\,\left(\dot{\phi}/\phi\right)^{-1}.\label{eq:timeScales}
\end{equation}
The time evolution of the background fields displaces the bath slightly
out of equilibrium. This displacement is then treated as a perturbation
to the state at $t_{0}$, and the time evolution of the thermal correlators
is computed in a perturbative expansion in this displacement. More
explicitly, the action governing the perturbative modes, which may
be obtained from eq.~(\ref{eq:MWIaction}) by expanding around a
background inflaton value, is

\begin{align}
S_{\mathrm{pert}}\!\left[\delta\phi,A^{a}_{\mu}\right] & =\int d^{4}x\,\sqrt{-g}\left[\frac{1}{2}\partial_{\mu}\delta\phi\partial^{\mu}\delta\phi-V''\!\left(\bar{\phi}\right)\frac{\delta\phi^{2}}{2}\right]\nonumber \\
 & \quad-\frac{1}{4}\int d^{4}x\,\sqrt{-g}\,G^{a}_{\mu\nu}G^{a,\mu\nu}+\frac{\alpha}{16\pi f}\int d^{4}x\,\bar{\phi}\,\epsilon^{\mu\nu\alpha\beta}G^{a}_{\mu\nu}G^{a}_{\alpha\beta}\nonumber \\
 & \quad+\frac{\alpha}{16\pi f}\int d^{4}x\,\delta\phi\!\left(\epsilon^{\mu\nu\alpha\beta}G^{a}_{\mu\nu}G^{a}_{\alpha\beta}-\langle\epsilon^{\mu\nu\alpha\beta}G^{a}_{\mu\nu}G^{a}_{\alpha\beta}\rangle\right).\label{eq:Spert}
\end{align}
Here the background EOM has been used to simplify terms linear in
$\bar{\phi}$. For $t\approx t_{0}$, when the system is taken to
be at equilibrium, the density matrix of the thermal bath is 
\begin{equation}
\rho\!\left(t\right)_{\vert t\approx t_{0}}=\frac{1}{\mathcal{Z}}\exp\!\left(-\frac{H_{0}}{T}\right)\label{eq:density_Mat_t_0}
\end{equation}
with $H_{0}$ the Hamiltonian for $t\approx t_{0}$, that is, the
Legendre transform of the Lagrangian density in eq.~(\ref{eq:Spert})
with background functions evaluated at $t=t_{0}$. When the background
evolves to a later time $t$, the Lagrangian density can be split
into a bare part for $t\approx t_{0}$ and a perturbation dependent
linearly on the displacements $\delta a\!\left(t\right)=a\!\left(t\right)-a\!\left(t_{0}\right)$
and $\delta\bar{\phi}\!\left(t\right)=\bar{\phi}\!\left(t\right)-\bar{\phi}\!\left(t_{0}\right)$.
Here, given eq.~(\ref{eq:timeScales}), the displacements are small,
satisfying $\delta a/a,\,\delta\bar{\phi}/\bar{\phi}\ll1$. Thus,
the free Lagrangian for the fluctuations can be written as 
\begin{align}
\mathcal{L}_{\mathrm{pert}}\!\left(\delta\phi,A^{a}_{\mu}\right) & =\sqrt{-g\left(t_{0}\right)}\left(\frac{1}{2}\partial_{\mu}\delta\phi\partial^{\mu}\delta\phi-\frac{1}{4}G^{a}_{\mu\nu}G^{a,\mu\nu}\right.\nonumber \\
 & \quad+\frac{\alpha}{16\pi f}\frac{\bar{\phi}\!\left(t_{0}\right)}{\sqrt{-g\left(t_{0}\right)}}\epsilon^{\mu\nu\alpha\beta}G^{a}_{\mu\nu}G^{a}_{\alpha\beta}-V''\!\left(\bar{\phi}\!\left(t_{0}\right)\right)\frac{\delta\phi^{2}}{2}\nonumber \\
 & \quad\left.+\frac{\alpha}{16\pi f}\frac{\delta\phi}{\sqrt{-g\left(t_{0}\right)}}\!\left(\epsilon^{\mu\nu\alpha\beta}G^{a}_{\mu\nu}G^{a}_{\alpha\beta}-\langle\epsilon^{\mu\nu\alpha\beta}G^{a}_{\mu\nu}G^{a}_{\alpha\beta}\rangle\!\left(t_{0}\right)\right)\right)\nonumber \\
 & \quad+\mathcal{L}_{\mathrm{int}}\!\left(\delta a,\delta\bar{\phi},\delta\phi,A^{a}_{\mu}\right)\label{eq:Lbatht>t0}
\end{align}

where the index contractions are carried out with the metric evaluated
at $t_{0}$, i.e.\ $g_{\mu\nu}\!\left(t_{0}\right)$, and the perturbation
denoted by $\mathcal{L}_{\mathrm{int}}$ is 
\begin{align}
\mathcal{L}_{\mathrm{int}} & =\sqrt{-g\left(t_{0}\right)}\left(\frac{\alpha}{16\pi f}\!\frac{\left(\bar{\phi}\!\left(t\right)-\bar{\phi}\!\left(t_{0}\right)\right)}{\sqrt{-g\left(t_{0}\right)}}\epsilon^{\mu\nu\alpha\beta}G^{a}_{\mu\nu}G^{a}_{\alpha\beta}\!\left(t\right)-\!\left(V''\!\left(\bar{\phi}\!\left(t\right)\right)-V''\!\left(\bar{\phi}\!\left(t_{0}\right)\right)\right)\frac{\delta\phi^{2}\!\left(t\right)}{2}\right.\nonumber \\
 & \quad\left.-\frac{\alpha}{16\pi f}\frac{\delta\phi}{\sqrt{-g\left(t_{0}\right)}}\!\left(\langle\epsilon^{\mu\nu\alpha\beta}G^{a}_{\mu\nu}G^{a}_{\alpha\beta}\!\left(t\right)\rangle-\langle\epsilon^{\mu\nu\alpha\beta}G^{a}_{\mu\nu}G^{a}_{\alpha\beta}\rangle\!\left(t_{0}\right)\right)\right)+\mathcal{L}_{\mathrm{int},\delta a}\!\left(\delta a,\delta\phi,A^{a}_{\mu}\right).
\end{align}
Here we have separated out the perturbation terms proportional to
$\delta a\!\left(t\right)$ in $\mathcal{L}_{\mathrm{int},a}$, since
these do not contribute to the $\langle\tilde{G}^{\mu\nu}_{a}G^{a}_{\mu\nu}\rangle\!\left(t,x\right)$
expectation value.\footnote{$\mathcal{L}_{\mathrm{int},a}$ characterizes how the bath evolves
with spacetime expansion and does not contribute to dissipative terms
characterizing energy exchange between the inflaton background and
the thermal bath to the leading order we are working in. This has
been explained in terms of CPT in section \ref{sec:Effective-friction-is}.} To find the time dependence of $\langle\tilde{G}^{\mu\nu}_{a}G^{a}_{\mu\nu}\rangle\!\left(t\right)$,
we may now rewrite it in the interaction picture as 
\begin{align}
\langle\tilde{G}^{\mu\nu}_{a}G^{a}_{\mu\nu}\rangle\!\left(t,x\right) & =\left\langle \exp\!\left(+i\int^{t}_{t_{0}}dt'\,H_{\mathrm{int},I}\!\left(t'\right)\right)\tilde{G}^{\mu\nu}_{I\,a}G^{a}_{I\,\mu\nu}\!\left(t,x\right)\exp\!\left(-i\int^{t}_{t_{0}}dt'\,H_{\mathrm{int},I}\!\left(t'\right)\right)\right\rangle _{0}\nonumber \\
 & \approx\langle\tilde{G}^{\mu\nu}_{I\,a}G^{a}_{I\,\mu\nu}\!\left(t,x\right)\rangle_{0}-i\int^{t}_{t_{0}}dt'\left\langle \left[\tilde{G}^{\mu\nu}_{I\,a}G^{a}_{I\,\mu\nu}\!\left(t,x\right),H_{\mathrm{int},I}\!\left(t'\right)\right]\right\rangle _{0}\label{eq:GGtExp}
\end{align}
where 
\begin{equation}
H_{\mathrm{pert}}=H_{0}+H_{\mathrm{int}}
\end{equation}
is the total Hamiltonian corresponding to eq.~(\ref{eq:Lbatht>t0})
and 
\begin{equation}
H_{\mathrm{int}}\equiv\int d^{3}x'\,\mathcal{H}_{\mathrm{int}}\!\left(x'\right)
\end{equation}
are the perturbative terms obtained from $\mathcal{L}_{\mathrm{int}}$.
The operators in the interaction picture have been denoted with the
subscripts $I$. The thermal expectation values are taken with respect
to the thermal state at $t_{0}$, i.e.\ eq.~(\ref{eq:density_Mat_t_0}),
and have been denoted by $\langle\ldots\rangle_{0}$. The first term
reduces to 
\begin{equation}
\langle\tilde{G}^{\mu\nu}_{I\,a}G^{a}_{I\,\mu\nu}\!\left(t,x\right)\rangle_{0}=\langle\tilde{G}^{\mu\nu}_{I\,a}G^{a}_{I\,\mu\nu}\!\left(t_{0},x\right)\rangle_{0}
\end{equation}
where we have used the evolution of interaction-picture operators
as 
\begin{equation}
\mathcal{O}_{I}\!\left(t\right)=\exp\!\left(+iH_{0}\!\left(t-t_{0}\right)\right)\mathcal{O}_{I}\!\left(t_{0}\right)\exp\!\left(-iH_{0}\!\left(t-t_{0}\right)\right).
\end{equation}
The term from $\mathcal{H}_{\mathrm{int}}$ which contributes non-trivially
to eq.~(\ref{eq:GGtExp}) is 
\begin{equation}
\mathcal{H}_{\mathrm{int}}\ni-\frac{\alpha}{16\pi f}\!\left(\bar{\phi}\!\left(t\right)-\bar{\phi}\!\left(t_{0}\right)\right)\epsilon^{\mu\nu\alpha\beta}G^{a}_{\mu\nu}G^{a}_{\alpha\beta},
\end{equation}
allowing the second term in eq.~(\ref{eq:GGtExp}) to be rewritten
as 
\begin{align}
\langle\tilde{G}^{\mu\nu}_{a}G^{a}_{\mu\nu}\rangle\!\left(t,x\right) & \ni+i\frac{\alpha}{16\pi f}\!\left(\bar{\phi}\!\left(t\right)-\bar{\phi}\!\left(t_{0}\right)\right)\nonumber \\
 & \qquad\times\int^{t}_{t_{0}}dt'\int d^{3}x'\left\langle \left[\tilde{G}^{\mu\nu}_{I\,a}G^{a}_{I\,\mu\nu}\!\left(t,x\right),\epsilon^{\alpha\beta\gamma\delta}G_{I\,b,\alpha\beta}G^{b}_{I\,\gamma\delta}\!\left(t',x'\right)\right]\right\rangle _{0}\nonumber \\
 & \quad-i\frac{\alpha}{16\pi f}\int^{t}_{t_{0}}dt'\,\!\left(\bar{\phi}\!\left(t\right)-\bar{\phi}\!\left(t'\right)\right)\times\nonumber \\
 & \qquad\int d^{3}x'\left\langle \left[\tilde{G}^{\mu\nu}_{I\,a}G^{a}_{I\,\mu\nu}\!\left(t,x\right),\epsilon^{\alpha\beta\gamma\delta}G_{I\,b,\alpha\beta}G^{b}_{I\,\gamma\delta}\!\left(t',x'\right)\right]\right\rangle _{0}.\label{eq:phiGGInt}
\end{align}
Here, the $\bar{\phi}$ dependencies have been rearranged to separate
the first term, corresponding to the partial derivative of the effective
inflaton potential~\cite{Morikawa:1984dz}. In particular, under
the approximations below, the integral coefficient in the first term
reduces to the usual axion mass generated non-perturbatively. This
mass is power-law suppressed for temperatures larger than the confinement
scale and is hence ignored below. Dissipative terms, on the other
hand, arise from the second term above. If the time scales associated
with gluon interactions in the thermal plasma ($\tau$) are much shorter
than the time scales associated with background evolution, that is
\begin{equation}
\left(\frac{\dot{\bar{\phi}}}{\bar{\phi}}\right)^{-1},\,H^{-1}\gg\tau\sim\left(\alpha^{2}T\right)^{-1}\gg\frac{1}{\Lambda_{\text{conf}}},\label{eq:LocalityAssum}
\end{equation}
(where $\Lambda_{\text{conf}}$ is the confinement scale) the correlation
functions in eq.~(\ref{eq:phiGGInt}) decay quickly, allowing a local/Markovian
approximation of $\langle\tilde{G}G\rangle\!\left(t\right)$. This
is obtained by Taylor expanding the $\bar{\phi}\!\left(t\right)$
dependence as follows: 
\begin{align}
\langle\tilde{G}^{\mu\nu}_{a}G^{a}_{\mu\nu}\rangle_{\mathrm{diss}}\!\left(t,x\right) & \equiv-i\frac{\alpha}{16\pi f}\int^{t}_{t_{0}}dt'\,\!\left(\bar{\phi}\!\left(t\right)-\bar{\phi}\!\left(t'\right)\right)\\
 & \qquad\quad\times\int d^{3}x'\left\langle \left[\tilde{G}^{\mu\nu}_{I\,a}G^{a}_{I\,\mu\nu}\!\left(t,x\right),\epsilon^{\alpha\beta\gamma\delta}G_{I\,b,\alpha\beta}G^{b}_{I\,\gamma\delta}\!\left(t',x'\right)\right]\right\rangle _{0}\nonumber \\
=-i\frac{\alpha}{16\pi f} & \int^{+\infty}_{-\infty}dt'\left\{ \left(\dot{\bar{\phi}}\!\left(t\right)\!\left(t-t'\right)-\ddot{\bar{\phi}}\!\left(t\right)\frac{\!\left(t-t'\right)^{2}}{2}+\dddot{\bar{\phi}}\!\left(t\right)\frac{\!\left(t-t'\right)^{3}}{3!}+\cdots\right)\right.\nonumber \\
\times & \left.\theta\!\left(t-t'\right)\int d^{3}x'\,\left\langle \left[\tilde{G}^{\mu\nu}_{I\,a}G^{a}_{I\,\mu\nu}\!\left(t,x\right),\epsilon^{\alpha\beta\gamma\delta}G_{I\,b,\alpha\beta}G^{b}_{I\,\gamma\delta}\!\left(t',x'\right)\right]\right\rangle _{0}\right\} \label{eq:TaylorCorr}
\end{align}
and carrying out the $t'$ integrals. In the above expression, the
equilibrium time $t_{0}$ has been taken to $-\infty$ since the thermal
correlators in eq.~(\ref{eq:phiGGInt}) decay within the interval
$\vert t-t_{0}\vert$. We have also introduced a theta function which
allows the upper integration limits to be extended to $+\infty$.
Finally note that eq.~(\ref{eq:LocalityAssum}) implies that the
correlator in eq.~(\ref{eq:RetCorr}) should reduce to that in flat
space, since the gauge fields see negligible background evolution
in time $\tau$. Explicitly, this may be seen by noting that under
the change of variables 
\begin{equation}
x\rightarrow\tilde{x}=a\!\left(t_{0}\right)x,\qquad t\rightarrow t\label{eq:VarChange}
\end{equation}
and 
\begin{equation}
\mathcal{A}^{a}_{i}\!\left(t,x\right)=a^{-1}\!\left(t_{0}\right)A^{a}_{i}\!\left(t,x\right),\quad\mathcal{A}^{a}_{0}\!\left(t,x\right)=A^{a}_{0}\!\left(t,x\right)\label{eq:FieldRedefMin}
\end{equation}
the action (Lagrangian density) where $\mathcal{L}_{\mathrm{pert}}$
defined on a background with constant scale factor $a\left(t_{0}\right)$,
reduces to it's flat-space analog, here forth denoted $\mathcal{\bar{L}}_{pert}$.
In terms of these redefined quantities 
\begin{align}
 & \int d^{3}x'\,\left\langle \left[\tilde{G}^{\mu\nu}_{I\,a}G^{a}_{I\,\mu\nu}\!\left(t,x\right),\epsilon^{\alpha\beta\gamma\delta}G_{I\,b,\alpha\beta}G^{b}_{I\,\gamma\delta}\!\left(t',x'\right)\right]\right\rangle _{0}\nonumber \\
 & \quad=\frac{a^{3}\!\left(t_{0}\right)}{a^{3}\!\left(t\right)}\int d^{3}\tilde{x}'\,\left\langle \left[\mathcal{\tilde{G}}_{I\,a,\mu\nu}\mathcal{G}^{a,\mu\nu}_{I}\!\left(t,\tilde{x}\right),\mathcal{\tilde{G}}_{I\,b,\alpha\beta}\mathcal{G}^{b,\alpha\beta}_{I}\!\left(t',\tilde{x}'\right)\right]\right\rangle _{0}\nonumber \\
 & \quad\approx\int d^{3}\tilde{x}'\,\left\langle \left[\mathcal{\tilde{G}}_{I\,a,\mu\nu}\mathcal{G}^{a,\mu\nu}_{I}\!\left(t,\tilde{x}\right),\mathcal{\tilde{G}}_{I\,b,\alpha\beta}\mathcal{G}^{b,\alpha\beta}_{I}\!\left(t',\tilde{x}'\right)\right]\right\rangle _{0}.
\end{align}
where the Hodge duals and metric contractions on the right-hand side
of the equation are defined with respect to the Minkowskian metric.
Consequently, the coefficients of the inflaton derivatives in eq.~(\ref{eq:TaylorCorr})
can now be expressed in terms of the Fourier transform of the flat-space
retarded correlator in the above, that is 
\begin{equation}
\Pi^{R}_{\tilde{\mathcal{G}}\mathcal{G}}\!\left(p\right)\equiv i\int d^{4}y\,\exp\!\left(ip_{\mu}y^{\mu}\right)\theta\!\left(y_{0}\right)\left\langle \left[\tilde{\mathcal{G}}^{\mu\nu}_{I\,a}\mathcal{G}^{a}_{I\,\mu\nu}\!\left(y\right),\tilde{\mathcal{G}}^{\alpha\beta}_{I\,b}\mathcal{G}^{b}_{I\:\alpha\beta}\!\left(0\right)\right]\right\rangle _{0}\label{eq:RetCorr}
\end{equation}
where 
\begin{equation}
p_{\mu}y^{\mu}=p_{0}y_{0}-p_{i}y_{i}.
\end{equation}
In the following, since we will only concern ourselves with flat-space
correlators, we will drop the stylized notation for the redefined
gauge fields and instead use $A^{a}_{\mu}$ and $G^{a}_{\mu\nu}$
as usual; similarly the redefined space-time coordinates will be denoted
by $\left(t,\vec{x}\right)$. Metric contractions will be carried
out with the Minkowski metric.

Dissipative terms in eq.~(\ref{eq:TaylorCorr}) should arise from
those with an odd number of time derivatives, since only these break
CPT symmetry as discussed in section \ref{sec:Effective-friction-is}.
The terms with an even number of time derivatives, on the other hand,
correspond to the effective potential or field renormalization. Considering
the terms proportional to $\dot{\bar{\phi}}\!\left(t\right)$ and
$\dddot{\bar{\phi}}\!\left(t\right)$, we have 
\begin{equation}
\Pi^{R}_{\tilde{G}G}(p)\approx T_{1}+T_{2}+...
\end{equation}
where 
\begin{align}
T_{1} & \equiv-i\frac{\alpha}{16\pi f}\dot{\bar{\phi}}\!\left(t\right)\int^{+\infty}_{-\infty}dt'\,\!\left(t-t'\right)\int d^{3}x'\,\theta\!\left(t-t'\right)\left\langle \left[\tilde{G}^{\mu\nu}_{a}G^{a}_{\mu\nu}\!\left(t,x\right),\tilde{G}^{\alpha\beta}_{b}G^{b}_{\alpha\beta}\!\left(t',x'\right)\right]\right\rangle _{0}\nonumber \\
 & =i\frac{\alpha}{16\pi f}\dot{\bar{\phi}}\!\left(t\right)\left[\frac{\partial}{\partial p_{0}}\Pi^{R}_{\tilde{G}G}\!\left(p_{0},0\right)\right]_{\vert p_{0}=0}\nonumber \\
 & =-\frac{\alpha}{16\pi f}\dot{\bar{\phi}}\!\left(t\right)\partial_{p_{0}}\rho_{\tilde{G}G}\!\left(p_{0},0\right)_{\vert p_{0}=0}\label{eq:T1}
\end{align}
and 
\begin{align}
T_{2} & \equiv-i\frac{\alpha}{16\pi f}\dddot{\bar{\phi}}\!\left(t\right)\frac{1}{3!}\int^{+\infty}_{-\infty}dt'\,\!\left(t-t'\right)^{3}\int d^{3}x'\,\theta\!\left(t-t'\right)\left\langle \left[\tilde{G}^{\mu\nu}_{a}G^{a}_{\mu\nu}\!\left(t,x\right),\tilde{G}^{\alpha\beta}_{b}G^{b}_{\alpha\beta}\!\left(t',x'\right)\right]\right\rangle \nonumber \\
 & =-i\frac{\alpha}{16\pi f}\dddot{\bar{\phi}}\!\left(t\right)\frac{1}{3!}\left[\frac{\partial^{3}}{\partial p^{3}_{0}}\Pi^{R}_{\tilde{G}G}\!\left(p_{0},0\right)\right]_{\vert p_{0}=0}\nonumber \\
 & =\frac{\alpha}{16\pi f}\dddot{\bar{\phi}}\!\left(t\right)\frac{1}{3!}\partial^{3}_{p_{0}}\rho_{\tilde{G}G}\!\left(p_{0},0\right)_{\vert p_{0}=0}\label{eq:PertDiss}
\end{align}
respectively. Here we have used that $\text{Re}\!\left(\Pi^{R}_{\tilde{G}G}\!\left(p_{0},\vec{p}\right)\right)$
is even in $p_{0}$ and the relation 
\begin{equation}
\text{Im}\!\left(\Pi^{R}_{\tilde{G}G}\!\left(p_{0},p\right)\right)=\rho_{\tilde{G}G}\!\left(p_{0},p\right)\label{eq:ImPiRho}
\end{equation}
between the retarded correlator and the spectral function.

$T_{1}$ does not have perturbative support, but is related to the
non-perturbative hot sphaleron diffusion rate 
\begin{equation}
\Gamma_{\mathrm{sphal}}\equiv\lim_{t\rightarrow\infty}\frac{\left(N_{CS}\left(t\right)-N_{CS}\left(0\right)\right)^{2}}{Vt}=\int d^{4}x\langle\frac{\alpha}{16\pi}\tilde{G}^{\mu\nu}_{a}G^{a}_{\mu\nu}\!\left(t,x\right)\frac{\alpha}{16\pi}\tilde{G}^{\alpha\beta}_{b}G^{b}_{\alpha\beta}\!\left(0\right)\rangle
\end{equation}
as 
\begin{equation}
\Gamma_{\mathrm{sphal}}=\left(\frac{\alpha}{16\pi}\right)^{2}2T\,\partial_{p_{0}}\rho_{\tilde{G}G}\!\left(p_{0},0\right)_{\vert p_{0}=0}\label{eq:Gammasphal}
\end{equation}
\cite{Moore:2010jd,Laine:2016hma}.\footnote{This may be seen by noting that the spectral density $\rho_{\tilde{G}G}\left(p_{0},\vec{p}\right)$
is odd under $p_{0}\rightarrow-p_{0}$. Therefore if $\Gamma_{\mathrm{sphal}}=\left(\alpha/16\pi\right)^{2}\lim_{p_{0}\rightarrow0}2T\rho_{\tilde{G}G}\left(p_{0},0\right)/p_{0}$
\cite{Laine:2016hma} is finite, then $\Gamma_{\mathrm{sphal}}\sim p_{0}$
for small $p_{0}$ and can be rewritten as the derivative in eq.
(\ref{eq:Gammasphal})} On the other hand, $T_{2}$ gains finite support from perturbative
processes in the thermal bath, corresponding to Landau damping and
plasmon decay of effective gluon modes in the thermal bath. Re-expressing
$\dddot{\bar{\phi}}$ in terms of slow-roll parameters and $\dot{\bar{\phi}}$,
we note that $T_{2}$ could lead to an additional perturbative contribution
to the friction coefficient in minimal warm inflation models. To derive
the corresponding friction coefficient we begin by computing $\rho_{\tilde{G}G}$
in the next section.

\section{$\rho_{\!\tilde{G}G}$ in terms of gluon spectral functions}

\label{sec:-Euc-corr}

In this
section, we simplify the spectral function $\rho_{\tilde{G}G}$ in
terms of the (1-loop) gluon spectral functions. Note that $\rho_{\tilde{G}G}$
is defined as the imaginary part of the retarded correlation function
in eq.~(\ref{eq:RetCorr}) found with respect to the equilibrium
thermal distribution at $t\approx t_{0}$. Correlation functions with
respect to an equilibrium distribution are more easily computed using
the imaginary-time formalism, where the resultant Euclidean correlators
are related to the corresponding Lorentzian ones by analytic continuations.
In particular, the spectral function of an operator $\mathcal{O}\!\left(t,x\right)$
can be obtained from the Euclidean two-point function $\langle\mathcal{O}\!\left(\tau,x\right)\mathcal{O}\!\left(0\right)\rangle$
as 
\begin{equation}
\rho_{\mathcal{O}}\!\left(p_{0},\vec{p}\right)=\text{Im}\!\left(\Pi^{R}_{\mathcal{O}}\!\left(p_{0},\vec{p}\right)\right)\label{eq:SpecEucProp}
\end{equation}
where 
\begin{align}
\Pi^{R}_{\mathcal{O}}\!\left(p_{0},\vec{p}\right) & =\frac{1}{2i}\mathrm{Disc}\!\left(\Pi^{E}_{\mathcal{O}}\!\left(p_{n}\rightarrow-ip_{0}+0^{+},\vec{p}\right)\right)\nonumber \\
 & \equiv\frac{1}{2i}\left\{ \Pi^{E}_{\mathcal{O}}\!\left(p_{n}\rightarrow-i\!\left(p_{0}+i0^{+}\right),\vec{p}\right)-\Pi^{E}_{\mathcal{O}}\!\left(p_{n}\rightarrow-i\!\left(p_{0}-i0^{+}\right),\vec{p}\right)\right\} \label{eq:retarded-corr}
\end{align}
and 
\begin{equation}
\Pi^{E}_{\mathcal{O}}\!\left(p_{n},\vec{p}\right)\equiv\int^{1/T}_{0}d\tau\int d^{3}x\,\exp\!\left(ip_{n}\tau-i\vec{p}\cdot\vec{x}\right)\langle\mathcal{O}\!\left(\tau,x\right)\mathcal{O}\!\left(0\right)\rangle.\label{eq:euclidean}
\end{equation}
Here, $p_{n}=2\pi nT$ are the Matsubara frequencies. In the following,
we denote Euclidean momenta by capitalized letters, e.g.\ $P_{\mu}\equiv\!\left(p_{n},\vec{p}\right)$,
and Minkowskian momenta by lower-case letters $p_{\mu}\equiv\!\left(p_{0},\vec{p}\right)$
(or explicitly as $\!\left(p_{0},\vec{p}\right)$). The Euclidean
correlation functions are in turn defined as partial derivatives (with
respect to sources) of the generating function $\mathcal{Z}\!\left(T^{-1};J_{\alpha}\right)$,
which is such that when the external sources are set to zero, 
\begin{equation}
\mathcal{Z}\!\left(T^{-1};0\right)=\mathrm{Tr}\!\left(\exp\!\left(-\frac{H_{0}}{T}\right)\right).
\end{equation}
The path-integral representation of $\mathcal{Z}\!\left(T^{-1};0\right)$
may be deduced from the Lorentzian path integral by the analytic continuations
$t\rightarrow-i\tau,\,A^{a}_{0}\rightarrow+i\bar{A}^{a}_{0}$ \footnote{Note that the analytic continuations are only a mnemonic to arrive
at the correct path integral derived from Hamiltonian formalism.}, and by compactifying the Euclidean time $\tau$ on a circle of perimeter
$T^{-1}$. For our model of interest, this gives 
\begin{equation}
\mathcal{Z}\!\left(T^{-1};0\right)=\int\mathcal{D}\delta\phi\,\mathcal{D}\bar{A}^{a}_{0}\exp\!\left(-\int d^{4}x\,\mathcal{L}_{E}\right)
\end{equation}
where 
\begin{equation}
\mathcal{L}_{E}\equiv-\bar{\mathcal{L}}_{\mathrm{pert}}\!\left(t\rightarrow-i\tau,\,A^{a}_{0}\!\left(t\rightarrow-i\tau\right)\rightarrow+i\bar{A}^{a}_{0}\!\left(\tau\right),\,x\right)
\end{equation}
is obtained from the analytic continuation of the Lagrangian density
describing the perturbative modes for $t\approx t_{0}$ ( the flat-space
analogue of the first three lines of eq.~(\ref{eq:Lbatht>t0})),
giving 
\begin{align}
\mathcal{L}_{E}\!\left(\delta\phi\!\left(\tau\right),\bar{A}^{a}_{\mu}\!\left(\tau\right)\right) & =\frac{1}{2}\partial_{\mu}\delta\phi\partial_{\mu}\delta\phi+V''\!\left(\bar{\phi}\!\left(t_{0}\right)\right)\frac{\delta\phi^{2}\!\left(t\right)}{2}+\frac{1}{4}\bar{G}^{a}_{\mu\nu}\bar{G}^{a}_{\mu\nu}\nonumber \\
 & \quad+\frac{\alpha i}{16\pi f}\bar{\phi}\!\left(t_{0}\right)\epsilon^{\mu\nu\alpha\beta}\bar{G}^{a}_{\mu\nu}\bar{G}^{a}_{\alpha\beta}\!\left(\tau\right)\nonumber \\
 & \quad+\frac{\alpha i}{16\pi f}\delta\phi\!\left(\epsilon^{\mu\nu\alpha\beta}\bar{G}^{a}_{\mu\nu}\bar{G}^{a}_{\alpha\beta}\!\left(\tau\right)-\langle\epsilon^{\mu\nu\alpha\beta}\bar{G}^{a}_{\mu\nu}\bar{G}^{a}_{\alpha\beta}\rangle\!\left(t_{0}\right)\right).\label{eq:EucLag}
\end{align}
Here, the repeated indices, both lowered, correspond to contraction
with the Euclidean metric $\delta_{\mu\nu}$, and $\bar{G}^{a}_{\mu\nu}$
denotes the field strength of the Euclidean gauge fields $\bar{A}^{a}_{\mu}\!\left(\tau,x\right)$.
These are related to the Minkowskian gauge fields as $\bar{A}^{a}_{0}\!\left(\tau\right)=-iA^{a}_{0}\!\left(t\rightarrow-i\tau\right)$
and $\bar{A}^{a}_{i}\!\left(\tau\right)=A^{a}_{i}\!\left(t\rightarrow-i\tau\right)$.
Note that the analytic continuation above takes $\tilde{G}G\rightarrow i\tilde{\bar{G}}\bar{G}$,
resulting in the additional factors of $i$ in eq.~(\ref{eq:EucLag}).
With this path integral, we may now compute the relevant Euclidean
two-point functions in eq.~(\ref{eq:SpecEucProp}) in the usual Feynman-diagram
expansion using bare (Euclidean) propagators obtained from the $g\rightarrow0$
limit of the above Lagrangian density.

The computation leading to $\rho_{\tilde{G}G}\!\left(p_{0},0\right)$
is presented in the following subsections. An important detail to
note is that applying eq.~(\ref{eq:SpecEucProp}) for the operator
$\tilde{\bar{G}}\bar{G}$ with the Euclidean gauge fields gives the
spectral function for $-\rho_{\tilde{G}G}$, since eq.~(\ref{eq:SpecEucProp})
does not account for the factor of $i$ in $\bar{A}^{a}_{0}\!\left(\tau\right)=-iA^{a}_{0}\!\left(t\rightarrow-i\tau\right)$,
that is 
\begin{align}
 & -\rho_{\!\tilde{G}G}\!\left(p_{0},\vec{p}\right)=\rho_{\!\tilde{\bar{G}}\bar{G}}\!\left(p_{0},\vec{p}\right)\nonumber \\
 & \quad=\frac{1}{2i}\!\left[\Pi^{E}_{\!\tilde{\bar{G}}\bar{G}}\!\!\left(p_{n}\rightarrow-i\!\left(p_{0}+i0^{+}\right),\vec{p}\right)-\Pi^{E}_{\!\tilde{\bar{G}}\bar{G}}\!\!\left(p_{n}\rightarrow-i\!\left(p_{0}-i0^{+}\right),\vec{p}\right)\right].\label{eq:SpecDensfromEucCorr}
\end{align}

\subsection{Simplifying $\Pi^{E}_{\tilde{\bar{G}}\bar{G}}$ in Terms of Gluon
Spectral Functions}

\label{subsec:PiGG2Pig}

In this section assuming perturbativity in $g$, we simplify eq.~(\ref{eq:SpecDensfromEucCorr})
to the leading order in terms of gluon spectral functions. At the
end of this subsection we also discuss the regime when perturbativity
fails in thermal Yang Mills theory and its importance for the friction
coefficient. At the leading perturbative order 
\begin{align}
\Pi^{E}_{\!\tilde{\bar{G}}\bar{G}}\!\left(P\right) & \equiv\int^{T^{-1}}_{0}d\tau_{y}\int d^{3}y\,\exp\!\left(-iP\cdot Y\right)\langle\tilde{\bar{G}}_{a\mu\nu}\bar{G}^{a}_{\mu\nu}\!\left(\tau,x\right)\tilde{\bar{G}}_{b,\alpha\beta}\bar{G}^{b}_{\alpha\beta}\!\left(\tau',x'\right)\rangle_{c}\nonumber \\
 & \approx16\int^{T^{-1}}_{0}d\tau_{y}\int d^{3}y\,\exp\!\left(-iP\cdot Y\right)\langle\epsilon^{\mu\nu\rho\sigma}\partial_{\mu}\bar{A}^{a}_{\nu}\partial_{\rho}\bar{A}^{a}_{\sigma}\!\left(\tau,x\right)\epsilon^{\alpha\beta\gamma\delta}\partial_{\alpha}\bar{A}^{d}_{\beta}\partial_{\gamma}\bar{A}^{d}_{\delta}\!\left(\tau',x'\right)\rangle_{c}
\end{align}
where $Y=\!\left(\tau_{y},\vec{y}\right)\equiv\!\left(\tau-\tau',\vec{x}-\vec{x}'\right)$,
$P\equiv\!\left(p_{n},\vec{p}\right)$, and $P\cdot Y=p_{n}\tau_{y}-\vec{p}\cdot\vec{y}$.
After Wick contractions, in terms gluon propagators in Euclidean momentum-space,
taking the form 
\begin{equation}
\langle\bar{A}^{a}_{\nu}\!\left(Q\right)\bar{A}^{d}_{\beta}\!\left(M\right)\rangle=\delta^{ad}\,\bar{\delta}\!\left(Q+M\right)\Pi_{\nu\beta}\!\left(Q\right)
\end{equation}
where 
\begin{equation}
\bar{\delta}\!\left(Q+M\right)\equiv T^{-1}\,\delta_{q_{n}+m_{n}}\!\left(2\pi\right)^{3}\delta^{\!\left(3\right)}\!\left(\vec{q}+\vec{m}\right),
\end{equation}
the above simplifies to 
\begin{align}
\Pi^{E}_{\!\tilde{\bar{G}}\bar{G}}\!\left(P\right) & \approx16\int^{T^{-1}}_{0}d\tau_{y}\int d^{3}y\,\exp\!\left(-iP\cdot Y\right)\langle\epsilon^{\mu\nu\rho\sigma}\partial_{\mu}\bar{A}^{a}_{\nu}\partial_{\rho}\bar{A}^{a}_{\sigma}\!\left(\tau,x\right)\epsilon^{\alpha\beta\gamma\delta}\partial_{\alpha}\bar{A}^{d}_{\beta}\partial_{\gamma}\bar{A}^{d}_{\delta}\!\left(\tau',x'\right)\rangle_{c}\nonumber \\
 & =32\,\delta^{aa}\int^{T^{-1}}_{0}d\tau_{y}\int d^{3}y\,\exp\!\left(-iP\cdot Y\right)\int_{\Sigma,Q}\int_{\Sigma,L}\exp\!\left(+i\!\left(Q+L\right)\cdot Y\right)\nonumber \\
 & \quad\quad\times\epsilon^{\mu\nu\rho\sigma}\epsilon^{\alpha\beta\gamma\delta}Q_{\mu}L_{\rho}Q_{\alpha}L_{\gamma}\Pi_{\nu\beta}\!\left(Q\right)\Pi_{\sigma\delta}\!\left(L\right).
\end{align}
Here, the additional factor of 2 arises from the two contractions
(for connected diagrams). To simplify further, we note that gauge
invariance implies the transversality of the exact Euclidean propagator
with respect to the (Euclidean) 4-momentum $Q$. This requires that
the propagator takes the form 
\begin{equation}
\Pi_{\nu\beta}\!\left(Q\right)=\Pi^{T}\!\left(Q\right)\mathbb{P}^{T}_{\nu\beta}\!\left(Q\right)+\Pi^{L}\!\left(Q\right)\mathbb{P}^{L}_{\nu\beta}\!\left(Q\right)+\frac{\xi Q_{\mu}Q_{\nu}}{\left(Q^{2}\right)^{2}}\label{eq:SlavanovTaylor}
\end{equation}
where 
\begin{equation}
\mathbb{P}^{T}_{\nu\beta}\!\left(Q\right)\equiv\delta_{\nu i}\delta_{\beta j}\!\left(\delta_{ij}-\frac{q_{i}q_{j}}{q^{2}}\right)\label{eq:transProj}
\end{equation}
\begin{equation}
\mathbb{P}^{L}_{\nu\beta}\!\left(Q\right)\equiv\delta_{\nu\beta}-\frac{Q_{\beta}Q_{\nu}}{Q^{2}}-\mathbb{P}^{T}_{\nu\beta}\!\left(Q\right)\label{eq:LongProj}
\end{equation}
in a general covariant gauge. Here, in addition to $\mathbb{P}^{T}_{\nu\beta}\!\left(Q\right)$
and $\mathbb{P}^{L}_{\nu\beta}\!\left(Q\right)$ being transverse
to $Q$, the former is also transverse to the spatial momentum $\vec{q}$.
The structure of the exact propagator is different from the zero-temperature
case since an equilibrium thermal bath at temperature $T$ breaks
Lorentz invariance in the system. Further, rewriting the above in
terms of transverse and longitudinal spectral densities, which are
defined as 
\begin{equation}
\rho^{T\!\left(L\right)}\left(q_{0},\vec{q}\right)=\frac{1}{2i}\mathrm{Disc}\!\left(\Pi^{T\!\left(L\right)}\!\left(q_{n}\rightarrow-iq_{0},\vec{q}\right)\right),
\end{equation}
we have 
\begin{align}
\Pi_{\nu\beta}\!\left(Q\right) & =\left(\int^{+\infty}_{-\infty}\frac{dq_{0}}{\pi}\frac{\rho^{T}\!\left(q_{0},\vec{q}\right)}{q_{0}-iq_{n}}\right)\mathbb{P}^{T}_{\nu\beta}\!\left(Q\right)+\left(\int^{+\infty}_{-\infty}\frac{dq_{0}}{\pi}\frac{\rho^{L}\!\left(q_{0},\vec{q}\right)}{q_{0}-iq_{n}}\right)\mathbb{P}^{L}_{\nu\beta}\!\left(Q\right)+\frac{\xi Q_{\mu}Q_{\nu}}{\left(Q^{2}\right)^{2}}.
\end{align}
Working in the Feynman gauge $\!\left(\xi=1\right)$, $\Pi^{E}_{\tilde{\bar{G}}\bar{G}}\!\left(P\right)$
can now be simplified in terms of the gluon spectral functions, after
carrying out index contractions and the Matsubara sums (details are
presented in appendix~\ref{sec:GGtoSpec}). Specializing to zero
external momentum $\vec{p}=0$ for eq.~(\ref{eq:PertDiss}) gives
\begin{align}
\Pi^{E}_{\tilde{\bar{G}}\bar{G}}\!\left(p_{n},0\right) & =32\,\delta^{aa}\int_{q}\int^{+\infty}_{-\infty}\frac{dq_{0}}{\pi}\int^{+\infty}_{-\infty}\frac{dl_{0}}{\pi}\,n_{B}\!\left(q_{0}\right)n_{B}\!\left(l_{0}\right)\frac{\!\left[\exp\!\left(T^{-1}\!\left(l_{0}+q_{0}\right)\right)-1\right]}{\!\left(l_{0}+q_{0}-ip_{n}\right)}\nonumber \\
 & \quad\times\left[-\rho^{T}\!\left(q_{0},\vec{q}\right)\rho^{T}\!\left(l_{0},-\vec{q}\right)2q^{2}\!\left(q_{0}+l_{0}\right)^{2}\right]\label{eq:EucCorrSimp}
\end{align}
from which eq.~(\ref{eq:SpecDensfromEucCorr}) implies the leading-order
contribution to $\rho_{\!\tilde{G}G}\!\left(p_{0},0\right)$ is 
\begin{align}
\rho_{\!\tilde{G}G}\!\left(p_{0},0\right) & =-\frac{1}{2i}\mathrm{Disc}\!\left[\Pi^{E}_{\!\tilde{\bar{G}}\bar{G}}\!\left(p_{n}\rightarrow-i\!\left(p_{0}+i0^{+}\right),0\right)\right]\nonumber \\
 & =4^{3}\,\delta^{aa}\,p^{2}_{0}\!\left(\exp\!\left(\frac{p_{0}}{T}\right)-1\right)\nonumber \\
 & \quad\times\left\{ \int_{q}\int^{+\infty}_{-\infty}\frac{dq_{0}}{\pi}\,n_{B}\!\left(q_{0}\right)n_{B}\!\left(p_{0}-q_{0}\right)\left[\rho^{T}\!\left(q_{0},q_{i}\right)\rho^{T}\!\left(p_{0}-q_{0},-q_{i}\right)q^{2}\right]\right\} .\label{eq:RhoGG}
\end{align}
Plugging into eq.~(\ref{eq:T1}) we see no contribution from perturbative
modes to leading order in the derivative expansion of the background,
that is, the coefficient of the $\dot{\bar{\phi}}$ term. 
An analogous computation
for axions derivatively coupled to other scalar fields was carried
out in \cite{Mishra:2011ph}. Unfortunately, a small error in distributing
derivatives of the interaction terms gives an incorrect dependence
on $p_{0}$ in the analogue of eq. (\ref{eq:RhoGG}) and consequently
a non-zero friction coefficient.

While the above implies that perturbative processes do not contribute
to the friction coefficient at least leading order in $g$, it doesn't
exclude non-perturbative contributions to the $\gamma_{1}\dot{\phi}$
(see sec.~\ref{sec:Effective-friction-is}). The latter corresponds
to gauge configurations for which the two terms in the covariant derivative
$\partial_{\mu}-igA_{\mu}$ are of the same order. In the Yang Mills
plasma this occurs for transverse modes for momenta $p\sim g^{2}T$
and amplitude $A^{T}\sim gT$ ; it can be seen from the partition
function that for the given momentum scale $A^{T}\sim gT$ is the
typical gauge field amplitude in the plasma. The time scales associated
with these momenta and amplitudes can be estimated from eq. (\ref{eq:RetardedSEYM})
below as $t\sim\left(g^{4}T\right)^{-1}$ - up to a logarithmic enhancement
that arises from accounting for interactions with plasma modes of
momenta $\sim T,gT$. The dynamics of these modes, including the effects
of interactions with the plasma, is described by B\"{o}deker's classical
EFT and takes the form of an overdamped Langevin equation~\cite{Bodeker:1998rg,Blaizot:2001nr,Blaizot:1993be,Arnold:1996dy}.
This dynamics drives the diffusion of the Chern-Simons number and
leads to the non-zero sphaleron diffusion constant in eq. (\ref{eq:Gammasphal}).

\subsection{1-Loop Gluon Spectral Density}

Returning to the analysis of the pert modes, we note that to see the finite
decay width of the correlators in eq.~(\ref{eq:phiGGInt}), we must
compute the transverse gluon spectral densities to at least 1-loop
order. In this section, we present the results of this computation,
deferring the details to appendix~(\ref{sec:Full-1-loop-comp}).
As we will see in section~\ref{sec:TransGluonSpecFunc}, this will
be insufficient for an exact leading-order computation of the dissipation
coefficient -- a consequence of subtleties related to IR divergences
and the breakdown of perturbativity of the bare loop expansion in
finite-temperature Yang--Mills theories. Nevertheless, it will provide
an understanding of the physical processes dominating perturbative
dissipation, and allow an order-of-magnitude estimate of the friction
coefficient from eq.~(\ref{eq:PertDiss}).

\begin{figure}[H]
a) \includegraphics[scale=0.52]{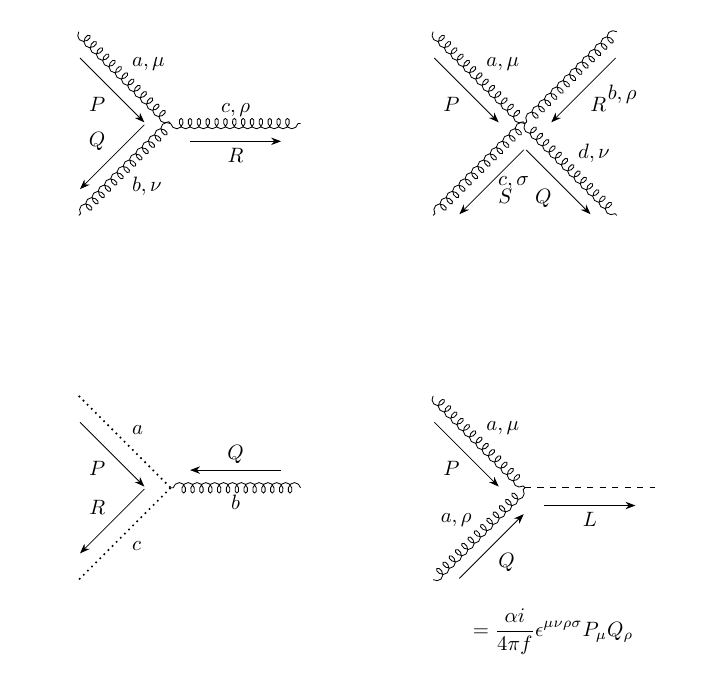} b) \includegraphics[scale=0.56]{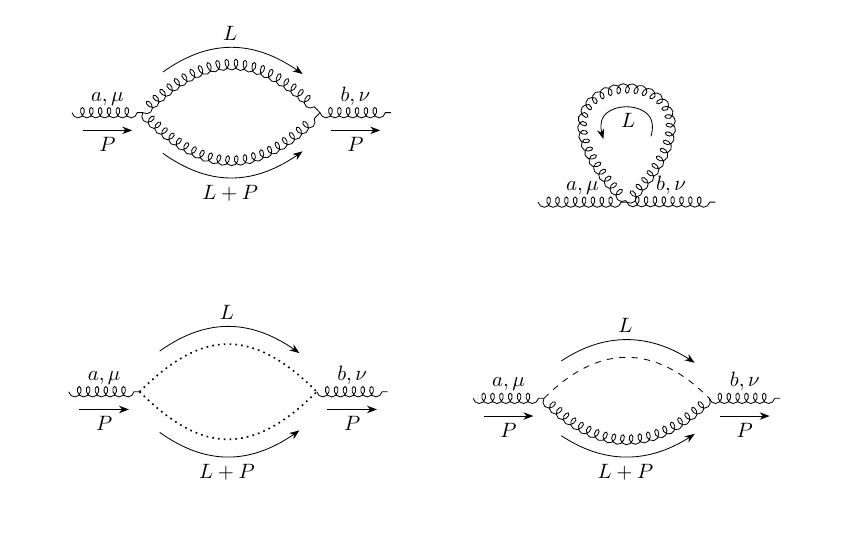}
\caption{Part a) Interaction vertices in the Euclidean theory, with curly,
dotted, and dashed lines denoting gluons, ghosts, and the inflaton,
respectively. Part b) 1-loop diagrams: clockwise from top left are
the two diagrams with gluons in the internal lines, followed by the
diagram with an inflaton-gluon exchange, and lastly the one with ghosts.}
\label{fig:verticesn!loopdgrm} 
\end{figure}

We begin by evaluating the relevant Euclidean 1-loop diagrams implied
by the interaction terms in eq.~(\ref{eq:EucLag}). These are shown
in Fig.~(\ref{fig:verticesn!loopdgrm}) and include, in addition
to the usual gluon and ghost diagrams, one involving an inflaton exchange.
The contribution from the latter is higher order in $\alpha/f$, and
therefore will be ignored for our leading-order estimates of the friction
coefficient. Nevertheless, it has been included here for completeness
(and ease of generalization to different models).

The combined contributions to the self-energy corrections, denoted
by $\bar{\Pi}_{\mu\nu}$, from the 1-loop diagrams, computed using
the bare propagators 
\begin{equation}
\langle A^{a}_{\mu}\!\left(Q\right)A^{b}_{\nu}\!\left(L\right)\rangle=\delta^{ab}\delta_{\mu\nu}\frac{\bar{\delta}\!\left(Q+L\right)}{Q^{2}}\qquad\!\left(\xi=1,\quad\text{Feynman gauge}\right)
\end{equation}
\begin{equation}
\langle\delta\phi\!\left(Q\right)\delta\phi\!\left(L\right)\rangle=\frac{\bar{\delta}\!\left(Q+L\right)}{Q^{2}+m^{2}_{\phi}},
\end{equation}
is given by 
\begin{align}
\bar{\Pi}_{\mu\nu}\!\left(Q\right) & =\left(\frac{\alpha}{2\pi f}\right)^{2}\epsilon^{\eta\mu\rho\beta}\epsilon^{\alpha\beta\gamma\nu}Q_{\eta}Q_{\gamma}\int_{\Sigma,L}\!\left(\frac{1}{L^{2}+m^{2}_{\phi}}\right)\frac{Q_{\rho}Q_{\alpha}}{\!\left(Q-L\right)^{2}}\nonumber \\
 & \quad+\frac{g^{2}N_{c}}{2}\int_{\Sigma,L}\frac{\delta_{\mu\nu}\!\left[-4Q^{2}+2\!\left(D-2\right)L^{2}\right]+\!\left(D+2\right)Q_{\mu}Q_{\nu}-4\!\left(D-2\right)L_{\mu}L_{\nu}}{L^{2}\!\left(U-L\right)^{2}}.\label{eq:1loopSE}
\end{align}
Here, $Q^{2}\equiv q^{2}_{n}+\vec{q}\cdot\vec{q}$, $m^{2}_{\phi}\equiv V''\!\left(\phi\!\left(t_{0}\right)\right)$
(assumed to satisfy $m_{\phi}\ll T$), and 
\begin{equation}
\int_{\Sigma,L}\equiv T\sum_{l_{n}}\int\frac{d^{3}l}{\!\left(2\pi\right)^{3}}.
\end{equation}
Once again gauge invariance implies that $\bar{\Pi}_{\mu\nu}\!\left(Q\right)$
takes the form 
\begin{equation}
\bar{\Pi}_{\mu\nu}\!\left(Q\right)=\bar{\Pi}^{T}\!\left(Q\right)\mathbb{P}^{T}_{\mu\nu}\!\left(Q\right)+\bar{\Pi}^{L}\!\left(Q\right)\mathbb{P}^{L}_{\mu\nu}\!\left(Q\right)\label{eq:transverse-long}
\end{equation}
where the transverse and longitudinal projectors 
are as in eqs.~(\ref{eq:transProj}) and (\ref{eq:LongProj}).
In terms of the projections $\bar{\Pi}^{T\!\left(L\right)}\!\left(Q\right)$,
the 1-loop corrected gluon propagators are 
\begin{equation}
\Pi_{\nu\beta}\!\left(Q\right)=\frac{1}{Q^{2}+\bar{\Pi}^{T}\!\left(Q\right)}\mathbb{P}^{T}_{\nu\beta}\!\left(Q\right)+\frac{1}{Q^{2}+\bar{\Pi}^{L}\!\left(Q\right)}\mathbb{P}^{L}_{\nu\beta}\!\left(Q\right)+\frac{Q_{\nu}Q_{\beta}}{\!\left(Q^{2}\right)^{2}}
\end{equation}
where the gauge parameter has been taken as $\xi=1$. Since the dissipation
coefficient eq.~(\ref{eq:PertDiss}) depends only on the transverse
spectral functions $\rho^{T}\!\left(q_{0},\vec{q}\right)$, here we
simplify and evaluate $\bar{\Pi}^{T}\!\left(Q\right)$. Taking the
transverse projection of eq.~(\ref{eq:1loopSE}), and carrying out
the loop integrals as detailed in appendix~\ref{sec:Full-1-loop-comp},
we see that $\bar{\Pi}^{T}\!\left(Q\right)$ is given by a sum of
vacuum and thermal contributions,\footnote{The Matsubara sum of any function of Euclidean momenta can be shown
to split into a vacuum and a thermal contribution.} 
the explicit expressions
for which are given in eqs.~(\ref{eq:FullVacSE}) and (\ref{eq:FullThermSE}).

The vacuum contributions are identical to those in the zero-temperature
Euclidean theory. This may be seen by taking the limit $\lim_{T^{-1}\rightarrow\infty}T\sum_{l_{n}}=\int\frac{dl_{0}}{2\pi}$
in eq.~(\ref{eq:1loopSE}). Consequently, these contributions involve
UV divergences which necessitate renormalization of the theory. On
the other hand, the thermal contributions involve loop momenta cut
off by the temperature and contribute no further divergences. Therefore,
to renormalize the theory we may proceed as in the zero-temperature
case. In eq.~(\ref{eq:FullVacSE}) we have used the $\overline{\mathrm{MS}}$
renormalization scheme to compute the vacuum contributions. For the
estimates of section \ref{sec:friction} the vacuum terms make $\sim g^{2}$
suppressed contributions relative to the thermal terms and hence will
be neglected.

To obtain the relevant Lorentzian quantities, we may now analytically
continue the Euclidean energies.\footnote{This analytic continuation involves two steps: first, from discrete
to continuous values of $u_{n}$, which is non-unique. The Matsubara
prescription as defined above provides a unique way to analytically
continue. Once continued to continuous Euclidean variables, the continuation
to real variables is given by $p_{n}\rightarrow-ip_{0}\pm0^{+}$.} Of particular relevance is the self-energy correction to the retarded
Green's function for soft external momenta $\!\left(q_{0},\vec{q}\right)\ll T$.
To leading order in $\left(q_{0}/T,q/T\right)$, the thermal contribution
is given by 
\begin{align}
\bar{\Pi}^{R,T}\!\left(q_{0},\vec{q}\right) & \equiv\bar{\Pi}^{T}\!\left(q_{n}\rightarrow-i\!\left(q_{0}+i0^{+}\right),\vec{q}\right)\nonumber \\
 & =\left(\frac{\alpha T^{2}}{24}\right)\left[\left(\frac{1}{2\pi f}\right)^{2}\!\frac{\alpha\!\left(3q^{4}+q^{4}_{0}\right)}{q^{2}}-\!\left(D-2\right)4\pi N_{c}\!\left(1-\frac{q^{2}_{0}}{q^{2}}\right)\right.\nonumber \\
 & \left.+4\pi N_{c}\left(D-2\right)^{2}-c\left(\alpha,g\right)\frac{\!\left(q^{2}-q^{2}_{0}\right)}{q^{3}}\!\left\{ q_{0}\ln\!\left(\frac{q_{0}+i0^{+}-q}{q_{0}+i0^{+}+q}\right)+q\right\} \right].\label{eq:RetardSE}
\end{align}
where 
\begin{equation}
c\left(\alpha,g\right)\equiv\left\{ -\!\left(\frac{1}{2\pi f}\right)^{2}\!\alpha\left(q^{2}-q^{2}_{0}\right)+\!\left(D-2\right)4\pi N_{c}\right\} 
\end{equation}
and $q\equiv\left(\vec{q}\cdot\vec{q}\right)^{1/2}$ is the magnitude
of the spatial momentum.\footnote{The dependence on $q$ is a result of carrying out angular parts of
the loop integrals.} The simple expression above results from approximating the loop
integrals to leading order in $q/l$, where $l$ is the magnitude
of the internal loop momentum and $l\sim T$. This is the so-called
hard thermal loop (HTL) approximation. Note that the above self-energy
correction is of order $\!\left(gT\right)^{2}$, and would be comparable
to the bare propagator if $q\sim gT$. Hence, for eq.~(\ref{eq:RetardSE})
to correspond to a perturbative correction to the bare propagator,
the gluon momentum should satisfy $gT\ll\!\left(q_{0},\vec{q}\right)\ll T$.
For `soft' external momenta, with $\!\left(q_{0},\vec{q}\right)\sim gT$,
the bare propagators must be resummed to include the 1-loop HTL correction
-- and similarly for interaction vertices with soft momenta on all
external legs -- to restore the perturbativity of loop expansions.
This is called HTL resummation~(\cite{Pisarski:1988vd,Braaten:1989mz,Laine:2016hma,Bellac:2011kqa,Gross:1980br}).

\section{Analytic Structure of the retarded gluon propagator}

\label{sec:TransGluonSpecFunc}

In this section, we review the analytic structure of the HTL approximated
\footnote{It can be seen from eq. (\ref{eq:FullThermSE}) that the analytic
structure of the 1-loop corrected gluon retarded Green's function
does not change for momenta $q_{0},q\sim T$.} retarded gluon propagator and describe the associated energy scales
and physical processes contributing to its imaginary part $\text{Im}\!\left(\bar{\Pi}^{R,T}\!\left(q_{0},q\right)\right)=\rho^{T}\!\left(q_{0},q\right)$.
Since the contribution from the loop with an inflaton exchange is
higher order in $\alpha/f$ 
, we ignore these terms in the following. Then, the self energy in
eq.~(\ref{eq:RetardSE}) reduces to that in finite-temperature Yang--Mills
theory, given by 
\begin{align}
\bar{\Pi}^{R,T}\!\left(q_{0},q\right) & =-\!\frac{\!\left(D-2\right)N_{c}g^{2}T^{2}}{24}\!\left(1-\frac{q^{2}_{0}}{q^{2}}\right)\!\left\{ \frac{q_{0}}{q}\ln\!\left(\frac{q_{0}+i0^{+}-q}{q_{0}+i0^{+}+q}\right)+2\right\} \nonumber \\
 & +g^{2}N_{c}\frac{\!\left(D-2\right)^{2}T^{2}}{24}.\label{eq:RetardedSEYM}
\end{align}
For an elaborate discussion of the analytic structure of gluon propagators
in this case, see (\cite{Kapusta:2006pm,Laine:2016hma,Bellac:2011kqa}).

\subsection{Screening Mass}

Defined as $\bar{\Pi}^{R,T\left(L\right)}\!\left(q_{0}=0,q\rightarrow0\right)$,
the screening mass corresponds to the inverse of length scale beyond
which static magnetic (electric) fields are screened in the thermal
plasma. For static electric fields, this is called the Debye mass,
and can be obtained from the 1-loop self-energy correction of the
longitudinal gluon modes as $m^{2}_{D}=N_{c}g^{2}T^{2}/3$. On the
other hand, from eq.~(\ref{eq:RetardSE}) we see 
\begin{align}
\bar{\Pi}^{R,T}\!\left(q_{0}=0,q\rightarrow0\right)=0,\label{eq:ScreenMass}
\end{align}
that is, to 1-loop order the screening mass for static magnetic fields
is zero. This continues to hold even if $\bar{\Pi}^{R,T}\left(q_{0},q\right)$
is corrected to include HTL-resummed propagators. The lack of a thermal
screening mass for the transverse gluons at this order results in
uncured IR divergences in finite-temperature Yang--Mills theory~\cite{Linde:1980ts,Gross:1980br};
to mitigate these, it is assumed that at higher orders a thermal mass
$m_{g}\sim g^{2}T$ is generated. Such a mass cannot be
computed in perturbation theory since at this scale thermal Yang Mills
looses perturbative computability as noted at the of section \ref{subsec:PiGG2Pig};
$n$- loop diagrams for $n\geq2$ contribute to $m_{g}$. Nevertheless,
this scale may be motivated by noting that the IR divergences here
arise from the static/zero energy transverse gluon modes and therefore,
can be mapped to the IR divergences in an equivalent three dimensional
non-Abelian gauge theory at high T. The corresponding 3$D$ theory
is a pure gauge theory with coupling $g^{2}T$. Since there are no
dimensionless small parameters in this theory, it cannot be treated
perturbatively. On the other hand, being a non-abelian theory it is
expected to confine and gain a mass gap, here at $\sim g^{2}T$ since
this is the only mass scale in the theory backed up by numerical work
\cite{Cucchieri:2000cy}. In the following, we will assume this to
hold, and treat $m_{g}$ as the IR cutoff scale in our computations
of the dissipation coefficient.

\subsection{Plasmons}

\begin{figure}[h]
\centering \includegraphics[scale=0.7]{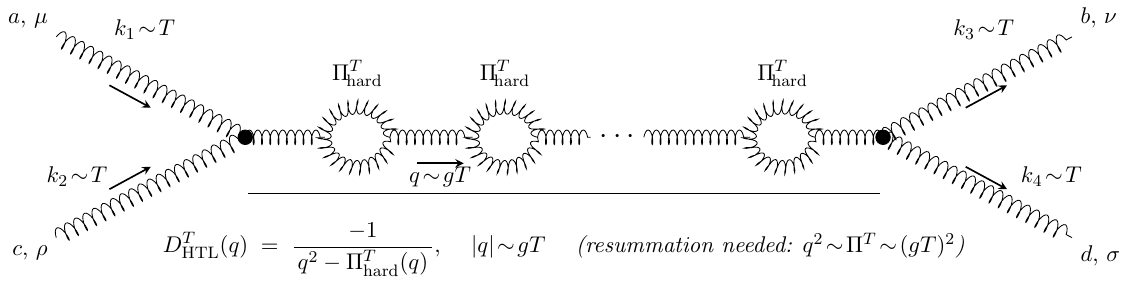}
\caption{Scattering of hard gluon modes with the exchange of a soft HTL resummed
effective gluon. This process gives the leading log contribution to
the plasmon decay rate. Note that although the momentum conservation naively looks impossible,
the vertex is an HTL-resummed vertex that allows this kinematics.\cite{PhysRevD.47.R769}}
\label{fig:Decay} 
\end{figure}

Plasmons are the on-shell effective gluon modes in the thermal plasma
with a dispersion relation $\omega$ 
given by 
\begin{equation}
-\omega^{2}+q^{2}+\bar{\Pi}^{R,T}\!\left(\omega,q\right)=0\label{eq:DispRel}
\end{equation}
and characterized by the scale $gT$. For $q\gg gT$, the dispersion
relation is modified from the free gluon one by a thermal mass $m^{2}_{p}=N_{c}\!\left(gT\right)^{2}/2$,
whereas for $q\lesssim gT$, in addition to a thermal mass $\sim m_{p}$,
$\omega$ also sees a modified $q$ dependence \cite{Laine:2016hma,Bellac:2011kqa}.
From eq.~(\ref{eq:RetardedSEYM}) we note that $\omega^{2}\!\left(0\right)=m^{2}_{D}/3$,
implying that it takes a finite amount of energy to create long-wavelength
plasmon oscillations in the medium.\footnote{The effect of terms proportional to $\alpha T/f$ from eq.~(\ref{eq:RetardSE})
is to shift the poles to the left, closer to the free dispersion relation.} Eq.~(\ref{eq:RetardedSEYM}) also seems to imply that the plasmon
decay rate determined by the imaginary part of $\bar{\Pi}^{R,T}\left(\omega,q\right)$is
zero (since $\omega>q$) at this order -- a consequence of on-shell
3-gluon scattering being forbidden by kinematics. $\text{Im}\left(\bar{\Pi}^{R,T}\right)$
is however non-zero at next-to-leading order in $g$, resulting in
a finite plasmon decay rate defined as 
\begin{equation}
\gamma\!\left(q\right)\equiv\frac{\text{Im}\!\left(\bar{\Pi}^{R,T}\!\left(\omega,q\right)\right)}{2\omega}\label{eq:DissRateDef}
\end{equation}
where, $\omega$ is the dispersion relation computed to leading order
in $g$. In fact, the imaginary part arises from soft loop momenta
and therefore to get an estimate we must resort to computing self-energy
corrections with HTL-resummed propagators and vertices. The leading
contribution to the imaginary part obtained this way for $q\sim T$
can be seen to correspond to scattering of hard effective gluons through
the exchange of a soft spacelike transverse gluon (Fig.~(\ref{fig:Decay})).
The corresponding amplitude computed with an HTL-resummed propagator
is given by~\cite{Pisarski:1992wqc,PhysRevLett.64.1338,Rebhan:1992ca,PhysRevD.47.R769,Burgess:1991wc,Braaten:1990it}:
\begin{equation}
\gamma=\alpha N_{c}T\ln\!\left(\frac{m_{D}}{m_{g}}\right)\sim\alpha N_{c}T\ln\!\left(\frac{1}{g}\right)\label{eq:DissRAte}
\end{equation}
where the dependence on $m_{D}$ arises from $\bar{\Pi}^{R,T}\!\left(q\right)$
and that on $m_{g}$ by imposing an IR cutoff on the otherwise divergent
amplitude.

\subsection{Landau Damping}

In contrast to the plasmon modes, effective gluons with spacelike
momenta acquire imaginary parts to their energy at 1-loop order, given
by (eq.~(\ref{eq:RetardedSEYM})): 
\begin{align}
\text{Im}\!\left(\bar{\Pi}^{R,T}\!\left(q_{0},q\right)\right) & =-\!\left[\frac{\pi\!\left(D-2\right)g^{2}T^{2}N_{c}}{24}\right]\!\left(1-\frac{q^{2}_{0}}{q^{2}}\right)\frac{q_{0}}{q}\qquad\forall\;\vert q_{0}\vert<q,\label{eq:LandauDamp}\\
 & =0\qquad\qquad\forall\;\vert q_{0}\vert>q.\nonumber 
\end{align}
The imaginary value arises from a net energy loss through scattering
events with hard gluons in the plasma. Momentum is exchanged between
the soft gluons and those hard gluons whose velocity along the wavevector
$\vec{q}$ matches the phase velocity of the soft mode. The fact that
$\text{Im}\!\left(\bar{\Pi}^{T}\!\left(q_{0},q\right)\right)$ is
non-zero only for $\vert q_{0}\vert<q$ may then be derived as 
\begin{equation}
\vec{q}\cdot\hat{k}-q_{0}=0\implies\cos\!\left(\theta\right)=\frac{q_{0}}{q},
\end{equation}
where $\hat{k}$ is the velocity vector of the hard mode. Since $\vert\cos\!\left(\theta\right)\vert\le1$,
loss of energy through this process occurs for $\vert q_{0}/q\vert\le1$.

\section{Friction Coefficient}

\label{sec:friction}

Using the results of the previous sections, here we estimate the coefficient
of $\dddot{\bar{\phi}}$, henceforth denoted as $\Gamma_{\mathrm{pert}}$,
in the effective inflaton EOM. From eqs.~(\ref{eq:PertDiss}) and
(\ref{eq:RhoGG}), the term proportional to $\dddot{\bar{\phi}}\left(t\right)$
in the inflaton EOM eq. (\ref{eq:phibackEOM}) is 
\begin{align}
\Gamma_{\mathrm{pert}}\dddot{\phi}\!\left(t\right) & =-\!\left(\frac{\alpha}{16\pi f}\right)^{2}\dddot{\bar{\phi}}\!\left(t\right)\frac{1}{3!}\partial^{3}_{p_{0}}\rho_{\tilde{G}G}\!\left(p_{0},0\right)_{\vert p_{0}=0}\nonumber \\
 & =-\!\left(\frac{\alpha}{2\pi f}\right)^{2}\dddot{\bar{\phi}}\!\left(t\right)\delta^{aa}T^{-1}\mathcal{I}\label{eq:GammpertExpr}
\end{align}
where 
\begin{align}
\mathcal{I} & \equiv\int_{q}\int^{+\infty}_{-\infty}\frac{dq_{0}}{\pi}\,n_{B}\!\left(q_{0}\right)\!\left(1+n_{B}\!\left(q_{0}\right)\right)\!\left[\rho^{T}\!\left(q_{0},q_{i}\right)\right]^{2}q^{2}\label{eq:GammaIntgrl}
\end{align}
and we have used $n_{B}\!\left(-q_{0}\right)=-\!\left(1+n_{B}\!\left(q_{0}\right)\right)$
and $\rho^{T}\!\left(-q_{0},q\right)=-\rho^{T}\!\left(q_{0},q\right)$.\footnote{Evaluating $\partial^{3}_{p_{0}}\rho_{\tilde{G}G}\!\left(p_{0},0\right)_{\vert p_{0}=0}$
reduces to eq.~(\ref{eq:GammpertExpr}) because the integrals are
regular at $p_{0}=0$.} The dominant contributions to $\mathcal{I}$ arise from Landau damping
and plasmon decay.

To compute the contributions from these processes, we begin by expressing
the gluon transverse spectral density, defined as 
\begin{equation}
\rho^{T}\!\left(q_{0},q\right)=\text{Im}\!\left(\Pi^{R,T}\!\left(q_{0},q\right)\right),
\end{equation}
in the form 
\begin{equation}
\rho^{T}\!\left(q_{0},q_{i}\right)=Z_{<}\!\left(q_{0},q\right)\theta\!\left(1-\frac{q^{2}_{0}}{q^{2}}\right)+Z_{>}\!\left(q_{0},q\right)\theta\!\left(\frac{q^{2}_{0}}{q^{2}}-1\right)
\end{equation}
where 
\begin{equation}
Z_{<\!\left(>\right)}\!\left(q_{0},q\right)=-\frac{\text{Im}\!\left(\bar{\Pi}^{T}_{<\!\left(>\right)}\!\left(q_{0},q\right)\right)}{\!\left(-q^{2}_{0}+q^{2}+\text{Re}\!\left(\bar{\Pi}^{T}_{<\!\left(>\right)}\!\left(q_{0},q\right)\right)\right)^{2}+\text{Im}\!\left(\bar{\Pi}^{T}_{<\!\left(>\right)}\!\left(q_{0},q\right)\right)^{2}}.
\end{equation}
The complete leading-order contributions to real and imaginary parts
of $\bar{\Pi}^{T}\!\left(q_{0},q\right)$ arise from 1-loop diagrams
computed with resummed propagators and vertices. Here, for an order-of-magnitude
estimate of the dissipation coefficient, we use the HTL-approximated
form of eq.~(\ref{eq:RetardedSEYM}) for $\bar{\Pi}^{T}_{<}\!\left(q_{0},q\right)$
and $\text{Re}\!\left(\bar{\Pi}^{T}_{>}\!\left(q_{0},q\right)\right)$,
and the leading-order decay rate from eq.~(\ref{eq:DissRAte}) for
$\text{Im}\!\left(\bar{\Pi}_{>}\!\left(q_{0},q\right)\right)$ near
the plasmon poles. As we will see below, the Landau damping contribution
is dominated by transverse spacelike gluons of soft momenta, for which
resummed propagators and vertices modify the momentum dependence in
eq.~(\ref{eq:RetardedSEYM}) by order-one factors. Therefore, our
estimates for contributions from Landau damping are only accurate
up to these factors.

Then, denoting the contribution to $\mathcal{I}$ from Landau damping
as $\mathcal{I}_{l}$, we have 
\begin{align}
\mathcal{I}_{l} & \equiv\int_{q}\int^{+q}_{-q}dq_{0}\,n_{B}\!\left(q_{0}\right)\!\left(1+n_{B}\!\left(q_{0}\right)\right)\!\left[Z_{<}\!\left(q_{0},q_{i}\right)\right]^{2}q^{2}\nonumber \\
 & \approx\left(\frac{\pi}{4}m^{2}_{D}\right)^{2}\frac{T^{2}}{2\pi^{3}}\left\{ \int^{\infty}_{m_{g}}dq\int^{+q}_{-q}dq_{0}\,\left[\frac{q^{2}-q^{2}_{0}}{\!\left(-q^{2}_{0}+q^{2}+m^{2}_{D}\frac{q^{2}_{0}}{q^{2}}\right)^{2}+\!\left(\pi\frac{m^{2}_{D}}{4}\frac{q_{0}}{q}\right)^{2}}\right]^{2}\frac{1}{q^{2}}\right\} 
\end{align}
where, realizing that the integrand is IR divergent with dominant
contributions arising from $m_{g}\lesssim q\lesssim m_{D}$, we have
approximated the Boltzmann factors as 
\begin{equation}
n_{B}\!\left(q_{0}\right)\!\left(1+n_{B}\!\left(q_{0}\right)\right)\approx\!\left(\frac{T}{q_{0}}\right)^{2}
\end{equation}
and the denominator of the spectral functions to leading order in
$q_{0}/q<1$. Finally, we have also introduced the IR cutoff $m_{g}$
to cut-off the integral at the softest perturbative modes. This simplifies
the integral to leading order in $g$ as 
\begin{align}
\mathcal{I}_{l}\approx\frac{T^{2}}{8\pi}\frac{m^{2}_{D}}{m^{2}_{g}}.
\end{align}

Turning to the contribution from plasmon decay -- in the parametric
range $\!\left(q_{0}>q\right)$ -- we see that eq.~(\ref{eq:GammaIntgrl})
is dominated by the region close to the plasmon poles, i.e.\ $-\omega^{2}+q^{2}+\text{Re}\!\left(\Pi^{T}\!\left(\omega,q\right)\right)+i\text{Im}\!\left(\Pi^{T}\!\left(\omega,q\right)\right)=0$.
We may then approximate the integrand near the pole in terms of the
variable $\delta q_{0}=\text{Re}\!\left(\omega\!\left(q\right)\right)-q_{0}$
as 
\begin{align}
\mathcal{I}_{p} & \equiv\int_{q}\int^{+\infty}_{-\infty}\frac{dq_{0}}{\pi}\,n_{B}\!\left(q_{0}\right)\!\left(1+n_{B}\!\left(q_{0}\right)\right)\!\left[Z_{2}\!\left(q_{0},q_{i}\right)\right]^{2}q^{2}\theta\!\left(q^{2}_{0}-q^{2}\right)\nonumber \\
 & \approx2\int_{q}\int^{+\infty}_{-\infty}\frac{d\,\delta q_{0}}{\pi}\,n_{B}\!\left(\bar{\omega}\right)\!\left(1+n_{B}\!\left(\bar{\omega}\right)\right)\!\left[\frac{\gamma\!\left(q\right)}{\delta q^{2}_{0}+\gamma^{2}\!\left(q\right)}\right]^{2}\nonumber \\
 & \approx\frac{T^{3}}{24\gamma}
\end{align}
where $\bar{\omega}=\text{Re}\!\left(\omega\!\left(q\right)\right)$
and $\gamma$ is the decay rate defined in eq.~(\ref{eq:DissRateDef})
and estimated in eq.~(\ref{eq:DissRAte}). Combining $\mathcal{I}_{p}$
and $\mathcal{I}_{l}$ in eq.~(\ref{eq:PertDiss}) we have 
\begin{align}
\Gamma_{\mathrm{pert}}\dddot{\phi}\!\left(t\right) & =-\!\left(\frac{\alpha}{2\pi f}\right)^{2}\dddot{\bar{\phi}}\!\left(t\right)\delta^{aa}T^{-1}\!\left(\mathcal{I}_{p}+\mathcal{I}_{l}\right)\nonumber \\
 & =-\frac{\alpha}{4\pi^{2}}\dddot{\bar{\phi}}\!\left(t\right)\frac{\!\left(N^{2}_{c}-1\right)}{24}\frac{T}{f^{2}}\!\left(\frac{1}{N_{c}\ln\!\left(\frac{m_{D}}{m_{g}}\right)}+\mathcal{O}\!\left(1\right)\frac{N_{c}}{4\pi^{2}}\right).\label{eq:Gammapert}
\end{align}
The above would correspond to an additional perturbative contribution
to the total friction seen by the background inflaton EOM if $\dddot{\bar{\phi}}\propto-\dot{\bar{\phi}}.$
On a generic inflationary background, this is not guaranteed and the
triple and first derivatives are related as 
\begin{equation}
\dddot{\bar{\phi}}=-s\!\left(\epsilon_{v},\eta_{v}\right)H^{2}\dot{\bar{\phi}}\label{eq:dddphi2dphi}
\end{equation}
where $s\left(\epsilon_{v},\eta_{v}\right)$ is a function of order
two in the slow roll parameters, and could take both positive and
negative values -- the latter corresponding to feedback to, rather
than dissipation from, the background. Both are, of course, back reaction
somewhat in contrast with what is typically said in the literature
as a confusing convention.

Combining these leading order perturbative contributions with the
non-perturbative contribution from hot sphaleron transitions, the total
effective friction term in the inflaton EOM (eq. (\ref{eq:phibackEOM}))
becomes 
\begin{align}
\Gamma_{\text{eff}}\dot{\bar{\phi}} & =\frac{\Gamma_{\text{sphal}}}{2f^{2}T}\dot{\bar{\phi}}+\Gamma_{\text{pert}}\dddot{\bar{\phi}}\nonumber \\
 & =\dot{\bar{\phi}}\left(\frac{\Gamma_{\text{sphal}}}{2f^{2}T}+\Gamma_{\text{pert}}s\!\left(\epsilon_{v},\eta_{v}\right)H^{2}\right).\label{eq:Gamma_eff_def}
\end{align}
\footnote{In terms of these, the background EOM in eq. (\ref{eq:phibackEOM})
can be written as 
\begin{equation}
\ddot{\bar{\phi}}+3H\dot{\bar{\phi}}+V_{\mathrm{eff}}'\!\left(\bar{\phi}\right)+\dot{\bar{\phi}}\,\Gamma_{\mathrm{tot}}\approx0
\end{equation}
where $V_{\text{eff}}\left(\bar{\phi}\right)$ includes the following
term in eq. (\ref{eq:phiGGInt}) 
\begin{equation}
\langle\tilde{G}^{\mu\nu}_{a}G^{a}_{\mu\nu}\rangle\!\left(t,x\right)\ni+i\frac{\alpha\left(\bar{\phi}\!\left(t\right)-\bar{\phi}\!\left(t_{0}\right)\right)}{16\pi f}\!\int^{t}_{t_{0}}dt'\int d^{3}x'\left\langle \left[\tilde{G}^{\mu\nu}_{I\,a}G^{a}_{I\,\mu\nu}\!\left(t,x\right),\epsilon^{\alpha\beta\gamma\delta}G_{I\,b,\alpha\beta}G^{b}_{I\,\gamma\delta}\!\left(t',x'\right)\right]\right\rangle _{0}.
\end{equation}
}
This corresponds to quantum corrections to the tree level inflaton
potential induced by interactions with the bath - here, this is just
the induced inflaton mass. The factor of $\bar{\phi}\left(t\right)-\bar{\phi}\left(t_{0}\right)$
accounts for the change in the background from $\bar{\phi}\left(t_{0}\right)$
to $\bar{\phi}\left(t\right)$. Here, to obtain the first term we have used eq. (\ref{eq:T1}) and
eq. (\ref{eq:Gammasphal}). The sphaleron diffusion rate $\Gamma_{\text{sphal}}$
was estimated using lattice simulations based on B\"{o}deker's effective
theory for non-perturbative the magnetic IR modes in \cite{Moore:2010jd}
giving 
\begin{equation}
\Gamma_{\text{sphal}}=\kappa\left(\alpha,N_{c},N_{f}\right)\alpha^{5}T^{4}.\label{eq:GammasphalExpre}
\end{equation}
Here $\kappa$ is an $\sim O\left(100\right)$ coefficient for $SU(3)$
Yang Mills theory with a weak logarithmic dependence on $\alpha$
and $\propto N^{5}_{c}$ for large $N_{c}$. Eqs. (\ref{eq:Gammapert},
\ref{eq:Gamma_eff_def}, and \ref{eq:GammasphalExpre}) then imply
\begin{equation}
\boxed{\Gamma_{\text{eff}}=\frac{\kappa\left(\alpha,N_{c},N_{f}\right)\alpha^{5}T^{3}}{2f^{2}}+\alpha\,s\!\left(\epsilon_{v},\eta_{v}\right)\frac{H^{2}T}{f^{2}}\,\frac{\!\left(N^{2}_{c}-1\right)}{96\pi^{2}}\!\left(\frac{1}{N_{c}\ln\!\left(\frac{m_{D}}{m_{g}}\right)}+\mathcal{O}\!\left(1\right)\frac{N_{c}}{4\pi^{2}}\right)}.
\end{equation}
In the following, we will refer to the first and second terms in the
above as the non-perturbative $\left(\Gamma_{\text{eff,np}}\right)$
and perturbative $\left(\Gamma_{\text{eff,p}}\right)$ contributions
to $\Gamma_{\text{eff}}$ respectively. The next section describes
the parametrics of these contributions.

\subsection{Parametrics of $\Gamma_{\mathrm{eff}}$}

Firstly note, due to slow-roll suppression, the perturbative contribution
to the dissipation coefficient $\Gamma_{\mathrm{eff,p}}$ is negligible
compared to Hubble drag, 
\begin{equation}
Q_{\mathrm{pert}}\equiv\frac{\Gamma_{\mathrm{eff,p}}}{3H}<1,
\end{equation}
implying that in the strong regime of warm inflation the dissipation
coefficient is dominated by Chern-Simons diffusion. Then, depending
on whether the background evolution is dominated by dissipation to
the thermal bath or Hubble drag, the functional form of $s\!\left(\epsilon_{v},\eta_{v}\right)$
takes the following limiting forms 
\begin{align}
s\!\left(\epsilon_{v},\eta_{v}\right) & =\frac{1}{49}\!\left(5\eta^{2}_{v}+24\epsilon^{2}_{v}-33\eta_{v}\epsilon_{v}\right) & Q\gg1,\label{eq:Qlarge}\\
 & =-\!\left(4\epsilon^{2}_{v}-5\epsilon_{v}\eta_{v}+\eta^{2}_{v}\right) & Q\ll1\label{eq:Qsmall}
\end{align}
for a $V\!\left(\phi\right)$
quadratic in $\phi$. This is derived by taking time derivatives of
the background inflaton EOM (eq. \ref{eq:phibackEOM}) and simplifying
in terms of the standard slow-roll parameters defined for warm inflationary
scenarios as 
\begin{equation}
\epsilon_{v}\equiv\frac{M^{2}_{\mathrm{pl}}}{2\!\left(1+Q\right)}\!\left(\frac{V'}{V}\right)^{2},\qquad\eta_{v}\equiv\frac{M^{2}_{\mathrm{pl}}}{\!\left(1+Q\right)}\frac{V''}{V}.
\end{equation}
Further, time derivatives of $Q$ can be related to $\epsilon_{v}$
and $\eta_{v}$ using energy conservation relating the dissipated
inflaton kinetic energy and the radiation energy density $\left(\rho_{r}\right)$.
In a quasi-steady state, this is captured by 
\begin{equation}
\Gamma_{\text{eff}}\dot{\phi}^{2}\approx4\rho_{r}H
\end{equation}
where $\dot{\rho}_{r}\approx0$ and in our scenario $\Gamma_{\text{eff}}\approx\Gamma_{\text{eff,np}}$.
A complete derivation is given in (\ref{sec:dddphi2dphi}). Here we
quote the $Q\gg1$ and $Q\ll1$ limits as these correspond to well
understood parametric regimes~\cite{Berera:2008ar,graham2009density,Bastero-Gil:2011rva,Bastero-Gil:2014jsa,BasteroGil:2009ec,Kamali:2023lzq}.

\begin{figure}[h]
\centering \includegraphics[scale=0.47]{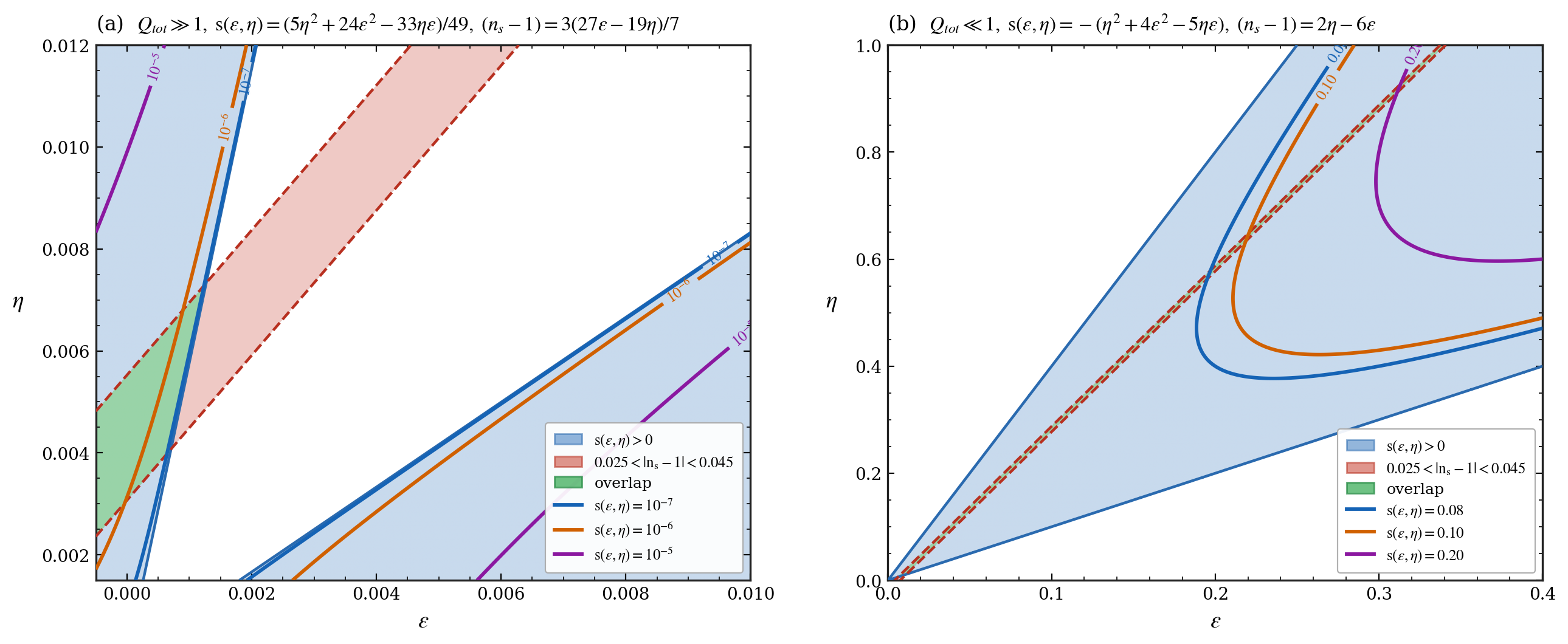} \caption{Plots of regions where $s\!\left(\epsilon,\eta\right)>0$ (blue) corresponding
to additional perturbative contributions to the friction coefficient,
and their overlap with the experimental bounds on the spectral index
$n_{s}$. The overlap region is shown in green, bounded by red dashed
lines. a) $Q_{\mathrm{tot}}\gg1$ and b) $Q_{\mathrm{tot}}\ll1$.
Note that the functional dependence of the spectral index on the slow-roll
parameters differs between the two backgrounds (\cite{Berghaus:2019whh,graham2009density,Hall:2003zp,Bastero-Gil:2011rva,Ramos:2013nsa}).
In the latter case, \cite{Berghaus:2019whh} notes that within the
presented computational framework, the observationally consistent
spectral index corresponds to the cold inflation limit.}
\label{fig:FricVsFeed} 
\end{figure}

With eqs.~(\ref{eq:Qsmall}) and (\ref{eq:Qlarge}), the slow roll
parameter ranges for $s\left(\epsilon_{v},\eta_{v}\right)>0$ and
$s\left(\epsilon_{v},\eta_{v}\right)<0$ corresponding to additional
perturbative dissipation and feedback respectively are shown in Fig.~(\ref{fig:FricVsFeed}).
Naively, within the regions where $s\left(\epsilon_{v},\eta_{v}\right)>0$,
it may seem possible that the lower suppression in $\alpha$ allows
a parameter region where the perturbative contributions dominate over
the nonperturbative diffusion. However, we note that the localization
approximation of the dissipation integrals required that the time
scales associated with background evolution are much longer than those
associated with interactions in the thermal bath, that is 
\begin{equation}
H\ll\alpha^{2}T.\label{eq:localAssump2}
\end{equation}
Comparing the magnitudes of the two contributions to $\Gamma_{\mathrm{eff}}$
for $N_{c}=3$ we see 
\begin{equation}
\frac{\Gamma_{\mathrm{eff,p}}}{\Gamma_{\mathrm{eff,np}}}\approx\frac{H^{2}}{\alpha^{4}T^{2}}\,s\!\left(\epsilon_{v},\eta_{v}\right)\,\mathcal{O}\!\left(10^{-5}\right)\ll1,
\end{equation}
implying that on generic slow-roll backgrounds the perturbative contributions
are negligible. Although, a workaround leading to significant perturbative
contributions seems plausible in scenarios where the localisation
constraint eq. (\ref{eq:localAssump2}) is relaxed and/or the slow-roll
suppression briefly violated. The former would require an understanding
of the full nonlocal dissipation kernels. Another interesting possibility
might be the inclusion of a large number of vector like fermion families.
These have been shown to suppress the sphaleron rate on an expanding
spacetime \cite{Berghaus:2025wdx,Berghaus:2024zfg,McLerran:1990de,Drewes:2023khq}while
increasing the perturbative contributions. We leave these directions
for future work. 

\begin{figure}[h]
\centering \includegraphics[scale=0.67]{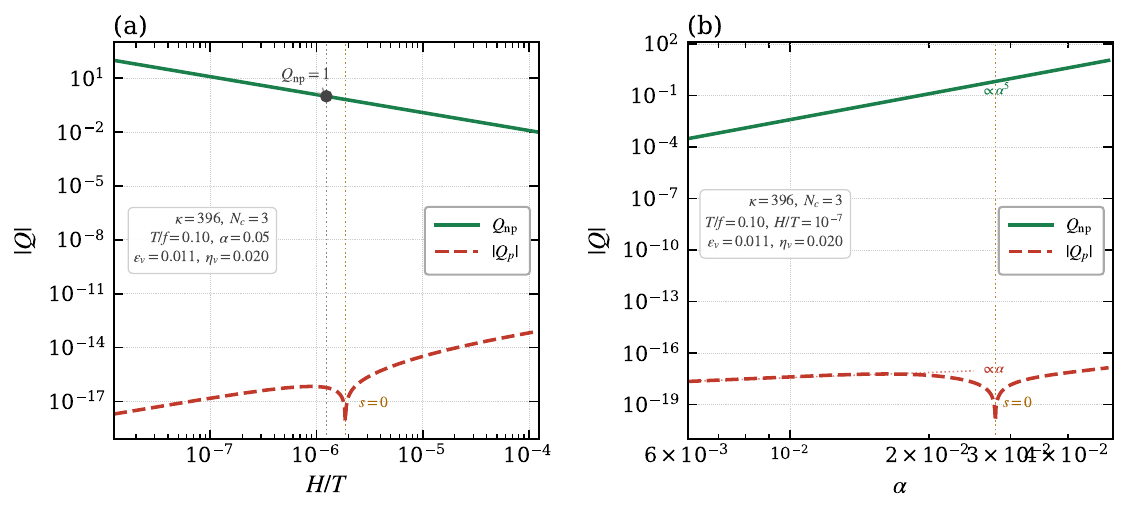} \caption{Parametrics of the perturbative ($Q_{p}$) and non-perturbative ($Q_{np}$)
contributions to the total friction coefficient $Q_{\mathrm{tot}}=\Gamma_{\mathrm{tot}}/H$.
The solid green and dashed red lines correspond to $Q_{np}$ and
$Q_{p}$ respectively. The parameters used are $\alpha=0.05$ (such
that $g=\sqrt{4\pi\alpha}<1$), 
$N_{c}=3$, $\kappa=396$ \cite{Moore:2010jd}, $T/f=0.1$, $\epsilon_{v}=0.011$,
and $\eta_{v}=0.020$. The choice of slow-roll parameters is such
that the spectral indices in the $Q_{\text{tot}}\gg1$ and $Q_{\text{tot}}\ll1$
regimes lie within $n_{s}\in(0.955,0.975)$. (In the latter case,
for the above set of parameters, the spectral index reduces to that
in cold inflation \cite{Berghaus:2019whh}.) a) Plots $\vert Q\vert$
as a function of $H/T$. As $H/T$ increases, $Q_{\mathrm{tot}}\approx Q_{np}$
decreases from $Q_{\mathrm{tot}}\gg1$ to $Q_{\mathrm{tot}}\ll1$ whereby
the background inflaton dynamics changes from being dominated by dissipative
friction to Hubble drag. The plot shows an overall power law dependence
in both $Q_{p}$ and $Q_{np}$. The dip in the red dashed line corresponds
to the parameter range where $Q_{\rm tot}\sim O\left(1\right)$ and the function
$s\left(\epsilon_{v},\eta_{v}\right)$ - due to its non-trivial dependence
on $Q_{\rm tot}$ passes through a zero. The range of $H/T$ is taken so as
to satisfy the bound eq. (\ref{eq:localAssump2}).
b) Plots $\vert Q\vert$
as a function of $\alpha$. The perturbative contribution shows a
nearly linear dependence with the dip arising from the $Q_{\rm tot}$ dependence
of $s\left(\epsilon_{v},\eta_{v}\right)$.}
\label{fig:parametrics} 
\end{figure}

\section{Summary}

We presented an analysis of the leading-order perturbative contributions
to dissipation in the minimal warm inflation (MWI) model in a regime
where the interaction rates within the plasma are much larger than
the rate of background inflaton evolution. This ensured that the thermal
bath is close to equilibrium throughout inflation. We first discussed
how a spontaneous CPT-violation of the non-conservative effective
equation of motion for $\bar{\phi}$ can be used to show that the
nonconservative terms will be of the form $\gamma_{b}\partial^{b}_{t}\bar{\phi}$
for $b\in$odd and the leading \emph{local }propagator contribution
to $\gamma_{1}\dot{\bar{\phi}}$ term is absent in theories with couplings
of the form $\phi\partial_{\mu}J^{\mu}$ regardless of the detailed
nature of $J^{\mu}$. The importance of the diffusion-related physics
that allows $\gamma_{1}\dot{\bar{\phi}}$ to exist in the case of
hot sphaleron and perhaps even helicity changing scattering channels
(eq.~(\ref{eq:estimatehere})) is also explained. The existence of
the $\gamma_{b}\partial^{b}_{t}\phi$ is also identified to be moments
of the same fundamental correlator (e.g.~see eq.~(\ref{eq:linearmoment})).

Using HTL-resummed EFT techniques valid in the momentum range $(\alpha T,T)$,
we estimated certain scattering-related IR mode corrections to dissipation
in MWI, given by eq.~(\ref{eq:PertDiss}), in terms of the spectral
density of the operator $\tilde{G}G$. Expressing this in terms of
the spectral density of transverse gluons to leading order in $\alpha$,
we note that contributions are dominated by Landau damping of soft
spacelike gluons and plasmon decay of hard (on-shell) modes. The latter
is dominated by scattering processes between hard thermal modes mediated
by a soft gluon. Since both these processes are IR-dominated, the
dissipation coefficient depends on the IR magnetic mass scale, leading
to an overall linear dependence of the perturbative friction contributions
($\Gamma_{\mathrm{eff,pert}}$) on $\alpha$.

Despite the smaller powers of $\alpha$ compared to the non-perturbative
sphaleron contribution, $\Gamma_{\mathrm{pert}}$ turns out to be
subdominant. This is primarily due to the slow-roll suppression in
\[
\dddot{\bar{\phi}}\sim\mathcal{O}\!\left(\epsilon^{2}\right)H^{2}\dot{\bar{\phi}},
\]
and additionally the constraint 
\begin{equation}
H\ll\alpha^{2}T,\label{eq:localize}
\end{equation}
which allowed the localization of the dissipation integrals in our
computation. Having understood the parametrics controlling the suppression
of the perturbative terms, which is 
\begin{equation}
\frac{\Gamma_{\mathrm{eff,p}}}{\Gamma_{\mathrm{eff,np}}}\approx\frac{H^{2}}{\alpha^{4}T^{2}}\,s\!\left(\epsilon_{v},\eta_{v}\right)\,\mathcal{O}\!\left(10^{-5}\right),
\end{equation}
we note that relaxing eq.~(\ref{eq:localize}) in scenarios where
the background is not slow-rolling could lead to significant contributions
from $\Gamma_{\mathrm{pert}}$. In such scenarios, one would have
to understand the time dependence of the non-local dissipation kernels.
Another interesting scenario would be the inclusion of vector like
fermions coupled to the gauge bosons. These have been shown to suppress
the effective friction from sphaleron diffusion in an expanding background~\cite{Broadberry:2025hep,Berghaus:2024zfg,Berghaus:2025wdx,Drewes:2023khq}
and will boost the perturbative contributions. Finally, it would be
interesting to validate or refute the speculation of other diffusing
modes coming from helicity changing scattering that can compete with
the sphaleron mode. We hope to explore these directions in future
works.

\acknowledgments
D.J.H.C. acknowledges the support by the U.S. Department of Energy,
Office of Science, Office of High Energy Physics under Award Number
DE-SC0017647.  N.S. acknowledges partial support from the Ray MacDonald
Fund and Wisconsin Alumni Research Fund at the University of
Wisconsin-Madison. The work of LTW was supported by the Department of Energy grant DE-SC0009924.

\appendix

\section{Diffusion effects}

\label{sec:Diffusion-effects}

In this section, we derive an alternative to eq.~(\ref{eq:answer})
obtained without integrating by parts in time, and then derive the
standard sphaleron result, carefully accounting for the self-consistency
of the diffusion model. We also argue that $2\rightarrow3$ helicity
changing amplitudes that contribute to $\Delta N$ (see Appendix \ref{sec:Comparison-to-Dissipation}
and reference \cite{Broadberry:2025hep}) may also be able to diffuse
to induce contributions to $\gamma_{1}$.

Start with eq.~(\ref{eq:firststep}) with $h_{I}(x)=-\partial_{\mu}K^{\,\mu}(x)/f$
and $\Omega_{I}(0)=\partial_{\mu}K^{\mu}(0,\vec{y})/f$ corresponding
to $\mathcal{L}\ni\phi\partial_{\mu}K^{\mu}/f$ (at least to leading
order): 
\begin{equation}
\gamma_{1}=\frac{-i}{f^{2}V_{3}}\int^{0}_{-\infty}dzz\langle[\int d^{3}x\left(-\partial_{\mu}K^{\,\mu}(z,\vec{x})\right),\int d^{3}y\partial_{\mu}K^{\mu}(0,\vec{y})]\rangle.
\end{equation}
In temporal gauge, we can throw out all spatial divergences without
issue: 
\begin{equation}
\gamma_{1}=\frac{i}{f^{2}V_{3}}\int^{0}_{-\infty}dzz\langle[\int d^{3}x\partial_{z}K^{0}(z,\vec{x}),\int d^{3}y\partial_{w}K^{0}(w,\vec{y})]\rangle.
\end{equation}
Using time-translation invariance of the perturbative density matrix,
rewrite this as 
\begin{equation}
\gamma_{1}=\frac{-1}{f^{2}V_{3}}\int^{\infty}_{-\infty}dzz\Theta(z)\partial^{2}_{z}\langle i[N(z),N(0)]\rangle\label{eq:twoderiv}
\end{equation}
where 
\begin{equation}
N(t)\equiv\int d^{3}xK^{0}(t,\vec{x})
\end{equation}
which contains both topological and non-topological contributions
for non-vacuum $A_{\mu}$ configurations. Because of the KMS condition
induced formula 
\begin{equation}
\left\langle [A(t_{A}),B(t_{B})]\right\rangle =\frac{i}{2T}\frac{d}{dt_{A}}\left(\left\langle A(t_{A})B(t_{B})\right\rangle +\left\langle B(t_{B})A(t_{A})\right\rangle \right)\label{eq:commutator}
\end{equation}
where the right hand side is related to the diffusion $\left\langle \left(N(t_{A})-N(t_{B})\right)^{2}\right\rangle $
through the formula 
\begin{equation}
\left\langle N^{2}(t_{A})\right\rangle +\left\langle N^{2}(t_{B})\right\rangle -\left\langle \left(N(t_{A})-N(t_{B})\right)^{2}\right\rangle =\left\langle \left\{ N(t_{A}),N(t_{B})\right\} \right\rangle \label{eq:identity}
\end{equation}
any diffusing variance $\left\langle \left(N(t_{A})-N(t_{B})\right)^{2}\right\rangle \propto|t_{A}-t_{B}|$
can lead to nontrivial support for $\left\langle \left[N(t_{A}),N(t_{B})\right]\right\rangle $
even as $t_{A}-t_{B}\rightarrow\infty$. Hence, it is the secular
buildup of multiple scattering that can yield a non-local time propagator
effect acting on nonlocal variables such as $N(t)$.

To see what is required for the effective diffusion equation of $N$,
consider 
\begin{equation}
\int d^{4}x'\frac{\alpha\phi}{16\pi f}\tilde{G}G\ni\frac{1}{f}\dot{\bar{\phi}}(t)\int dt'\frac{d}{dt'}\left\{ (t'-t)N(t')\right\} -\frac{1}{f}\dot{\bar{\phi}}(t)\int dt'N(t').
\end{equation}
The first term on the right can be interpreted as initial condition
dependent contribution. The second term generates a potential force
term for $N$: 
\begin{equation}
-V_{N}(N)=C_{N}N
\end{equation}
where 
\begin{equation}
C_{N}\equiv\dot{\bar{\phi}}(t)/f
\end{equation}
which we can take to be a constant on the time scales of interest
for this illustrative estimate. Hence, the force on $N$ is 
\begin{align}
F_{N} & =-\frac{\partial V_{N}}{\partial N}=C_{N}.
\end{align}
Just like there is a friction term for $\phi$, there will generically
be a friction term for $N$ which we call $\Psi$.\footnote{The variable $\Psi$ would be a very difficult object to compute from
first principles.} Langevin approximation occurs when the friction balances the force
(without the noise term): 
\begin{equation}
\Psi\dot{N}=F_{N}
\end{equation}
Hence we arrive at the Langevin equation for $N$ which is 
\begin{equation}
\dot{N}=\frac{F_{N}}{\Psi}+\xi
\end{equation}
where $\xi$ is the noise term for $N$: 
\begin{equation}
\left\langle \xi(t_{1})\xi(t_{2})\right\rangle _{\xi}=g_{L}\delta(t_{1}-t_{2})\label{eq:collision}
\end{equation}
with $\left\langle \xi\right\rangle =0$.

Now we can compute 
\begin{equation}
\left\langle \dot{N}\right\rangle =\frac{F_{N}}{\Psi}
\end{equation}
and 
\begin{equation}
N(s)=\int^{s}_{0}d\bar{s}\xi(\bar{s})+\left\langle \dot{N}\right\rangle s+N_{0}
\end{equation}
Since 
\begin{equation}
\left\langle N(s)\right\rangle =\left\langle \dot{N}\right\rangle s+N_{0}.,
\end{equation}
we arrive at 
\begin{align}
\left\langle \left(\Delta N(s)\right)^{2}\right\rangle \equiv\left\langle \left(N(s)-\left\langle N(s)\right\rangle \right)^{2}\right\rangle  & =\int^{s}_{0}d\bar{s}\int^{s}_{0}d\bar{s}'\left\langle \xi(\bar{s})\xi(\bar{s}')\right\rangle =g_{L}s\label{eq:diffusion}
\end{align}
for $s\geq0$ which is the well-known diffusion equation. In the constant
$\dot{\bar{\phi}}$ limit, we have the same steady state variance
as without introducing $F_{N}/\Psi$. For the estimates below, we
will work on a time scale where $\left\langle \dot{N}\right\rangle s+N_{0}$
is small.

Using eq.~(\ref{eq:commutator}) and (\ref{eq:identity}), we arrive
at 
\begin{equation}
\left\langle [N(t_{A}),N(t_{B})]\right\rangle =\frac{-i}{2T}\frac{d}{dt_{A}}\left\langle \left(N(t_{A})-N(t_{B})\right)^{2}\right\rangle \label{eq:commutator-1}
\end{equation}
which allows us to conclude for $t\geq t'$ 
\begin{equation}
\left\langle \left[N(t),N(t')\right]\right\rangle =\frac{-ig_{L}}{2T}+\frac{ig_{L}}{2T}\Theta(t'-t)\label{eq:finalresolution}
\end{equation}
with the step-function $\Theta(z)$ defined with the additional property
$\lim_{\epsilon\rightarrow0^{+}}\Theta(0-\epsilon)\equiv1$ to allow
$\left\langle \left[N(t),N(t)\right]\right\rangle =0$ to be satisfied
and integration over delta-functions arising from derivatives of $\Theta$
to have support. Note that Langevin approximation cannot resolve small
time-differences as $\delta(t_{1}-t_{2})$ represents a collision
that in reality takes finite time. One can readily evaluate 
\begin{equation}
\gamma_{1}\frac{V_{3}f^{2}}{i}=\int^{\epsilon}_{-\infty}dss\partial_{s}\left[\frac{ig_{L}}{2T}\delta(s)\right]=-\frac{ig_{L}}{2T}
\end{equation}
where we used eqs.~(\ref{eq:twoderiv}) and (\ref{eq:finalresolution}).
We thus found 
\begin{equation}
\boxed{\gamma_{1}=\frac{g_{L}}{2TV_{3}f^{2}}}\label{eq:simpleeq}
\end{equation}
with $g_{L}$ read off from eq.~(\ref{eq:diffusion}). For example,
the hot sphalerons diffusion rate can be read off from \cite{Moore:2010jd}
as 
\begin{equation}
g^{\text{sphaleron}}_{L}=V_{3}\kappa(\ln T)\alpha^{5}T^{4}
\end{equation}
that gives $\gamma_{1}$ which matches the result used in \cite{Berghaus:2019whh}.

For any generic macroscopic quantity $\Delta N(t)=\sum_{i}\chi(i$)
driven by localized collisions, the amplitude of the noise spectrum
\begin{equation}
\left\langle \Xi(t)\Xi(t')\rangle\right\rangle =g_{L}\delta(t-t')
\end{equation}
in $\partial_{t}\Delta N=\Xi$ is 
\begin{align}
\frac{g_{L}}{V_{3}} & =\frac{1}{2}\frac{1}{S!}\sum^{\text{inc}}_{a,b}\sum^{\text{out}}_{c,d,...}\int\prod_{j}\frac{d^{3}p_{j}}{(2\pi)^{3}2E_{j}}\left(2\pi\right)^{4}\delta^{(4)}(\sum p_{i}-\sum p_{f})\left|\mathcal{M}_{2\rightarrow n}\right|^{2}\times\nonumber \\
 & \prod f_{\text{init}}\prod(1\pm f_{\text{fin}})\left[\Delta\chi\right]^{2}\label{eq:scatter}
\end{align}
where $S!$ accounts for identical particle stat and $\Delta\chi$
is the change in $\chi$ per collision $ab\rightarrow cd...$: 
\begin{equation}
\Delta\chi=(\chi_{a}+\chi_{b})-(\chi_{c}+\chi_{d}+...)
\end{equation}
Hence, we can estimate the $2\rightarrow3$ scattering that changes
the polarization (see Appendix \ref{sec:Comparison-to-Dissipation}
and reference \cite{Broadberry:2025hep}) as 
\begin{align}
\frac{g_{L}}{V_{3}} & \sim\alpha^{4}T^{4}\left(\frac{\left\langle \sigma_{2\rightarrow3}v\right\rangle T^{2}}{\alpha^{3}}\right)\left(\frac{n_{g}}{T^{3}}\right)^{2}
\end{align}
where $n_{g}=$gluon density. Using eq.~(\ref{eq:simpleeq}), we
estimate 
\begin{equation}
\boxed{\gamma^{\text{scatter}}_{1}=\frac{\alpha^{4}T^{3}}{2f^{2}}\left(\frac{n_{g}}{T^{3}}\right)^{2}\left(\frac{\left\langle \sigma_{2\rightarrow3}v\right\rangle T^{2}}{\alpha^{3}}\right)}.\label{eq:estimatehere}
\end{equation}
It is important to note that each of these $2\rightarrow3$ polarization
changing scattering events lead to $\Delta N\sim O(\alpha)$ that
do not induce topological shifts in the gauge field configurations
and therefore do not couple to anomalies if one introduced them as
probes.\footnote{Also see \cite{KHLEBNIKOV1988885}.} Indeed, it is
tempting to speculate that these are related to the UV sensitive noise
that shifts the CS diffusion rate reported in \cite{Laine:2022ytc}.
As long as the density of the helicity asymmetry is in dilute free
gluons, there will be no energy cost in each $\Delta N$ event, allowing
the variance to grow. Further work will be needed to verify this picture.

\section{$\rho_{\tilde{\bar{G}}G}$ in terms of $\rho^{T}$}

\label{sec:GGtoSpec}

In this appendix, we detail the computations leading to eq.~(\ref{eq:EucCorrSimp}).As
noted in the main text, the Euclidean correlator

\begin{align}
\Pi^{E}_{\!\tilde{\bar{G}}\bar{G}}\!\left(P\right) & \equiv\int^{T^{-1}}_{0}d\tau_{y}\int d^{3}y\,\exp\!\left(-iP\cdot Y\right)\langle\tilde{\bar{G}}_{a\mu\nu}\bar{G}^{a}_{\mu\nu}\!\left(\tau\right)\tilde{\bar{G}}_{b\alpha\beta}\bar{G}^{b}_{\alpha\beta}\!\left(\tau'\right)\rangle_{c}\nonumber \\
 & \approx16\int^{T^{-1}}_{0}d\tau_{y}\int d^{3}y\,\exp\!\left(-iP\cdot Y\right)\langle\epsilon^{\mu\nu\rho\sigma}\partial_{\mu}\bar{A}^{a}_{\nu}\partial_{\rho}\bar{A}^{a}_{\sigma}\!\left(\tau\right)\epsilon^{\alpha\beta\gamma\delta}\partial_{\alpha}\bar{A}^{d}_{\beta}\partial_{\gamma}\bar{A}^{d}_{\delta}\!\left(\tau'\right)\rangle_{c}
\end{align}
simplifies, after Wick contractions, to 
\begin{align}
\Pi^{E}_{\!\tilde{\bar{G}}\bar{G}}\!\left(P\right) & \approx16\int^{T^{-1}}_{0}d\tau_{y}\int d^{3}y\,\exp\!\left(-iP\cdot Y\right)\langle\epsilon^{\mu\nu\rho\sigma}\partial_{\mu}\bar{A}^{a}_{\nu}\partial_{\rho}\bar{A}^{a}_{\sigma}\!\left(\tau\right)\epsilon^{\alpha\beta\gamma\delta}\partial_{\alpha}\bar{A}^{d}_{\beta}\partial_{\gamma}\bar{A}^{d}_{\delta}\!\left(\tau'\right)\rangle_{c}\nonumber \\
 & =16\int^{T^{-1}}_{0}d\tau_{y}\int d^{3}y\,\exp\!\left(-iP\cdot Y\right)\int_{\Sigma,Q}\exp\!\left(+iQ\cdot X\right)\int_{\Sigma,L}\exp\!\left(+iL\cdot X\right)\nonumber \\
 & \quad\times\int_{\Sigma,M}\exp\!\left(+iM\cdot X'\right)\int_{\Sigma,N}\exp\!\left(+iN\cdot X'\right)\epsilon^{\mu\nu\rho\sigma}\epsilon^{\alpha\beta\gamma\delta}Q_{\mu}L_{\rho}M_{\alpha}N_{\gamma}\nonumber \\
 & \quad\times\!\left\{ \langle\bar{A}^{a}_{\nu}\!\left(Q\right)\bar{A}^{d}_{\beta}\!\left(M\right)\rangle\langle\bar{A}^{a}_{\sigma}\!\left(L\right)\bar{A}^{d}_{\delta}\!\left(N\right)\rangle+\langle\bar{A}^{a}_{\nu}\!\left(Q\right)\bar{A}^{d}_{\delta}\!\left(N\right)\rangle\langle\bar{A}^{a}_{\sigma}\!\left(L\right)\bar{A}^{d}_{\beta}\!\left(M\right)\rangle\right\} \nonumber \\
 & =32\,\delta^{aa}\int^{T^{-1}}_{0}d\tau_{y}\int d^{3}y\,\exp\!\left(-iP\cdot Y\right)\int_{\Sigma,Q}\int_{\Sigma,L}\exp\!\left(+i\!\left(Q+L\right)\cdot Y\right)\nonumber \\
 & \quad\times\epsilon^{\mu\nu\rho\sigma}\epsilon^{\alpha\beta\gamma\delta}Q_{\mu}L_{\rho}Q_{\alpha}L_{\gamma}\Pi_{\nu\beta}\!\left(Q\right)\Pi_{\sigma\delta}\!\left(L\right)
\end{align}
with the gluon propagators taking the form 
\begin{equation}
\langle\bar{A}^{a}_{\nu}\!\left(Q\right)\bar{A}^{d}_{\beta}\!\left(L\right)\rangle=\delta^{ad}\,\bar{\delta}\!\left(Q+L\right)\Pi_{\nu\beta}\!\left(Q\right)
\end{equation}
in momentum space. Here, we have denoted Euclidean position and momenta
by capitalized letters ($X,X',Y$ and $P,Q,L,M,N\ldots$ respectively),
and defined $Y\equiv X-X'=\!\left(\tau_{y},\vec{y}\right)$. The additional
factor of 2 in the last equality accounts for the two contractions
(for connected diagrams). To relate this to the gluon spectral functions,
we first note that gauge invariance (Slavanov--Taylor identities)
implies transversality of the Euclidean propagator with respect to
the Euclidean 4-momentum $Q$ and therefore it should take the form
\begin{equation}
\Pi_{\nu\beta}\!\left(Q\right)=\Pi^{T}\!\left(Q\right)\mathbb{P}^{T}_{\nu\beta}\!\left(Q\right)+\Pi^{L}\!\left(Q\right)\mathbb{P}^{L}_{\nu\beta}\!\left(Q\right)+\frac{\xi Q_{\mu}Q_{\nu}}{\!\left(Q^{2}\right)^{2}}
\end{equation}
where 
\begin{equation}
\mathbb{P}^{T}_{\nu\beta}\!\left(Q\right)\equiv\delta_{\nu i}\delta_{\beta j}\!\left(\delta_{ij}-\frac{q_{i}q_{j}}{q^{2}}\right),
\end{equation}
\begin{equation}
\mathbb{P}^{L}_{\nu\beta}\!\left(Q\right)\equiv\delta_{\nu\beta}-\frac{Q_{\beta}Q_{\nu}}{Q^{2}}-\mathbb{P}^{T}_{\nu\beta}\!\left(Q\right)
\end{equation}
in a general covariant gauge. We then combine the above with the following
spectral representation: 
\begin{align}
\Pi_{\nu\beta}\!\left(Q\right) & =\!\left(\int^{+\infty}_{-\infty}\frac{dq_{0}}{\pi}\frac{\rho^{T}\!\left(q_{0},q_{i}\right)}{q_{0}-iq_{n}}\right)\mathbb{P}^{T}_{\nu\beta}\!\left(Q\right)+\!\left(\int^{+\infty}_{-\infty}\frac{dq_{0}}{\pi}\frac{\rho^{L}\!\left(q_{0},q_{i}\right)}{q_{0}-iq_{n}}\right)\mathbb{P}^{L}_{\nu\beta}\!\left(Q\right)+\frac{\xi Q_{\mu}Q_{\nu}}{\left(Q^{2}\right)^{2}}
\end{align}
and obtain $\Pi^{E}_{\!\tilde{\bar{G}}\bar{G}}\!\left(P\right)$ in
the Feynman gauge $\!\left(\xi=1\right)$ as 
\begin{align}
\Pi^{E}_{\!\tilde{\bar{G}}\bar{G}}\!\left(P\right) & =32\,\delta^{aa}\int^{T^{-1}}_{0}d\tau_{y}\int d^{3}y\,\exp\!\left(-iP\cdot Y\right)\int_{\Sigma,Q}\int_{\Sigma,L}\exp\!\left(+i\!\left(Q+L\right)\cdot Y\right)\\
 & \times\epsilon^{\mu\nu\rho\sigma}\epsilon^{\alpha\beta\gamma\delta}Q_{\mu}L_{\rho}Q_{\alpha}L_{\gamma}\nonumber \\
 & \times\!\left(\!\left(\int^{+\infty}_{-\infty}\frac{dq_{0}}{\pi}\frac{\rho^{T}\!\left(q_{0},q_{i}\right)}{q_{0}-iq_{n}}\right)\mathbb{P}^{T}_{\nu\beta}\!\left(Q\right)+\!\left(\int^{+\infty}_{-\infty}\frac{dq_{0}}{\pi}\frac{\rho^{L}\!\left(q_{0},q_{i}\right)}{q_{0}-iq_{n}}\right)\mathbb{P}^{L}_{\nu\beta}\!\left(Q\right)\right)\nonumber \\
 & \times\!\left(\!\left(\int^{+\infty}_{-\infty}\frac{dl_{0}}{\pi}\frac{\rho^{T}\!\left(l_{0},l_{i}\right)}{l_{0}-il_{n}}\right)\mathbb{P}^{T}_{\sigma\delta}\!\left(L\right)+\!\left(\int^{+\infty}_{-\infty}\frac{dl_{0}}{\pi}\frac{\rho^{L}\!\left(l_{0},l_{i}\right)}{l_{0}-il_{n}}\right)\mathbb{P}^{L}_{\sigma\delta}\!\left(L\right)\right).\label{eq:ECorr_SpecDen_B4contrac-1}
\end{align}
The different index contractions in the above may be simplified as
\begin{align}
\text{PT}_{1} & =\epsilon^{\mu\nu\rho\sigma}\epsilon^{\alpha\beta\gamma\delta}Q_{\mu}L_{\rho}Q_{\alpha}L_{\gamma}\mathbb{P}^{T}_{\sigma\delta}\!\left(L\right)\mathbb{P}^{T}_{\nu\beta}\!\left(Q\right)\nonumber \\
 & =\!\left\{ l^{2}q^{2}_{n}+l^{2}_{n}q^{2}-4\!\left(l\cdot q\right)l_{n}q_{n}\right\} +\frac{l^{2}_{n}(l\cdot q)^{2}}{l^{2}}+\frac{q^{2}_{n}}{q^{2}}(l\cdot q)^{2}
\end{align}
\begin{equation}
\text{PT}_{2}=\epsilon^{\mu\nu\rho\sigma}\epsilon^{\alpha\beta\gamma\delta}Q_{\mu}L_{\rho}Q_{\alpha}L_{\gamma}\mathbb{P}^{T}_{\sigma\delta}\!\left(L\right)\mathbb{P}^{L}_{\nu\beta}\!\left(Q\right)=-\frac{Q^{2}\!\left(l\cdot q\right)^{2}}{q^{2}}+l^{2}Q^{2}
\end{equation}
\begin{equation}
\text{PT}_{3}=\epsilon^{\mu\nu\rho\sigma}\epsilon^{\alpha\beta\gamma\delta}Q_{\mu}L_{\rho}Q_{\alpha}L_{\gamma}\mathbb{P}^{L}_{\sigma\delta}\!\left(L\right)\mathbb{P}^{T}_{\nu\beta}\!\left(Q\right)=-\frac{L^{2}\!\left(l\cdot q\right)^{2}}{l^{2}}+L^{2}q^{2}
\end{equation}
\begin{equation}
\text{PT}_{4}=\epsilon^{\mu\nu\rho\sigma}\epsilon^{\alpha\beta\gamma\delta}Q_{\mu}L_{\rho}Q_{\alpha}L_{\gamma}\mathbb{P}^{L}_{\sigma\delta}\!\left(L\right)\mathbb{P}^{L}_{\nu\beta}\!\left(Q\right)=0
\end{equation}
where $M^{2}\equiv m^{2}_{n}+\vec{m}^{2}$, $m^{2}\equiv\vec{m}\cdot\vec{m}$
and $n\cdot m\equiv\vec{n}\cdot\vec{m}$ for Euclidean momenta $M=\!\left(m_{n},\vec{m}\right)$
and $N=\!\left(n_{n},\vec{n}\right)$. Plugging these back into eq.~(\ref{eq:ECorr_SpecDen_B4contrac-1})
gives 
\begin{align}
\Pi^{E}_{\!\tilde{\bar{G}}\bar{G}}\!\left(P\right)= & 32\,\delta^{aa}\int^{T^{-1}}_{0}d\tau_{y}\int d^{3}y\,\exp\!\left(-iP\cdot Y\right)\int_{\Sigma,Q}\int_{\Sigma,L}\exp\!\left(+i\!\left(Q+L\right)\cdot Y\right)\nonumber \\
 & \times\Bigg(\underbrace{\!\left(\int^{+\infty}_{-\infty}\frac{dq_{0}}{\pi}\frac{\rho^{T}\!\left(q_{0},q\right)}{q_{0}-iq_{n}}\right)\!\left(\int^{+\infty}_{-\infty}\frac{dl_{0}}{\pi}\frac{\rho^{T}\!\left(l_{0},l\right)}{l_{0}-il_{n}}\right)f(L,Q)}_{MT1}\nonumber \\
 & +\underbrace{\!\left(\int^{+\infty}_{-\infty}\frac{dq_{0}}{\pi}\frac{\rho^{L}\!\left(q_{0},q\right)}{q_{0}-iq_{n}}\right)\!\left(\int^{+\infty}_{-\infty}\frac{dl_{0}}{\pi}\frac{\rho^{T}\!\left(l_{0},l\right)}{l_{0}-il_{n}}\right)\!\left[-\frac{Q^{2}\!\left(l\cdot q\right)^{2}}{q^{2}}+l^{2}Q^{2}\right]}_{MT2}\nonumber \\
 & +\left.\underbrace{\!\left(\int^{+\infty}_{-\infty}\frac{dq_{0}}{\pi}\frac{\rho^{T}\!\left(q_{0},q\right)}{q_{0}-iq_{n}}\right)\!\left(\int^{+\infty}_{-\infty}\frac{dl_{0}}{\pi}\frac{\rho^{L}\!\left(l_{0},l\right)}{l_{0}-il_{n}}\right)\!\left[-\frac{L^{2}\!\left(l\cdot q\right)^{2}}{l^{2}}+L^{2}q^{2}\right]}_{MT3}\right)\label{eq:SpecDEn_B4Matsub-1}
\end{align}
where 
\begin{equation}
f(L,Q)\equiv\!\!l^{2}q^{2}_{n}+l^{2}_{n}q^{2}-4\!\left(l\cdot q\right)l_{n}q_{n}+\frac{l^{2}_{n}(l\cdot q)^{2}}{l^{2}}+\frac{q^{2}_{n}(l\cdot q)^{2}}{q^{2}}.
\end{equation}
This can be simplified further by computing the Matsubara sums in
the underbraced terms above. Using the identity 
\begin{equation}
T\sum_{\omega_{n}}\frac{\exp\!\left(i\omega_{n}\tau\right)}{\omega^{2}_{n}+\omega^{2}}=\frac{n_{B}\!\left(\omega\right)}{2\omega}\!\left[\exp\!\left\{ \!\left(T^{-1}-\tau\right)\omega\right\} +\exp\!\left\{ \tau\omega\right\} \right]\qquad\text{where}\;0<\tau<T^{-1}\label{eq:MatsubaraIden}
\end{equation}
and its derivatives with respect to $\tau$ we have 
\begin{align}
\text{MT}_{1} & =\sum_{l_{n}}\sum_{q_{n}}\frac{\exp\!\left(i\!\left(q_{n}+l_{n}\right)\tau_{y}\right)}{\!\left(q_{0}-iq_{n}\right)\!\left(l_{0}-il_{n}\right)}\!\left[\!\left\{ l^{2}q^{2}_{n}+l^{2}_{n}q^{2}-4\!\left(l\cdot q\right)l_{n}q_{n}\right\} +\frac{l^{2}_{n}(l\cdot q)^{2}}{l^{2}}+\frac{q^{2}_{n}}{q^{2}}(l\cdot q)^{2}\right]\nonumber \\
 & =-n_{B}\!\left(q_{0}\right)n_{B}\!\left(l_{0}\right)\exp\!\left(\tau_{y}\!\left(l_{0}+q_{0}\right)\right)\!\left\{ \!\left(\bar{l}^{2}+\frac{(l\cdot q)^{2}}{q^{2}}\right)q^{2}_{0}+\!\left(\bar{q}^{2}+\frac{(l\cdot q)^{2}}{l^{2}}\right)l^{2}_{0}-4\!\left(l\cdot q\right)l_{0}q_{0}\right\} 
\end{align}
\begin{align}
\text{MT}_{2} & =\sum_{l_{n}}\sum_{q_{n}}\frac{\exp\!\left(i\!\left(q_{n}+l_{n}\right)\tau_{y}\right)}{\!\left(q_{0}-iq_{n}\right)\!\left(l_{0}-il_{n}\right)}\!\left[-\frac{Q^{2}\!\left(l\cdot q\right)^{2}}{q^{2}}+l^{2}Q^{2}\right]\nonumber \\
 & =n_{B}\!\left(q_{0}\right)n_{B}\!\left(l_{0}\right)\exp\!\left(\tau_{y}\!\left(l_{0}+q_{0}\right)\right)\!\left[q^{2}_{0}\frac{\!\left(l\cdot q\right)^{2}}{q^{2}}-q^{2}_{0}l^{2}-\!\left(l\cdot q\right)^{2}+l^{2}q^{2}\right]
\end{align}
\begin{align}
\text{MT}_{3} & =\sum_{l_{n}}\sum_{q_{n}}\frac{\exp\!\left(i\!\left(q_{n}+l_{n}\right)\tau_{y}\right)}{\!\left(q_{0}-iq_{n}\right)\!\left(l_{0}-il_{n}\right)}\!\left[-\frac{\!\left(l^{2}_{n}+l^{2}\right)\!\left(l\cdot q\right)^{2}}{l^{2}}+\!\left(l^{2}_{n}+l^{2}\right)q^{2}\right]\nonumber \\
 & =n_{B}\!\left(q_{0}\right)\exp\!\left(\tau_{y}q_{0}\right)n_{B}\!\left(l_{0}\right)\exp\!\left(\tau_{y}l_{0}\right)\!\left\{ l^{2}_{0}\frac{\!\left(l\cdot q\right)^{2}}{l^{2}}-l^{2}_{0}q^{2}-\!\left(l\cdot q\right)^{2}+l^{2}q^{2}\right\} .
\end{align}
Plugging these back into eq.~(\ref{eq:SpecDEn_B4Matsub-1}) and simplifying
the spatial integrals we have 
\begin{align}
\Pi^{E}_{\!\tilde{\bar{G}}\bar{G}}\!\left(P\right) & =32\,\delta^{aa}\int_{q}\int_{l}\!\left\{ \!\left(2\pi\right)^{3}\delta^{3}\!\left(-\vec{p}+\vec{q}+\vec{l}\right)\right\} \int^{+\infty}_{-\infty}\frac{dq_{0}}{\pi}\int^{+\infty}_{-\infty}\frac{dl_{0}}{\pi}\,\nonumber \\
 & \qquad\qquad\times n_{B}\!\left(q_{0}\right)n_{B}\!\left(l_{0}\right)\frac{\!\left[\exp\!\left(\left(l_{0}+q_{0}\right)/T\right)-1\right]}{\!\left(l_{0}+q_{0}-ip_{n}\right)}\nonumber \\
 & \times\Bigg(-\rho^{T}\!\left(q_{0},q\right)\rho^{T}\!\left(l_{0},l\right)\underbrace{\!\left\{ \!\left(l^{2}+\frac{(l\cdot q)^{2}}{q^{2}}\right)q^{2}_{0}+\!\left(q^{2}+\frac{(l\cdot q)^{2}}{l^{2}}\right)l^{2}_{0}-4\!\left(l\cdot q\right)l_{0}q_{0}\right\} }_{MT1}\nonumber \\
 & \quad\quad+\rho^{L}\!\left(q_{0},q\right)\rho^{T}\!\left(l_{0},l\right)\underbrace{\!\left[q^{2}_{0}\frac{\!\left(l\cdot q\right)^{2}}{q^{2}}-q^{2}_{0}l^{2}-\!\left(l\cdot q\right)^{2}+l^{2}q^{2}\right]}_{MT2}\nonumber \\
 & \quad\quad+\left.\rho^{T}\!\left(q_{0},q\right)\rho^{L}\!\left(l_{0},l\right)\underbrace{\!\left\{ l^{2}_{0}\frac{\!\left(l\cdot q\right)^{2}}{l^{2}}-l^{2}_{0}q^{2}-\!\left(l\cdot q\right)^{2}+l^{2}q^{2}\right\} }_{MT3}\right).
\end{align}
Since from eq.~(\ref{eq:SpecDensfromEucCorr}) we only need to compute
$\Pi^{E}_{\!\tilde{\bar{G}}\bar{G}}\!\left(P\right)$ for $\vec{p}=0$,
we can set the spatial part of the external momentum to zero in the
above. The terms arising from $\text{MT}_{2}$ and $\text{MT}_{3}$
vanish when $\vec{p}=0$, giving 
\begin{align}
\Pi^{E}_{\!\tilde{\bar{G}}\bar{G}}\!\left(p_{n},0\right) & =-32\,\delta^{aa}\int_{q}\int^{+\infty}_{-\infty}\frac{dq_{0}}{\pi}\int^{+\infty}_{-\infty}\frac{dl_{0}}{\pi}\,n_{B}\!\left(q_{0}\right)n_{B}\!\left(l_{0}\right)\frac{\!\left[\exp\!\left(\left(l_{0}+q_{0}\right)/T\right)-1\right]}{\!\left(l_{0}+q_{0}-ip_{n}\right)}\nonumber \\
 & \quad\times\!\left[\rho^{T}\!\left(q_{0},q\right)\rho^{T}\!\left(l_{0},q\right)2q^{2}\!\left(q_{0}+l_{0}\right)^{2}\right].
\end{align}
Using eq.~(\ref{eq:SpecDensfromEucCorr}) we then have eq.~(\ref{eq:RhoGG}):
\begin{align}
\rho_{\!\tilde{G}G}\!\left(p_{0},0\right) & =4^{3}\,\delta^{aa}\,p^{2}_{0}\!\left(\exp\!\left(p_{0}/T\right)-1\right)\nonumber \\
 & \quad\times\!\left\{ \int_{q}\int^{+\infty}_{-\infty}\frac{dq_{0}}{\pi}\,n_{B}\!\left(q_{0}\right)n_{B}\!\left(p_{0}-q_{0}\right)\!\left[\rho^{T}\!\left(q_{0},\vec{q}\right)\rho^{T}\!\left(p_{0}-q_{0},-\vec{q}\right)q^{2}\right]\right\} .
\end{align}

\section{1-Loop self energy correction to transverse gluon propagator}

\label{sec:Full-1-loop-comp}

In this appendix we review the computation of the 1-loop contributions
from the gauge and ghost loops, and in addition compute the contribution
from the loop with an inflaton exchange (Fig.~(\ref{fig:verticesn!loopdgrm})).
For reviews of the relevant finite-temperature techniques, see~\cite{Kapusta:2006pm,Bellac:2011kqa,Laine:2016hma}.
The combined contributions from these diagrams are 
\begin{align}
\bar{\Pi}_{\mu\nu}\!\left(Q\right) & =\frac{g^{2}N_{c}}{2}\int_{\Sigma,L}\frac{\delta_{\mu\nu}\!\left[-4Q^{2}+2\!\left(D-2\right)L^{2}\right]+\!\left(D+2\right)Q_{\mu}Q_{\nu}-4\!\left(D-2\right)L_{\mu}L_{\nu}}{L^{2}\!\left(Q-L\right)^{2}}\nonumber \\
 & \quad+\!\left(\frac{\alpha}{2\pi f}\right)^{2}\epsilon^{\mu\eta\rho\beta}\epsilon^{\alpha\beta\gamma\chi}Q_{\mu}Q_{\gamma}\int_{\Sigma,L}\!\left(\frac{1}{L^{2}+m^{2}_{\phi}}\right)\frac{L_{\rho}L_{\alpha}}{\!\left(Q-L\right)^{2}}
\end{align}
where the contributions from the gauge and ghost loops may be found
in say~\cite{Laine:2016hma}. Since we need the self-energy correction
of the transverse gluons, projecting the above as 
\begin{equation}
\bar{\Pi}^{T}\!\left(Q\right)=\frac{1}{D-2}\mathbb{P}^{T}_{\nu\beta}\!\left(Q\right)\Pi^{T}_{\beta\nu}\!\left(Q\right)
\end{equation}
we get 
\begin{align}
\bar{\Pi}^{T}\!\left(Q\right) & =-\!\left(D-2\right)g^{2}N_{c}\!\left(\int_{\Sigma,L}\frac{\!\left(Q^{2}+l^{2}-\frac{\!\left(l\cdot q\right)^{2}}{q^{2}}\right)}{L^{2}\!\left(Q-L\right)^{2}}-\frac{\!\left(D-2\right)}{2}\int_{\Sigma,P}\frac{1}{P^{2}}\right)\nonumber \\
 & \quad+\frac{1}{D-2}\!\left(\frac{\alpha}{2\pi f}\right)^{2}\int_{\Sigma,L}\frac{\!\left[\!\left(l\cdot u\right)^{2}\!\left(1-\frac{q^{2}_{n}}{q^{2}}\right)-l^{2}q^{2}\!\left(1+\frac{q^{2}_{n}}{q^{2}}\right)+4l_{n}q_{n}\!\left(l\cdot q\right)-2l^{2}_{n}q^{2}\right]}{\!\left(L^{2}+m^{2}_{\phi}\right)\!\left(Q-L\right)^{2}}.\label{eq: SEtranProjApp}
\end{align}
To simplify the above integrals, we first carry out the Matsubara
sums using the identity eq. (\ref{eq:MatsubaraIden})and its derivatives
with respect to $-i\partial_{\tau}$. Terms where the Matsubara sum
is over a product of propagators, as in the first and last terms of
eq.~(\ref{eq: SEtranProjApp}), may be rewritten as 
\begin{align}
 & T\sum_{l_{n}}\frac{1}{\!\left(l^{2}_{n}+\epsilon^{2}_{1}\right)\!\left(\!\left(q_{n}-l_{n}\right)^{2}+\epsilon^{2}_{2}\right)}\nonumber \\
 & \quad=\!\left[T\sum_{l_{n}}\frac{1}{\!\left(l^{2}_{n}+\epsilon^{2}_{1}\right)}\right]\!\left[T\sum_{r_{n}}\frac{T^{-1}\bar{\delta}\!\left(r_{n}-\!\left(q_{n}-l_{n}\right)\right)}{\!\left(r^{2}_{n}+\epsilon^{2}_{2}\right)}\right]\nonumber \\
 & \quad=\!\left[T\sum_{l_{n}}\frac{1}{\!\left(l^{2}_{n}+\epsilon^{2}_{1}\right)}\right]\!\left[T\sum_{r_{n}}\frac{\int^{T^{-1}}_{0}d\tau\,\exp\!\left(i\!\left(r_{n}-\!\left(q_{n}-l_{n}\right)\right)\tau\right)}{\!\left(r^{2}_{n}+\epsilon^{2}_{2}\right)}\right]\nonumber \\
 & \quad=\int^{T^{-1}}_{0}d\tau\,\exp\!\left(-iq_{n}\tau\right)\!\left[T\sum_{l_{n}}\frac{\exp\!\left(-il_{n}\tau\right)}{\!\left(l^{2}_{n}+\epsilon^{2}_{1}\right)}\right]\!\left[T\sum_{r_{n}}\frac{\exp\!\left(ir_{n}\tau\right)}{\!\left(r^{2}_{n}+\epsilon^{2}_{2}\right)}\right]
\end{align}
(Saclay method), where eq.~(\ref{eq:MatsubaraIden}) can be used
for simplification. Here 
\begin{equation}
T^{-1}\,\bar{\delta}\!\left(r_{n}-\!\left(q_{n}-l_{n}\right)\right)=\int^{T^{-1}}_{0}d\tau\,\exp\!\left(i\!\left(r_{n}-\!\left(q_{n}-l_{n}\right)\right)\tau\right).
\end{equation}
Computing the sums over the Matsubara frequencies and the $\tau$
integrals arising from the above identity, eq.~(\ref{eq: SEtranProjApp})
simplifies to a sum of the zero-temperature and thermal contributions,
that is, 
\begin{equation}
\bar{\Pi}^{T}\!\left(Q\right)=\bar{\Pi}^{T}_{vac}\!\left(Q\right)+\bar{\Pi}^{T}_{th}\!\left(Q\right).
\end{equation}
The vacuum (or temperature-independent) contribution is identical
to that from the zero-temperature Euclidean theory, as noted in section
\ref{subsec:PiGG2Pig}. In particular, as $T\rightarrow0$, the self-energy
corrections of the transverse and longitudinal modes become equal
and $\bar{\Pi}_{\mu\nu}\!\left(Q\right)$ reduces to 
\begin{equation}
\bar{\Pi}_{\mu\nu}\!\left(Q\right)=\bar{\Pi}\!\left(Q\right)\!\left(\delta_{\mu\nu}-\frac{Q_{\mu}Q_{\nu}}{Q^{2}}\right).
\end{equation}
Consequently, this contribution is more easily evaluated using standard
techniques for vacuum Feynman diagrams, and is given by 
\begin{align}
\bar{\Pi}^{T}_{vac}\!\left(Q\right) & =\frac{Q^{2}}{D-1}\left\{ -\!\left(\frac{\alpha}{2\pi f}\right)^{2}\int^{1}_{0}dx\,\frac{D_{1}}{\!\left(4\pi\right)^{2}}\!\left\{ -3\ln\!\left(\frac{D_{1}}{\bar{\mu}^{2}}\right)+1\right\} \right.\nonumber \\
 & \qquad\qquad\left.+\frac{g^{2}N_{c}}{\!\left(4\pi\right)^{2}}\int^{1}_{0}dx\,\!\left\{ 5\ln\!\left(\frac{D_{2}}{\bar{\mu}^{2}}\right)+3\right\} \right\} \label{eq:FullVacSE}
\end{align}
where 
\begin{align}
D_{1} & =Q^{2}x\!\left(1-x\right)+\!\left(1-x\right)m^{2}_{\phi},\qquad\text{and}\qquad D_{2}=Q^{2}x\!\left(1-x\right).
\end{align}
The UV divergences have been renormalized in the $\overline{\mathrm{MS}}$
scheme. The thermal contributions are 
\begin{align}
\bar{\Pi}^{T}_{th}\!\left(Q\right) & =\!\left(\frac{\alpha}{2\pi f}\right)^{2}\int_{l}\!\left[\!\left(\!\left(l\cdot q\right)^{2}-l^{2}q^{2}\right)\!\left(1-\frac{q^{2}_{n}}{q^{2}}\right)\right]\frac{1}{2l}\frac{1}{2\epsilon_{ql}}\nonumber \\
 & \quad\times\left\{ n_{B}\!\left(\epsilon_{ql}\right)\!\left(\frac{\!\left(l+\epsilon_{ql}\right)}{\!\left(q^{2}_{n}+\!\left(l+\epsilon_{ql}\right)^{2}\right)}+\frac{\!\left(l-\epsilon_{ql}\right)}{\!\left(q^{2}_{n}+\!\left(l-\epsilon_{ql}\right)^{2}\right)}\right)\right.\nonumber \\
 & \quad\quad\left.+n_{B}\!\left(l\right)\!\left(\frac{\!\left(l+\epsilon_{ql}\right)}{\!\left(q^{2}_{n}+\!\left(l+\epsilon_{ql}\right)^{2}\right)}-\frac{\!\left(l-\epsilon_{ql}\right)}{\!\left(q^{2}_{n}+\!\left(l-\epsilon_{ql}\right)^{2}\right)}\right)\right\} \nonumber \\
 & \quad+4q^{2}_{n}\!\left(\frac{\alpha}{2\pi f}\right)^{2}\int_{l}\frac{1}{4\epsilon_{ql}}\!\left(l\cdot q\right)\left[n_{B}\!\left(\epsilon_{ql}\right)\!\left\{ \frac{1}{\!\left(q^{2}_{n}+\!\left(l-\epsilon_{ql}\right)^{2}\right)}+\frac{1}{\!\left(q^{2}_{n}+\!\left(l+\epsilon_{ql}\right)^{2}\right)}\right\} \right.\nonumber \\
 & \quad\left.+n_{B}\!\left(l\right)\!\left\{ \frac{1}{\!\left(q^{2}_{n}+\!\left(l+\epsilon_{ql}\right)^{2}\right)}-\frac{1}{\!\left(q^{2}_{n}+\!\left(l-\epsilon_{ql}\right)^{2}\right)}\right\} \right]+\mathcal{O}\!\left(m_{\phi}\right)\nonumber \\
 & \quad-2\!\left(D-2\right)g^{2}N_{c}\int_{l}\!\left(Q^{2}+l^{2}-\frac{\!\left(l\cdot q\right)^{2}}{q^{2}}\right)\frac{1}{2l}\frac{1}{2\epsilon_{ql}}\left(n_{B}\!\left(l\right)\!\left(\frac{\!\left(l+\epsilon_{ql}\right)}{\!\left(q^{2}_{n}+\!\left(l+\epsilon_{ql}\right)^{2}\right)}\right)\right.\nonumber \\
 & \quad\left.+n_{B}\!\left(l\right)\frac{\!\left(\epsilon_{ql}-l\right)}{\!\left(q^{2}_{n}+\!\left(l-\epsilon_{ql}\right)^{2}\right)}+n_{B}\!\left(\epsilon_{ql}\right)\!\left(\frac{\!\left(l+\epsilon_{ql}\right)}{\!\left(q^{2}_{n}+\!\left(l+\epsilon_{ql}\right)^{2}\right)}-\frac{\!\left(\epsilon_{ql}-l\right)}{\!\left(q^{2}_{n}+\!\left(l-\epsilon_{ql}\right)^{2}\right)}\right)\right)\nonumber \\
 & \quad+\frac{g^{2}N_{c}}{4}\!\left(D-2\right)^{2}\int_{l}\frac{2n_{B}\!\left(l\right)}{l}.\label{eq:PiTAfterMatsu}
\end{align}
Here 
\begin{equation}
\epsilon^{2}_{l}=l^{2}+m^{2}_{\phi},\quad\text{and}\quad\epsilon^{2}_{ql}=q^{2}+l^{2}-2q\cdot l,
\end{equation}
and the integrals are over the spatial part of the loop momenta, that
is, for a loop momentum $Q=\!\left(q_{n},\vec{q}\right)$ 
\begin{equation}
\int_{q}\equiv\int\frac{d^{3}q}{\!\left(2\pi\right)^{3}}.
\end{equation}
In the thermal contribution, with internal loop momenta weighted by
the Boltzmann distribution, the $m_{\phi}$ dependence may be approximated
in a Taylor expansion, with eq.~(\ref{eq:PiTAfterMatsu}) being the
leading term. Note that the integrals of the terms proportional to
the Boltzmann factor $n_{B}\!\left(\epsilon_{ql}\right)$ can be simplified
with a change of integration variable to $\vec{y}=\vec{l}-\vec{q}$.\footnote{This implies $l^{2}=y^{2}+2y\cdot u+u^{2}=\epsilon^{2}_{uy}$ and
$\epsilon^{2}_{ul}=y^{2}$. Using this, in the first terms proportional
to $\alpha^{2}$ and $g^{2}$, we see that the contributions from
the terms proportional to $n_{B}\!\left(\epsilon_{ul}\right)$ and
$n_{B}\!\left(l\right)$ are identical.} Further, we note that using the identity 
\begin{align}
\!\left(\!\left(l-\epsilon_{ql}\right)^{2}+q^{2}_{n}\right)\!\left(\!\left(l+\epsilon_{ql}\right)^{2}+q^{2}_{n}\right) & =\!\left(q^{2}_{n}+q^{2}-2q\cdot l\right)^{2}+4l^{2}q^{2}_{n},
\end{align}
factors in the integrand can be reduced to simpler forms, such as
\begin{align}
 & \frac{\!\left(l+\epsilon_{ql}\right)}{\!\left(q^{2}_{n}+\!\left(l+\epsilon_{ql}\right)^{2}\right)}+\frac{\!\left(l-\epsilon_{ql}\right)}{\!\left(q^{2}_{n}+\!\left(l-\epsilon_{ql}\right)^{2}\right)}\nonumber \\
 & \quad=\frac{2l\!\left(q^{2}_{n}+l^{2}-\epsilon^{2}_{ql}\right)}{\!\left(q^{2}_{n}+q^{2}-2q\cdot l\right)^{2}+4l^{2}q^{2}_{n}}=\frac{2l\!\left(q^{2}_{n}-q^{2}+2q\cdot l\right)}{\!\left(q^{2}_{n}+q^{2}-2q\cdot l\right)^{2}+4l^{2}q^{2}_{n}}
\end{align}
and so on. Such simplifications allow the angular integrals to be
done easily, yielding 
\begin{align}
\bar{\Pi}^{T}_{th}\!\left(q_{n,}q\right) & =\!\left(\frac{\alpha}{2\pi f}\right)^{2}\!\left(\frac{T^{2}}{24}\right)\frac{1}{q^{2}}\!\left(3q^{4}+q^{4}_{n}\right)\nonumber \\
 & \quad+\!\left(\frac{\alpha}{2\pi f}\right)^{2}\int\frac{dy}{\!\left(2\pi\right)^{2}}\,n_{B}\!\left(y\right)\frac{1}{8q^{3}}\left\{ \!\left(q^{2}+q^{2}_{n}\right)^{2}8q_{n}y\!\left[\cot^{-1}\!\!\left(\frac{2q_{n}y}{q^{2}+q^{2}_{n}-2qy}\right)\right.\notag\right\} \nonumber \\
 & \quad-4\!\left(D-2\right)g^{2}N_{c}\frac{T^{2}}{24\times4}\!\left(1+\frac{q^{2}_{n}}{q^{2}}\right)\nonumber \\
 & \quad-4\!\left(D-2\right)g^{2}N_{c}\int\frac{dy}{\!\left(2\pi\right)^{2}}\,n_{B}\!\left(y\right)\frac{\!\left(q^{2}+q^{2}_{n}\right)}{32q^{3}}\left[8yq_{n}\!\left(\cot^{-1}\!\!\left(\frac{2yq_{n}}{-2yq+q^{2}+q^{2}_{n}}\right)\right.\notag\right]\nonumber \\
 & \qquad+g^{2}N_{c}\frac{\!\left(D-2\right)^{2}}{2}\frac{T^{2}}{12}.\label{eq:FullThermSE}
\end{align}
The self-energy correction to the retarded Green's function then follows
from the analytic continuation $q_{n}\rightarrow-i\!\left(q_{0}+i0^{+}\right)$
of the above. For $\!\left(q_{0},q\right)\ll T\sim y$, computing
the loop integrals in a hard thermal loop approximation, where the
integrand is approximated by the leading-order term in the $q/y$
expansion, gives 
\begin{align}
\bar{\Pi}^{T}_{th}\!\left(q_{0},q\right) & =\!\left(\frac{\alpha}{2\pi f}\right)^{2}\!\left(\frac{T^{2}}{24}\right)\frac{1}{q^{2}}\!\left(3q^{4}+q^{4}_{0}\right)-\!\left(D-2\right)g^{2}N_{c}\frac{T^{2}}{24}\!\left(1-\frac{q^{2}_{0}}{q^{2}}\right)\nonumber \\
 & +g^{2}T^{2}N_{c}\frac{\!\left(D-2\right)^{2}}{24}-\!\left[\frac{T^{2}}{24}\right]\!\left\{ -\!\left(\frac{\alpha}{2\pi f}\right)^{2}\!\left(q^{2}-q^{2}_{0}\right)+\!\left(D-2\right)g^{2}N_{c}\right\} \times\nonumber \\
 & \frac{\!\left(q^{2}-q^{2}_{0}\right)}{q^{3}}\!\left\{ q_{0}\ln\!\left(\frac{q_{0}+i0^{+}-q}{q_{0}+i0^{+}+q}\right)+q\right\} .\label{eq:FullHTL_alpha_g}
\end{align}

\section{Relating $\dddot{\bar{\phi}}$ to $\dot{\bar{\phi}}$}

\label{sec:dddphi2dphi}

Here we compute $s\left(\epsilon_{v},\eta_{v}\right)$ in eq. (\ref{eq:dddphi2dphi})
relating the triple time derivative $\dddot{\bar{\phi}}$ to $\dot{\bar{\phi}}$.
In warm inflation scenarios, the background inflaton equation under
slow roll is 
\begin{equation}
3H\left(1+Q\right)\dot{\phi}+V'\left(\phi\right)\approx0\label{eq:Sloroll2}
\end{equation}
and corresponding slow roll parameters are taken to be 
\begin{align}
\epsilon_{H} & \equiv-\frac{\dot{H}}{H^{2}}\approx+\frac{M^{2}_{pl}}{2}\left(\frac{V'\left(\phi\right)}{V\left(\phi\right)}\right)^{2}\frac{1}{\left(1+Q\right)}\equiv\epsilon_{v},\qquad\eta_{v}\equiv\frac{M^{2}_{pl}}{\left(1+Q\right)}\frac{V''\left(\phi\right)}{V\left(\phi\right)}.
\end{align}
In terms of these, $\dddot{\bar{\phi}}$ may be related to $\dot{\bar{\phi}}$
by taking time derivatives of eq. (\ref{eq:Sloroll2}) and noting
that $\dot{Q}$ can also be expressed in terms of the slow roll parameters
using the energy conservation equation 
\begin{equation}
\Gamma\dot{\bar{\phi}}^{2}\approx4H\rho_{R}.\label{eq:QuasiECon}
\end{equation}
In the above we have set $\dot{\rho}_{R}\approx0$ given the quasi-static
evolution of the thermal bath. For MWI within the parameter regime
considered here 
\begin{equation}
\Gamma_{\text{eff}}\approx\Gamma_{\text{sphal}}=\frac{\kappa\alpha^{5}T^{3}}{f^{2}}.\label{eq:SphaRateDiff}
\end{equation}
Combining the logarithmic time derivatives of eq. (\ref{eq:Sloroll2},
\ref{eq:SphaRateDiff}, and \ref{eq:QuasiECon}), which gives 
\begin{equation}
\frac{d\ln\left(\dot{\bar{\phi}}\right)}{dN}=\epsilon_{v}-\eta_{v}-\frac{Q}{1+Q}\frac{d\ln Q}{dN}\label{eq:lnphidN}
\end{equation}
\begin{equation}
4\frac{d\ln T}{dN}\approx2\frac{d\ln\dot{\bar{\phi}}}{dN}+\frac{d\ln Q}{dN}
\end{equation}
and 
\begin{equation}
\frac{d\ln Q}{dN}=\frac{\dot{Q}}{HQ}\approx\frac{\dot{Q}_{sphal}}{HQ_{sphal}}=\epsilon_{v}+3\frac{d\ln T}{dN}
\end{equation}
we get 
\begin{equation}
\frac{d\ln Q}{dN}=\frac{2\left(1+Q\right)}{1+7Q}\left(5\epsilon_{v}-3\eta_{v}\right).
\end{equation}
Using these in the derivatives of the slow roll parameters we get
\begin{equation}
\dot{\epsilon}_{v}=-2H\left[\left(\frac{1+4Q}{1+7Q}\right)\eta_{v}\epsilon_{v}-\left(\frac{2+9Q}{1+7Q}\right)\epsilon^{2}_{v}\right]
\end{equation}
and 
\begin{equation}
\dot{\eta}_{v}=2H\left[\eta_{v}\epsilon_{v}\left(\frac{1+2Q}{1+7Q}\right)+3\frac{Q}{\left(1+7Q\right)}\eta^{2}_{v}\right]
\end{equation}
which may be used to simplify the time derivative of eq. (\ref{eq:lnphidN})
to get 
\begin{align}
\dddot{\phi} & =\frac{H^{2}\dot{\phi}}{\left(1+7Q\right)^{3}}\left[(1+Q)^{2}\left(1-35Q\right)\eta^{2}_{v}+\left(231Q^{3}+181Q^{2}+89Q-5\right)\eta_{v}\epsilon_{v}\right.\\
 & \left.-4\left(42Q^{3}+38Q^{2}+19Q-1\right)\epsilon^{2}_{v}\right].
\end{align}
$s\left(\epsilon_{v},\eta_{v}\right)$ can be seen to limit to eq.
(\ref{eq:Qlarge}) and eq. (\ref{eq:Qsmall}) for $Q\gg1$ and $Q\ll1$
respectively.

\section{Comparison to Dissipation in Out-of-Equilibrium Distributions}

\label{sec:Comparison-to-Dissipation}

When the system is far from equilibrium, it evolves towards it through
helicity-violating interactions as noted in \cite{Broadberry:2025hep}
on time scales $\sim\left(\dot{\phi}/f\right)^{-1}$.~ This may be
understood to be driven by an applied chemical potential for the helicity
implied by the leading order term in the axion gauge coupling, that
is $\propto-\dot{\phi}A\partial A$. This appendix clarifies what
is meant by considering helicity as a charge: unlike conserved charges
corresponding to internal symmetries, helicity is associated with
spacetime transformations, and naturally defined in the momentum space.
Consequently, its representation is non-local in position space.

The charge 
\begin{equation}
\bar{Q}=\int d^{3}x\,\epsilon^{ijk}A^{a}_{i}\partial_{j}A^{a}_{k}
\end{equation}
is not conserved by the free $\!\left(g\rightarrow0\right)$ gauge
field equations of motion. On the other hand, the quantized operator
$\hat{\bar{Q}}$ does measure the total helicity of Hamiltonian eigenstates.
Therefore, we may attempt to find a symmetry transformation of the
free (abelian) gauge theory (without the CS interaction term) whose
conserved current is the total helicity and is related to $\bar{Q}$.
In the Coulomb gauge, where the free Lagrangian is 
\begin{equation}
\mathcal{L}_{g=0}=\frac{1}{2}\!\left(\dot{A}^{a,T}\dot{A}^{a,T}-\!\left(\nabla\times A^{aT}\right)\cdot\!\left(\nabla\times A^{aT}\right)\right),\label{eq:freeQuad}
\end{equation}
such a symmetry transformation is given by 
\begin{equation}
\delta A^{a}_{i}=-\beta\nabla^{-2}\nabla\times\!\left(\dot{A}^{a}_{i}\right),\label{eq:SymmTrans}
\end{equation}
which, importantly, is non-local in position. The associated conserved
current can be found to be 
\begin{equation}
Q=\frac{1}{2}\int d^{3}x\,\!\left(\dot{\vec{A}}^{a,T}\cdot\nabla^{-2}\nabla\times\dot{\vec{A}}^{a,T}-A^{a,T}\cdot\nabla\times A^{a,T}\right).
\end{equation}

Now we may attempt to remove the $\dot{\phi}$ interaction term in
eq.~(\ref{eq:GGtExp}) with the following field redefinition of the
gauge fields 
\begin{equation}
A^{a}_{i}\rightarrow A^{a}_{i}-\frac{\alpha\dot{\phi}t}{f}\nabla^{-2}\nabla\times\!\left(\dot{A}^{a}_{i}\right).\label{eq:FieldRedef}
\end{equation}
where we promoted $\beta$ in eq.~(\ref{eq:SymmTrans}) to a function
of time taking $\beta\!\left(t\right)=\alpha\phi\!\left(t\right)/f=\alpha\dot{\phi}t/f$.
This transforms the quadratic terms of eq.~(\ref{eq:GGtExp}) (in
Coulomb gauge) to 
\begin{align}
S_{(2)} & =\int d^{4}x\,\left\{ \frac{1}{2}\!\left[\!\left(\partial_{0}A^{a}_{i}\right)^{2}-\!\left(\partial_{j}A^{a}_{i}\right)^{2}\right]-\frac{\alpha\dot{\phi}}{2f}\!\left\{ A^{a}\cdot\!\left(\nabla\times A^{a}\right)+\dot{A}^{a}\nabla^{-2}\!\left(\nabla\times\dot{A}^{a}\right)\right\} \right\} \label{eq:TranAct}
\end{align}
and we note that the $\dot{\phi}$-dependent terms above cancel only
when the gauge fields satisfy 
\begin{equation}
\ddot{A}^{a}-\nabla^{2}A^{a}=0,\label{eq:Onshe}
\end{equation}
that is, are on shell. This is easily seen by using integration by
parts on the second term proportional to $\dot{\phi}$ in eq.~(\ref{eq:TranAct}).
Therefore, in this scenario, because the quadratic term in the Chern-Simons
interaction proportional to $\bar{Q}$ does not by itself correspond
to a conserved current, the $\dot{\phi}$ dependence cannot be removed
off-shell. In addition, the field redefinition in eq.~(\ref{eq:FieldRedef})
now introduces non-local factors in the gluon interaction terms. While
these naively seem to pose a threat to the argument in \cite{Broadberry:2025hep},
we note that on the helicity eigenstates, defined in position space
as 
\begin{equation}
A^{\pm}=A\pm\frac{\nabla\times A}{\sqrt{-\nabla^{2}}}\qquad\text{such that}\qquad\frac{\nabla\times A^{\pm}}{\sqrt{-\nabla^{2}}}=\pm A^{\pm},
\end{equation}
the field redefinition takes the form 
\begin{equation}
A^{\pm}\!\left(x\right)\rightarrow\!\left(1-\frac{\alpha\dot{\phi}t}{f}\frac{d}{dt}\nabla^{-2}\nabla\times\right)A^{\pm}\!\left(x\right)=\!\left(1\pm is\frac{\alpha\dot{\phi}t}{f}\right)A^{\pm}\!\left(x\right)\approx\exp\!\left(\pm is\frac{\alpha\dot{\phi}t}{f}\right)A^{\pm}\!\left(x\right)
\end{equation}
for on-shell gauge fields satisfying eq.~(\ref{eq:Onshe}): 
\begin{equation}
\ddot{A}^{a}-\nabla^{2}A=0\implies\!\left(\frac{d}{dt}+i\sqrt{-\nabla^{2}}\right)\!\left(\frac{d}{dt}-i\sqrt{-\nabla^{2}}\right)A=0\implies\dot{A}=s\sqrt{-\nabla^{2}}A,\quad s=\pm1.
\end{equation}
That is, for on-shell gluon states, the effect of the field redefinition
is a phase rotation as in the scalar example of \cite{Broadberry:2025hep}.

As an interesting aside, we note that on on-shell field configurations,
the above symmetry is equivalent to the `duality symmetry' of electromagnetism.
The duality symmetry acts on the field strengths $F_{\mu\nu}$ and
$\tilde{F}_{\mu\nu}$ as 
\begin{equation}
F^{\pm}_{\mu\nu}\rightarrow F^{\pm}_{\mu\nu}\exp\!\left(\pm i\alpha\right)
\end{equation}
where 
\begin{equation}
F^{\pm}_{\mu\nu}=F_{\mu\nu}\pm i\tilde{F}_{\mu\nu},\qquad\tilde{F}_{\mu\nu}=\frac{1}{2}\epsilon_{\mu\nu\rho\sigma}F^{\rho\sigma}
\end{equation}
and thereby on the electric and magnetic fields as 
\begin{equation}
\begin{pmatrix}E'\\
B'
\end{pmatrix}=\begin{pmatrix}\cos\alpha & -\sin\alpha\\
\sin\alpha & \cos\alpha
\end{pmatrix}\begin{pmatrix}E\\
B
\end{pmatrix}.
\end{equation}

\bibliographystyle{jhep}
\bibliography{MWI_refs}

@article{KHLEBNIKOV1988885,
	author = {S.Yu. Khlebnikov and M.E. Shaposhnikov},
	doi = {https://doi.org/10.1016/0550-3213(88)90133-2},
	issn = {0550-3213},
	journal = {Nuclear Physics B},
	number = {4},
	pages = {885-912},
	title = {The statistical theory of anomalous fermion number non-conservation},
	url = {https://www.sciencedirect.com/science/article/pii/0550321388901332},
	volume = {308},
	year = {1988}}

@article{Laine:2022ytc,
	archiveprefix = {arXiv},
	author = {Laine, M. and Niemi, L. and Procacci, S. and Rummukainen, K.},
	doi = {10.1007/JHEP11(2022)126},
	eprint = {2209.13804},
	journal = {JHEP},
	pages = {126},
	primaryclass = {hep-ph},
	title = {{Shape of the hot topological charge density spectral function}},
	volume = {11},
	year = {2022}}

@article{Cucchieri:2000cy,
	archiveprefix = {arXiv},
	author = {Cucchieri, A. and Karsch, F. and Petreczky, P.},
	doi = {10.1016/S0370-2693(00)01331-9},
	eprint = {hep-lat/0004027},
	journal = {Phys. Lett. B},
	pages = {80--84},
	reportnumber = {BI-TP-2000-11},
	title = {{Magnetic screening in hot nonAbelian gauge theory}},
	volume = {497},
	year = {2001}}

@article{Bastero-Gil:2014jsa,
	archiveprefix = {arXiv},
	author = {Bastero-Gil, Mar and Berera, Arjun and Moss, Ian G. and Ramos, Rudnei O.},
	doi = {10.1088/1475-7516/2014/05/004},
	eprint = {1401.1149},
	journal = {JCAP},
	pages = {004},
	primaryclass = {astro-ph.CO},
	title = {{Cosmological fluctuations of a random field and radiation fluid}},
	volume = {05},
	year = {2014}}

@article{Berghaus:2024zfg,
	archiveprefix = {arXiv},
	author = {Berghaus, Kim V. and Forslund, Matthew and Guevarra, Mark Vincent},
	doi = {10.1088/1475-7516/2024/10/103},
	eprint = {2402.13535},
	journal = {JCAP},
	pages = {103},
	primaryclass = {hep-ph},
	title = {{Warm inflation with a heavy QCD axion}},
	volume = {10},
	year = {2024}}

@article{PhysRevLett.64.1338,
	author = {Braaten, Eric and Pisarski, Robert D.},
	doi = {10.1103/PhysRevLett.64.1338},
	issue = {12},
	journal = {Phys. Rev. Lett.},
	month = {Mar},
	numpages = {0},
	pages = {1338--1341},
	publisher = {American Physical Society},
	title = {Resummation and gauge invariance of the gluon damping rate in hot QCD},
	url = {https://link.aps.org/doi/10.1103/PhysRevLett.64.1338},
	volume = {64},
	year = {1990}}

@article{Pisarski:1992wqc,
	archiveprefix = {arXiv},
	author = {Pisarski, Robert D.},
	eprint = {hep-ph/9302242},
	month = {12},
	reportnumber = {BNL-P-1-92, BNL-48816, BNL-P-92-01},
	title = {{Damping rates for moving particles in hot QCD}},
	year = {1992}}

@article{PhysRevD.47.R769,
	author = {Heiselberg, H. and Pethick, C. J.},
	doi = {10.1103/PhysRevD.47.R769},
	issue = {3},
	journal = {Phys. Rev. D},
	month = {Feb},
	numpages = {0},
	pages = {R769(R)--R771(R)},
	publisher = {American Physical Society},
	title = {Quark and gluon relaxation in quark-gluon plasmas},
	url = {https://link.aps.org/doi/10.1103/PhysRevD.47.R769},
	volume = {47},
	year = {1993}}

@article{Braaten:1990it,
	author = {Braaten, Eric and Pisarski, Robert D.},
	doi = {10.1103/PhysRevD.42.2156},
	journal = {Phys. Rev. D},
	pages = {2156--2160},
	reportnumber = {BNL-44553, NUHEP-TH-90-15},
	title = {{Calculation of the gluon damping rate in hot QCD}},
	volume = {42},
	year = {1990}}

@article{Rebhan:1992ca,
	archiveprefix = {arXiv},
	author = {Rebhan, Anton},
	doi = {10.1103/PhysRevD.46.482},
	eprint = {hep-ph/9203211},
	journal = {Phys. Rev. D},
	pages = {482--483},
	reportnumber = {CERN-TH-6375-92},
	title = {{Comment on `Damping of energetic gluons and quarks in high temperature QCD'}},
	volume = {46},
	year = {1992}}

@article{Burgess:1991wc,
	archiveprefix = {arXiv},
	author = {Burgess, C. P. and Marini, A. L.},
	doi = {10.1103/PhysRevD.45.R17},
	eprint = {hep-th/9109051},
	journal = {Phys. Rev. D},
	pages = {17--20},
	reportnumber = {MCGILL-91-24},
	title = {{The Damping of energetic gluons and quarks in high temperature QCD}},
	volume = {45},
	year = {1992}}

@article{Blaizot:1993be,
	archiveprefix = {arXiv},
	author = {Blaizot, Jean Paul and Iancu, Edmond},
	doi = {10.1016/0550-3213(94)90486-3},
	eprint = {hep-ph/9306294},
	journal = {Nucl. Phys. B},
	pages = {608--673},
	reportnumber = {SACLAY-SPH-T-93-064},
	title = {{Soft collective excitations in hot gauge theories}},
	volume = {417},
	year = {1994}}

@inproceedings{Bodeker:1998rg,
	archiveprefix = {arXiv},
	author = {Bodeker, Dietrich},
	booktitle = {{5th International Workshop on Thermal Field Theories and Their Applications}},
	eprint = {hep-ph/9810265},
	month = {10},
	reportnumber = {HD-THEP-98-35, NBI-HE-98-29},
	title = {{An effective theory for hot nonAbelian dynamics}},
	year = {1998}}

@article{Morikawa:1984dz,
	author = {Morikawa, Masahiro and Sasaki, Misao},
	doi = {10.1143/PTP.72.782},
	journal = {Prog. Theor. Phys.},
	pages = {782},
	reportnumber = {KUNS 735},
	title = {{Entropy Production in the Inflationary Universe}},
	volume = {72},
	year = {1984}}

@article{Moss:2006gt,
	archiveprefix = {arXiv},
	author = {Moss, Ian G and Xiong, Chun},
	eprint = {hep-ph/0603266},
	month = {3},
	title = {{Dissipation coefficients for supersymmetric inflatonary models}},
	year = {2006}}

@article{Berera:1995ie,
	archiveprefix = {arXiv},
	author = {Berera, Arjun},
	eprint = {astro-ph/9509049},
	journal = {Phys. Rev. Lett.},
	pages = {3218--3221},
	title = {Warm inflation},
	volume = {75},
	year = {1995}}

@article{Berera:1995wh,
	archiveprefix = {arXiv},
	author = {Berera, Arjun and Fang, Li-Zhi},
	eprint = {astro-ph/9501024},
	journal = {Phys. Rev. Lett.},
	pages = {1912--1915},
	title = {Thermally induced density perturbations in the inflation era},
	volume = {74},
	year = {1995}}

@article{Berera:1996nv,
	archiveprefix = {arXiv},
	author = {Berera, Arjun},
	eprint = {hep-th/9601134},
	journal = {Phys. Rev. D},
	pages = {2519--2534},
	title = {Thermal properties of an inflationary universe},
	volume = {54},
	year = {1996}}

@article{Berera:2000xe,
	archiveprefix = {arXiv},
	author = {Berera, Arjun},
	eprint = {hep-ph/9904409},
	journal = {Nucl. Phys. B},
	pages = {666--714},
	title = {Warm inflation at arbitrary adiabaticity: a model, an existence proof for inflationary dynamics in quantum field theory},
	volume = {585},
	year = {2000}}

@article{Berera:2008ar,
	archiveprefix = {arXiv},
	author = {Berera, Arjun and Moss, Ian G. and Ramos, Rudnei O.},
	eprint = {0808.1855},
	journal = {Rept. Prog. Phys.},
	pages = {026901},
	title = {Warm inflation and its microphysical basis},
	volume = {72},
	year = {2009}}

@article{Berghaus:2019whh,
	archiveprefix = {arXiv},
	author = {Berghaus, Kim V. and Graham, Peter W. and Kaplan, David E.},
	eprint = {1910.07525},
	journal = {JCAP},
	note = {Erratum: JCAP 10, E02 (2023)},
	pages = {034},
	title = {Minimal warm inflation},
	volume = {03},
	year = {2020}}

@article{Moore:2010jd,
	archiveprefix = {arXiv},
	author = {Moore, Guy D. and Tassler, Marcus},
	eprint = {1011.1167},
	journal = {JHEP},
	pages = {105},
	title = {The sphaleron rate in {SU}($N$) gauge theory},
	volume = {02},
	year = {2011}}

@article{Arnold:1987mh,
	author = {Arnold, Peter B. and McLerran, Larry},
	journal = {Phys. Rev. D},
	pages = {581},
	title = {Sphalerons, small fluctuations, and baryon number violation in electroweak theory},
	volume = {36},
	year = {1987}}

@article{Laine:2016hma,
	archiveprefix = {arXiv},
	author = {Laine, Mikko and Vuorinen, Aleksi},
	eprint = {1701.01554},
	journal = {Lect. Notes Phys.},
	pages = {1--281},
	title = {Basics of thermal field theory},
	volume = {925},
	year = {2016}}

@article{Berera:2002sp,
	archiveprefix = {arXiv},
	author = {Berera, Arjun and Ramos, Rudnei O.},
	eprint = {hep-ph/0210301},
	journal = {Phys. Lett. B},
	pages = {294--304},
	title = {Construction of a robust warm inflation mechanism},
	volume = {567},
	year = {2003}}

@article{Bastero-Gil:2010pb,
	archiveprefix = {arXiv},
	author = {Bastero-Gil, Mar and Berera, Arjun and Ramos, Rudnei O.},
	eprint = {1008.1929},
	journal = {JCAP},
	pages = {033},
	title = {Dissipation coefficients from scalar and fermion quantum field interactions},
	volume = {09},
	year = {2011}}

@article{graham2009density,
	archiveprefix = {arXiv},
	author = {Graham, Chris and Moss, Ian G},
	eprint = {0905.3500},
	journal = {Journal of Cosmology and Astroparticle Physics},
	pages = {013},
	title = {Density fluctuations from warm inflation},
	volume = {07},
	year = {2009}}

@article{Bastero-Gil:2011rva,
	archiveprefix = {arXiv},
	author = {Bastero-Gil, Mar and Berera, Arjun and Ramos, Rudnei O.},
	eprint = {1106.0701},
	journal = {JCAP},
	pages = {030},
	title = {Shear viscous effects on the primordial power spectrum from warm inflation},
	volume = {07},
	year = {2011}}

@article{Yokoyama:1998ju,
	archiveprefix = {arXiv},
	author = {Yokoyama, Jun'ichi and Linde, Andrei D.},
	eprint = {hep-ph/9809409},
	journal = {Phys. Rev. D},
	pages = {083509},
	title = {Is warm inflation possible?},
	volume = {60},
	year = {1999}}

@article{Bastero-Gil:2016qru,
	archiveprefix = {arXiv},
	author = {Bastero-Gil, Mar and Berera, Arjun and Ramos, Rudnei O. and Rosa, Jo\~ao G.},
	eprint = {1604.08838},
	journal = {Phys. Rev. Lett.},
	number = {15},
	pages = {151301},
	title = {Warm little inflaton},
	volume = {117},
	year = {2016}}

@article{Bastero-Gil:2019gao,
	archiveprefix = {arXiv},
	author = {Bastero-Gil, Mar and Berera, Arjun and Ramos, Rudnei O. and Rosa, Jo\~ao G.},
	eprint = {1907.13410},
	journal = {Phys. Lett. B},
	pages = {136055},
	title = {Towards a reliable effective field theory of inflation},
	volume = {813},
	year = {2021}}

@article{Visinelli:2011jy,
	archiveprefix = {arXiv},
	author = {Visinelli, Luca},
	eprint = {1107.3523},
	journal = {JCAP},
	pages = {013},
	title = {Natural warm inflation},
	volume = {09},
	year = {2011}}

@article{Kamali:2019ppi,
	archiveprefix = {arXiv},
	author = {Kamali, Vahid},
	eprint = {1901.01897},
	journal = {Phys. Rev. D},
	number = {4},
	pages = {043520},
	title = {Warm pseudoscalar inflation},
	volume = {100},
	year = {2019}}

@article{Ferreira:2017lnd,
	archiveprefix = {arXiv},
	author = {Ferreira, Ricardo Z. and Notari, Alessio},
	eprint = {1706.00373},
	journal = {JCAP},
	pages = {007},
	title = {Thermalized axion inflation},
	volume = {09},
	year = {2017}}

@article{Broadberry:2025hep,
	archiveprefix = {arXiv},
	author = {Broadberry, Edward and Hook, Anson and Mondal, Sagnik},
	eprint = {2505.07943},
	primaryclass = {hep-ph},
	title = {Warm inflation with pseudo-scalar couplings},
	year = {2025}}

@article{Berghaus:2025wdx,
	archiveprefix = {arXiv},
	author = {Berghaus, Kim V. and Drewes, Marco and Zell, Sebastian},
	eprint = {2503.18829},
	primaryclass = {hep-ph},
	title = {Warm inflation with the {S}tandard {M}odel},
	year = {2025}}

@article{Gross:1980br,
	author = {Gross, David J. and Pisarski, Robert D. and Yaffe, Laurence G.},
	journal = {Rev. Mod. Phys.},
	pages = {43},
	title = {QCD and instantons at finite temperature},
	volume = {53},
	year = {1981}}

@article{Frison:2016vuc,
	archiveprefix = {arXiv},
	author = {Frison, Julien and Kitano, Ryuichiro and Matsufuru, Hideo and Mori, Satsuki and Yamada, Norikazu},
	eprint = {1606.07175},
	journal = {JHEP},
	pages = {021},
	title = {Topological susceptibility at high temperature on the lattice},
	volume = {09},
	year = {2016}}

@article{Ramos:2013nsa,
	archiveprefix = {arXiv},
	author = {Ramos, Rudnei O. and da Silva, L. A.},
	eprint = {1302.3544},
	journal = {JCAP},
	pages = {032},
	title = {Power spectrum for inflation models with quantum and thermal noises},
	volume = {03},
	year = {2013}}

@article{Mishra:2011ph,
	archiveprefix = {arXiv},
	author = {Mishra, Hiranmaya and Mohanty, Subhendra and Nautiyal, Akhilesh},
	eprint = {1106.3039},
	journal = {Phys. Lett. B},
	pages = {245--250},
	title = {Warm natural inflation},
	volume = {710},
	year = {2012}}

@article{Guth:1980zm,
	author = {Guth, Alan H.},
	journal = {Phys. Rev. D},
	pages = {347--356},
	title = {The inflationary universe: A possible solution to the horizon and flatness problems},
	volume = {23},
	year = {1981}}

@article{Linde:1981mu,
	author = {Linde, Andrei D.},
	journal = {Phys. Lett. B},
	pages = {389--393},
	title = {A new inflationary universe scenario: A possible solution of the horizon, flatness, homogeneity, isotropy and primordial monopole problems},
	volume = {108},
	year = {1982}}

@article{BasteroGil:2009ec,
	archiveprefix = {arXiv},
	author = {Bastero-Gil, Mar and Berera, Arjun},
	eprint = {0902.0521},
	journal = {Int. J. Mod. Phys. A},
	pages = {2207--2240},
	title = {Warm inflation model building},
	volume = {24},
	year = {2009}}

@article{Kamali:2023lzq,
	archiveprefix = {arXiv},
	author = {Kamali, Vahid and Motaharfar, Meysam and Ramos, Rudnei O.},
	eprint = {2302.02827},
	journal = {Universe},
	number = {3},
	pages = {124},
	title = {Recent developments in warm inflation},
	volume = {9},
	year = {2023}}

@article{Hall:2003zp,
	archiveprefix = {arXiv},
	author = {Hall, Lisa M. H. and Moss, Ian G. and Berera, Arjun},
	eprint = {astro-ph/0305015},
	journal = {Phys. Rev. D},
	pages = {083525},
	title = {Scalar perturbation spectra from warm inflation},
	volume = {69},
	year = {2004}}

@article{Laine:2021ego,
	archiveprefix = {arXiv},
	author = {Laine, Mikko and Procacci, Sacha},
	eprint = {2102.09913},
	journal = {JCAP},
	pages = {031},
	title = {Minimal warm inflation with complete medium response},
	volume = {06},
	year = {2021}}

@article{Drewes:2023khq,
	archiveprefix = {arXiv},
	author = {Drewes, Marco and Zell, Sebastian},
	eprint = {2312.13739},
	journal = {JCAP},
	pages = {038},
	title = {On sphaleron heating in the presence of fermions},
	volume = {06},
	year = {2024}}

@article{McLerran:1990de,
	author = {McLerran, Larry D. and Mottola, Emil and Shaposhnikov, Mikhail E.},
	journal = {Phys. Rev. D},
	pages = {2027--2035},
	title = {Sphalerons and axion dynamics in high temperature {QCD}},
	volume = {43},
	year = {1991}}

@article{Arnold:1996dy,
	archiveprefix = {arXiv},
	author = {Arnold, Peter Brockway and Son, Dam and Yaffe, Laurence G.},
	eprint = {hep-ph/9609481},
	journal = {Phys. Rev. D},
	pages = {6264--6273},
	title = {The hot baryon violation rate is {$O(\alpha_w^5 T^4)$}},
	volume = {55},
	year = {1997}}

@article{Linde:1980ts,
	author = {Linde, Andrei D.},
	journal = {Phys. Lett. B},
	pages = {289--292},
	title = {Infrared problem in the thermodynamics of the {Y}ang-{M}ills gas},
	volume = {96},
	year = {1980}}

@article{Pisarski:1988vd,
	author = {Pisarski, Robert D.},
	journal = {Phys. Rev. Lett.},
	pages = {1129},
	title = {Scattering amplitudes in hot gauge theories},
	volume = {63},
	year = {1989}}

@article{Braaten:1989mz,
	author = {Braaten, Eric and Pisarski, Robert D.},
	journal = {Nucl. Phys. B},
	pages = {569--634},
	title = {Soft amplitudes in hot gauge theories: A general analysis},
	volume = {337},
	year = {1990}}

@article{Blaizot:2001nr,
	archiveprefix = {arXiv},
	author = {Blaizot, Jean-Paul and Iancu, Edmond},
	eprint = {hep-ph/0101103},
	journal = {Phys. Rept.},
	pages = {355--528},
	title = {The quark-gluon plasma: Collective dynamics and hard thermal loops},
	volume = {359},
	year = {2002}}

@book{Kapusta:2006pm,
	author = {Kapusta, Joseph I. and Gale, Charles},
	publisher = {Cambridge University Press},
	series = {Cambridge Monographs on Mathematical Physics},
	title = {Finite-temperature field theory: Principles and applications},
	year = {2006}}

@book{Bellac:2011kqa,
	author = {Le Bellac, Michel},
	publisher = {Cambridge University Press},
	series = {Cambridge Monographs on Mathematical Physics},
	title = {Thermal field theory},
	year = {1996}}

\end{document}